\documentclass[pdflatex,sn-mathphys-num,referee]{sn-jnl}

\usepackage{graphicx}%
\usepackage{upgreek}%
\usepackage{multirow}%
\usepackage{amsmath,amssymb,amsfonts}%
\usepackage{amsthm}%
\usepackage{mathrsfs}%
\usepackage[title]{appendix}%
\usepackage{xcolor}%
\usepackage{textcomp}%
\usepackage{manyfoot}%
\usepackage{rotating}%
\usepackage{booktabs}%
\usepackage{algorithm}%
\usepackage{algorithmicx}%
\usepackage{algpseudocode}%
\usepackage{listings}%
\usepackage{placeins}%
\usepackage{nameref}

\theoremstyle{thmstyleone}%
\theoremstyle{thmstyletwo}%

\theoremstyle{thmstylethree}%

\begin{document}

\title[Article Title]{Quantum Limits of Electronic Transport in Nanostructured
	Macroscopic Conductors}


\author*[1]{\fnm{Agnieszka E.} \sur{Lekawa-Raus}}\email{agnieszka.raus@pw.edu.pl}

\author[2,3,4]{\fnm{John S.} \sur{Bulmer}}

\author[5]{\fnm{Teresa} \sur{Kulka}}

\author[6,7]{\fnm{Magdalena} \sur{Marganska}}

\author[8]{\fnm{Nick} \sur{Papior}}

\author[9]{\fnm{Dwight G.} \sur{Rickel}}

\author[9]{\fnm{Fedor F.} \sur{Balakirev}}

\author[5,10]{\fnm{Jacek A.} \sur{Majewski}}

\author[2,11]{\fnm{Krzysztof} \sur{Koziol}}

\author*[12,13]{\fnm{Karolina Z.} \sur{Milowska}}\email{karolina.milowska@gmail.com}




\affil*[1]{\orgdiv{Centre for Advanced Materials and Technologies (CEZAMAT)}, \orgname{Warsaw University of Technology}, \orgaddress{\street{ul. Poleczki 19}, \city{Warsaw}, \postcode{02-822},  \country{Poland}}}

\affil[2]{\orgdiv{Department of Materials Science and Metallurgy}, \orgname{University of Cambridge}, \orgaddress{\street{27 Charles Babbage Rd}, \city{Cambridge}, \postcode{CB3 0FS}, \country{United Kingdom}}}

\affil[3]{\orgdiv{Aerospace Systems Directorate}, \orgname{Air Force Research Laboratory, Wright-Patterson Air Force Base}, \orgaddress{\postcode{45433}, \state{OH},  \country{USA}}}

\affil[4]{\orgname{University of Cincinnati}, \orgaddress{\street{598 Rhodes Hall}, \city{Cincinnati}, \postcode{P.O. Box 210072}, \state{OH},  \country{USA}}}

\affil[5]{\orgdiv{Institute of Theoretical Physics, Faculty of Physics}, \orgname{University of Warsaw}, \orgaddress{\street{ul. Pasteura 5}, \city{Warsaw}, \postcode{02-093},  \country{Poland}}}

\affil[6]{\orgdiv{Institute for Theoretical Physics}, \orgname{University of Regensburg},  \orgaddress{\city{Regensburg}, \postcode{93040},  \country{Germany}}}

\affil[7]{\orgdiv{Institute of Theoretical Physics}, \orgname{Wroclaw University of Science and Technology}, \orgaddress{\street{ul. Wybrzeze Wyspianskiego 27}, \city{Wroclaw}, \postcode{50-370},  \country{Poland}}}

\affil[8]{\orgdiv{DTU Computing Center, Department of Applied Mathematics and Computer Science}, \orgname{Technical University of Denmark},  \orgaddress{\city{Kongens Lyngby}, \postcode{DK-2800},  \country{Denmark}}}

\affil[9]{\orgname{National High Magnetic Field Laboratory, Los Alamos National Laboratory}, \orgaddress{\city{Los Alamos}, \postcode{87545}, \state{NM},  \country{USA}}}

\affil[10]{\orgdiv{Terahertz Research and Application Center CENTERA2, Center of Advanced Materials and Technologies CEZAMAT }, \orgname{Warsaw University of Technology}, \orgaddress{\street{ul. Poleczki 19}, \city{Warsaw}, \postcode{02-822},  \country{Poland}}}

\affil[11]{\orgdiv{Enhanced Composites and Structures Centre}, \orgname{Cranfield University}, \orgaddress{\street{Bld 56}, \city{Cranfield, Bedfordshire}, \postcode{MK43 0AL}, \country{United Kingdom}}}

\affil*[12]{\orgname{CIC nanoGUNE}, \orgaddress{\street{Tolosa Hiribidea 76}, \city{Donostia-San Sebastian}, \postcode{20018}, \country{Spain}}}

\affil*[13]{\orgname{Ikerbasque, Basque Foundation for Science}, \orgaddress{\street{Plaza Euskadi 5}, \city{Bilbao}, \postcode{48009}, \country{Spain}}}





\abstract{%
Macroscopic assemblies of one- and two-dimensional materials promise to translate nanoscale electronic properties into device-scale performance, yet the microscopic principles governing charge transport in such networks remain unresolved. In these systems, conductivity is often interpreted using phenomenological models that do not explicitly connect electronic structure to macroscopic magnetotransport. Here we develop a unified atomistic framework that links quantum-coherent transport, thermal disorder and magnetic-field effects, and combine it with ultrahigh-field magnetotransport measurements up to 60 T over a broad temperature range on carbon nanotube fibres. We show that positive magnetoresistance is controlled by junction overlap length, whereas negative magnetoresistance arises predominantly from lattice-mismatched heterojunctions rather than weak localisation alone. Statistical analysis of a large-scale numerical dataset reveals that the experimentally observed positive quadratic magnetoresistance originates from junction transport. These results show that macroscopic transport in disordered low-dimensional networks is governed primarily by junction-level quantum interference rather than solely by defects or doping.	
}




\maketitle


Understanding electronic transport in nanostructured networks remains a central challenge in nanoscience. Macroscopic conductors assembled from nanoscale building blocks--fibres, films, printed tracks, and three-dimensional composites--promise lightweight, flexible, and multifunctional components for applications spanning energy conversion, wearable electronics, aerospace systems, or neuromorphic circuits\cite{vahid2021, fang2020, dong2024, kamyshny2019}  Such conductors are commonly based on networks of carbon nanotubes (CNTs), graphene, MXenes, transition-metal dichalcogenides (TMDs), metal nanowires, and others. Despite rapid progress in synthesis and processing, a fundamental question persists: which structural and electronic factors control charge transport and limit conductivity? Identifying these parameters is essential for designing high-performance conductors and tailoring functionality.

The challenge is twofold. Experimentally, characterising current pathways through complex, disordered networks is difficult; theoretically, interpreting the data is even harder. Recent models treat these materials as resistor networks, with each junction or segment represented as a lumped element and the total resistance computed via percolation or effective-medium approximations. While these approaches capture connectivity and scale to macroscopic dimensions\cite{tarasevich2023,gabnett2024,jagota2020}, they neglect quantum interference and cannot easily incorporate temperature or magnetic-field effects, thereby missing critical physics rooted in junction-level interactions.

A second group of studies relies on semi-empirical fitting of temperature-dependent resistance or magnetoresistance data to classical models that obscure the underlying nanoscopic mechanisms\cite{yu2025, zhang2016, neal2013, bulmer2017}. Among network materials, carbon-nanotube assemblies provide perhaps the most intricate example\cite{lekawa2014}. Depending on synthesis and post-processing, a CNT network can contain single-, double-, and multi-walled nanotubes with broad distributions of length, chirality, defect density, and doping. Additional variability arises from residual catalyst carbonaceous impurities and differences in alignment and densification. These factors affect both intra-tube transport and inter-tube coupling, which together determine the resistance--temperature (R--T) dependence and magnetoresistance (MR) behaviour\cite{yu2025, zorn2021, bulmer2017, lekawa2015a, lekawa2015b}.

At zero magnetic field, R--T curves are generally classified relative to the metal-insulator transition \cite{vavro2005}. Low-conductivity samples--often with short or misaligned nanotubes, insufficient densification, or high multi-walled tube and impurity content--lie on the insulating side and are fitted using one-, two-, or three-dimensional variable-range-hopping (VRH) models. More aligned or doped networks with longer nanotubes usually appear metallic and are interpreted via weak-localisation (WL) or fluctuation-induced-tunnelling (FIT) models\cite{bulmer2017, yanagi2010, yu2025}. Although these models describe certain temperature trends, they are largely phenomenological and provide little microscopic insight into junction physics.

Interpreting magnetotransport is even more demanding. Depending on material, field strength, and temperature, CNT assemblies can exhibit positive MR, negative MR, sign changes, or saturation\cite{yu2025, bulmer2017, bulmer2026}. Classical models associated with VRH or WL often fail to reproduce this diversity, forcing researchers to invoke multiple mechanisms or choose between equally plausible fits. This hampers the ability to draw general conclusions. Most experiments are also limited to $\le$ 15\,T, missing key features that emerge only in ultrahigh fields\cite{bulmer2017}. Finally, differences in synthesis conditions lead to variations in morphology and purity across samples, making cross-study comparisons challenging\cite{bulmer2026, bulmer2017}.

Comparable issues arise in other disordered low-dimensional conductors. Graphene films, MXenes, and TMDs exhibit similarly complex MR--sign changes, non-saturating responses, and strong temperature and field dependence\cite{neal2013, zhang2016, halim2016, xin2023}. These effects reflect a delicate interplay of hopping, quantum interference, and orbital effects, highlighting the limits of classical transport models. Their convergence across materials underscores the need for frameworks that link nanoscale disorder and junction physics to macroscopic conductivity.

To address these challenges, we develop a Landauer--Peierls--Molecular--Dynamics framework that unifies quantum-coherent transport, thermal disorder, and magnetic-field effects within a single atomistic description.  Combining tight-binding transport calculations and Peierls magnetic coupling with molecular-dynamics sampling, the approach captures both quantum interference and structural fluctuations. We apply this method to a model system: CNT fibres fabricated via floating-catalyst chemical-vapour deposition (FC-CVD) using varied feedstocks. Atomistic modelling shows that distinct forms of lattice mismatch at inter-tube junctions govern MR sign and magnitude--beyond the reach of semi-empirical models such as VRH or WL. Consistently, MR measurements up to 60\,T (perpendicular field) across a wide temperature range reveal a quadratic field dependence of positive magnetoresistance, which statistical analysis of the simulations attributes to junction transport. These results not only rationalise the diverse transport signatures observed in CNT networks but also establish principles applicable to other nanostructured conductors. More broadly, they point toward junction-level design of heterogeneous conductors, where local electronic interactions and structural statistics determine macroscopic performance.

\section*{Transport behaviour in experimental CNT fibres }\label{experimental}

This study compares three types of CNT fibres produced by direct spinning from a floating-catalyst chemical vapour deposition (FC-CVD) process (see Section~\nameref{methods} for details). Fibre morphology and properties were varied with the synthesis conditions and post-processing. The first type, previously described in ref.\cite{sundaram2011}, used carbon disulphide and methane to produce fibres composed mainly of single-wall CNTs (SWCNTs) with minimal impurities, as confirmed by clear RBM peaks and a high G:D ratio in Raman spectra (Supplementary Fig.~\ref{FigS1}a,b). These fibres had low linear density ($\sim$0.4\,dtex) and were poorly condensed, as seen in SEM and Raman alignment metrics (Supplementary Figs.~\ref{FigS2}a,b; Supplementary Fig.~\ref{FigS1}d). Two samples from this batch (SWCNT\,1 and SWCNT\,2) were measured in duplicate. After initial MR and R--T measurements, SWCNT\,2 was doped with nitric acid and retested. Doping induced a clear G peak upshift\cite{lepak2018} and pore filling, as seen in Raman and SEM data (Supplementary Figs.~\ref{FigS2}c,d; Supplementary Fig.~\ref{FigS1}c).

\begin{figure}[h!]
	\centering
	\includegraphics[width=0.95\textwidth]{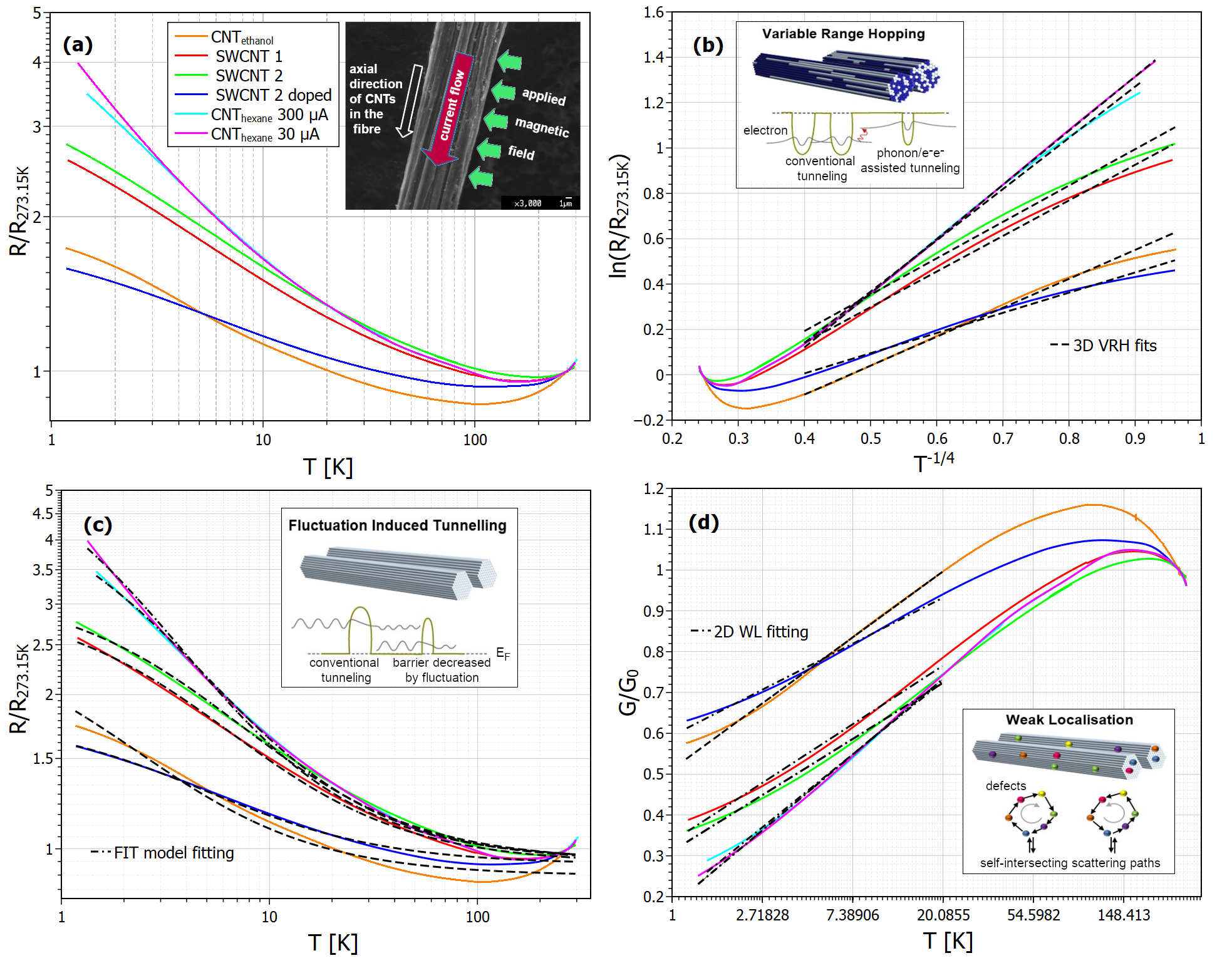}
	\caption{a) Relative resistance vs. temperature plots (measured at zero field) for SWCNT\,1, SWCNT\,2, doped SWCNT\,2, CNTethanol and CNThexane measured using 30~$\upmu$A and 300~$\upmu$A current. Fits of the resistance vs. temperature data to b) 3D variable range hopping (VRH), c) fluctuation-induced tunnelling (FIT) and d) 2D weak localisation (WL) models. All fitting equations and temperature ranges are provided in Supplementary Section~\ref{secFit}}
	\label{Fig1}
\end{figure}

The other fibres were synthesised using thiophene with either ethanol or hexane. The ethanol-derived fibres (CNT$_{\mathrm{ethanol}}$), described earlier in ref.\cite{koziol2007}, contained long single-, double-, and triple-walled CNTs with diverse chiralities and dense packing ($\sim$1\,dtex; Supplementary Figs.~\ref{FigS2}e,f; Supplementary Fig.~\ref{FigS1}b). The hexane-based fibres (CNT$_{\mathrm{hexane}}$) were expected to be of lower quality, with high impurity content, a naturally varying chirality distribution, and primarily multi-walled CNTs (Supplementary Figs.~\ref{FigS2}g,h; Supplementary Fig.~\ref{FigS1}b). Magnetoresistance (MR) was measured for one sample from each batch.

\begin{figure}[h!]
	\centering
	\includegraphics[width=0.98\textwidth]{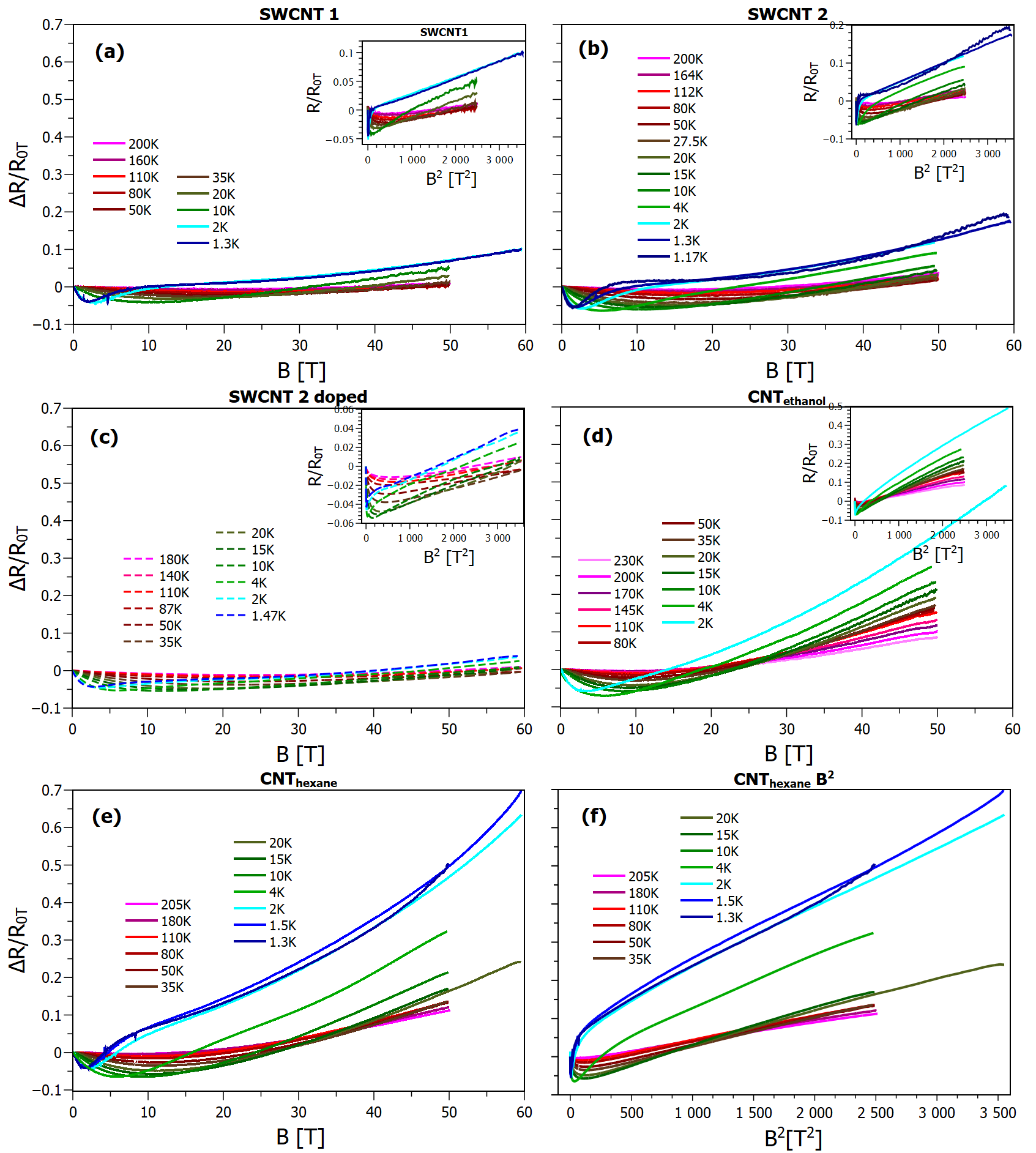}
	\caption{ Magnetoresistance of the a) SWCNT\,1, b) SWCNT\,2, c) doped SWCNT\,2, d) CNT$_\mathrm{ethanol}$ and e) CNT$_\mathrm{hexane}$ samples. Insets and f) represent MR against B$^2$ for respective samples. The MR is calculated as the relative values $\Delta \mathrm{R/R}_{\mathrm{0T}}$, where $\Delta$R stands for R$_\mathrm{XT}$-R$_\mathrm{0T}$. Here: R$_\mathrm{0T}$ is the resistance of the sample measured at a given temperature and 0\,T field, while  R$_\mathrm{X\,T}$ stands for the resistance of the sample measured at this temperature and applied field.}
	\label{Fig2}
\end{figure}

Figure~\ref{Fig1} presents resistance vs. temperature (R--T) curves for all samples. Each showed a characteristic U-shaped profile--resistance decreased with cooling, then rose again at low temperatures--common among CNT-based conductors\cite{lekawa2015a, bulmer2026}. The CNT$_{\mathrm{hexane}}$ sample, measured at 30\,$\upmu$A to avoid internal heating, exhibited the highest resistance ratio (R$_\mathrm{1.6K}$/R$_\mathrm{273.15K}$) and was the only sample to show a clear signature of 3D variable-range hopping (VRH) in fitting (Fig.~\ref{Fig1}b). All other samples lay on the metallic side of the metal--insulator transition, with finite resistance extrapolated to 0 K and clear divergence from VRH fits. Notably, pristine SWCNTs had higher resistance ratios than CNT$_{\mathrm{ethanol}}$, likely due to their loose packing; nitric acid doping compensated for this, reducing the ratio to CNT$_{\mathrm{ethanol}}$-like values. Following a classical approach, we fitted all R--T curves using fluctuation-induced tunnelling (FIT; Fig.~\ref{Fig1}c) and weak localisation (WL; Fig.~\ref{Fig1}d) models. However, none produced satisfactory fits across a meaningful temperature range.

Figure~\ref{Fig2} shows MR curves for each fibre. At every testing temperature, all samples displayed a negative MR at low fields, followed by a positive contribution at higher fields. While the amplitude of the negative MR was similar across samples (Extended Data Fig.~\ref{FigE1}a), its relative contribution was larger in more conductive fibres (Extended Data Fig.~\ref{FigE1}b). The field at which MR changed sign--from negative to positive--grew with decreasing temperature, then dropped again at the lowest temperatures (Extended Data Fig.~\ref{FigE1}c). SWCNT samples exhibited higher crossover fields than CNT$_{\mathrm{ethanol}}$ and CNT$_{\mathrm{hexane}}$. For the doped SWCNT\,2 sample, the crossover field could not be directly observed within the 60\,T range but was estimated by extrapolation. These findings indicate that positive MR emerges even in highly conductive fibres, a feature that remained undetected in many studies limited to $\le$15\,T\cite{ghanbarri2018,vavro2005}. Maximum MR at 50\,T (Extended Data Fig.~\ref{FigE1}d) increased with decreasing temperature, peaking for CNT$_{\mathrm{hexane}}$, followed by CNT$_{\mathrm{ethanol}}$, SWCNT, and doped SWCNT\,2--consistent with a larger negative MR area diminishing the net positive component in more conductive fibres. Given the poor fits of R--T curves to the WL model (Fig.\ref{Fig1}), and the lack of an MR extension for the FIT model, we did not pursue these frameworks further. Instead, all samples showed high-field MR that follows a quadratic (B$^{2}$) dependence (Fig.~\ref{Fig2} insets, Fig.~\ref{Fig2}f). Similar temperature-driven MR sign crossovers and quadratic high-field behaviour have also been observed in materials as diverse as EuSe$_{2}$ and ZrTe$_{5}$, suggesting a possible universal delocalisation mechanism\cite{dong2025,pi2024}. However, deviations from B$^{2}$ behaviour below 1.3\,K--and the continued growth of positive MR in all samples, including CNT$\mathrm{hexane}$, which exhibits 3D-VRH-like R--T behaviour--contrast with our previous results, where CNT films showing 3D-VRH also exhibited MR saturation at low temperature and high field\cite{bulmer2017}. These discrepancies highlight the limitations of semi-empirical fits for explaining electronic transport in macroscopic nanostructured conductors. 

\section*{Key factors governing magnetoresistance in CNT assemblies}\label{modelling}

To capture electron transport in CNT networks under magnetic fields, thermal fluctuations, and structural disorder, we developed a multiscale TB--NEGF--MD--Peierls framework. It combines quantum-coherent tight-binding transport (TB--NEGF), orbital magnetic field effects (via Peierls substitution), and thermally disordered atomic configurations generated from molecular dynamics (MD) in a single workflow. By computing and averaging transmission over many MD snapshots, the method naturally includes anharmonic phonon scattering, long-wavelength structural fluctuations, and defect-induced disorder--without relying on the perturbative or harmonic approximations that limit Boltzmann transport theories\cite{markussen2017}. Although zero-point motion and explicit inelastic processes are omitted, this approach realistically captures magnetotransport in disordered CNT assemblies (see Supplementary Sec.~\ref{secM}). 

As shown in the literature\cite{nakai2014} and confirmed by our calculations, 
junctions dominate charge transport in nanostructured networks and modelling isolated CNTs alone is insufficient. Magnetic-field response of a single SWCNT of moderate diameter (Extended Data Fig.~\ref{FigE2}a,b) under perpendicular magnetic fields up to 60\,T remained below 10$^{-5}$ across all dopings, which is consistent with earlier predictions\cite{nemec2006, cresti2021} that strong perpendicular-field responses for such nanotubes occur only at fields far beyond experimental reach. This suggests that the observed MR in nanostructured networks must arise from junction-level effects. Realistic fibre is a network of overlapping CNTs connected to electrodes, with transport relying on inter-tube tunnelling. We therefore focus on simple CNT junctions (Extended Data Fig.~\ref{FigE2}c) and their structural and electronic parameters. The model further includes Stone-Wales defects -- the common topological defects in CNTs with low formation energy\cite{collins2017}  (Extended Data Fig.~\ref{FigE2}h), and extended geometries such as overlapping bundles and multi-junction configurations to capture the hierarchical morphology of realistic fibres (Extended Data Fig.~\ref{FigE2}d-g).

\begin{figure}[h!]
	\centering
	\includegraphics[width=0.95\textwidth]{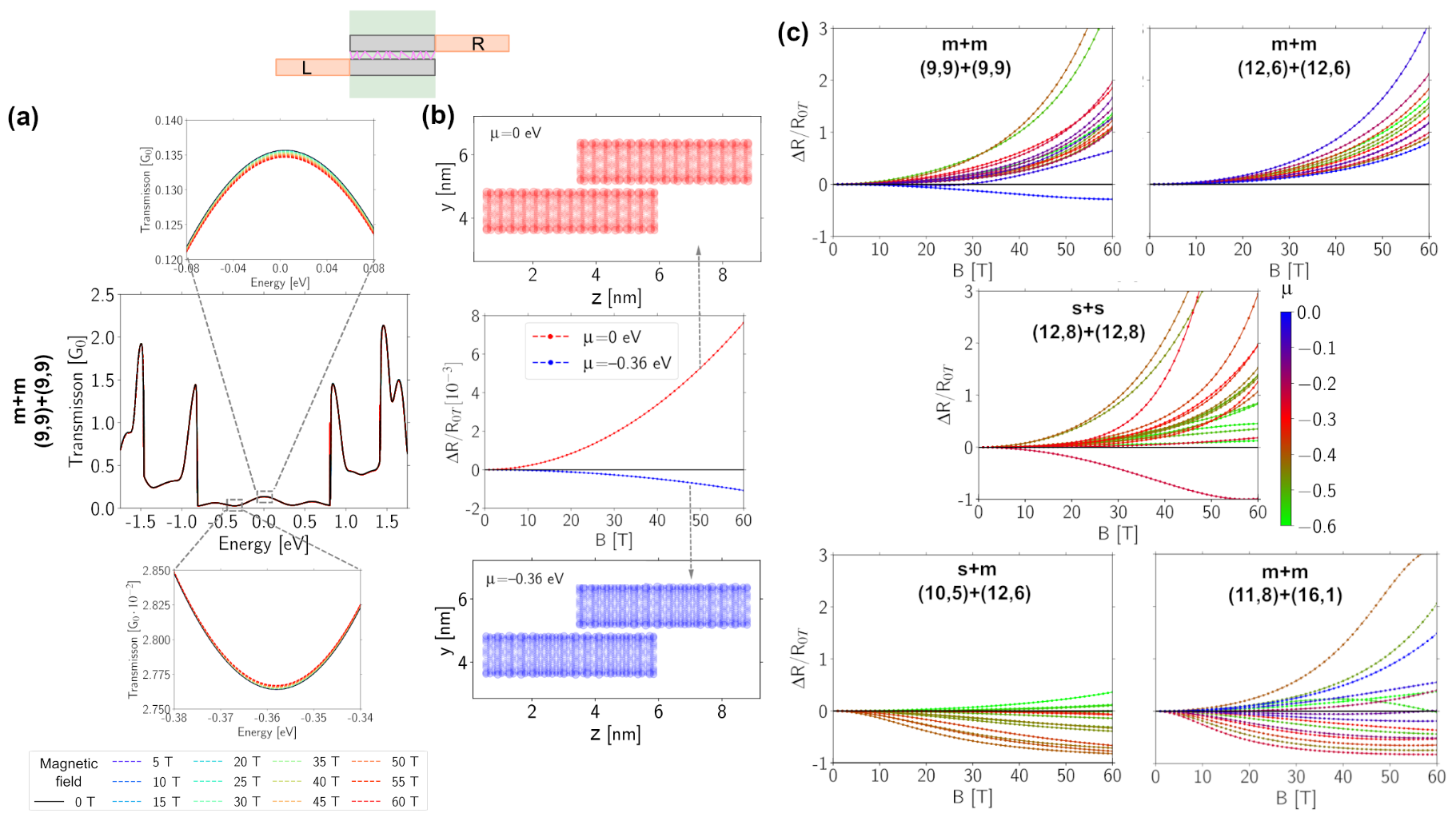}
	\caption{Magnetotransport properties of simple CNT junction: the role of doping, chirality and
		metallicity. a) Computed zero-bias transmission spectra of a short (9,9)+(9,9) metallic-metallic CNT junction with an overlap region (z) of 2.46\,nm, consisting of ten units of a (9,9) CNT. The zoomed-in transmission spectra shown above and below highlight energy ranges where the transmissions under an external perpendicular magnetic field (coloured lines) are lower or higher than those without the applied field (black line). This leads to positive (red) or negative (blue) magnetoresistance ($\Delta$R/R$_\mathrm{0T}$, also denoted as MR) obtained for different doping levels ($\upmu$) as shown on panel b). The corresponding local density of states (LDOS) maps, plotted at B = 0\,T, illustrate the system's state before applying any magnetic field. These maps are shown above and below the two exemplary magnetoresistance functions, highlighting the differences between positive and negative MR regimes for the (9,9)+(9,9) metallic-metallic CNT junction. c) Computed magnetoresistance of simple CNT junctions with different metallic characters and chiralities as functions of the external perpendicular magnetic field (B). Representative MR functions are shown for metallic-metallic (9,9)+(9,9) (z = 2.46\,nm), metallic-metallic (12,6)+(12,6) (z = 22.26\,nm), semiconducting-semiconducting (12,8)+(12,8) (z = 22.23\,nm), semiconducting-metallic (10,5)+(12,6) (z = 22.26\,nm), and metallic-metallic (11,8)+(16,1) CNT junctions. The MR functions were plotted for doping levels ($\upmu$) within the range [-0.6, 0.0]\,eV, sampled at 0.001\,eV intervals. The number of MR functions plotted in each panel was selected to reflect different observed trends while maintaining clarity.}
	\label{Fig3}
\end{figure}

\textbf{Zero Magnetic Field}. Our simulated junctions consist of CNTs with differing representative chiralities (Extended Data Fig.~\ref{FigE2}c,a). Energy-resolved transmission across a CNT-CNT junction is strongly suppressed and fragmented compared to the step-like transmission profile of a pristine tube (cf. Fig~\ref{Fig3}a and Supplementary Fig.~\ref{FigS6}a). Instead of broad plateaus, sharp resonances emerge near van Hove singularities, separated by low and oscillatory transmission regions. This reflects reduced inter-tube coupling, required for conduction across the junction, relative to coherent propagation in a continuous nanotube.

At zero field, junction conductance is maximised at commensurate orientations of nanotubes, and suppressed by lattice mismatch. Parallel or nearly parallel junctions maximise $\uppi$-$\uppi$ coupling, while crossing angle and overlap length modulate conductance through interference effects--most strongly in aligned homojunctions (CNTs with the same chirality)\cite{xu2013, tripathy2016,adinehloo2023}. However, the transmission oscillations (Supplementary Fig.~\ref{FigS8}) are not only related to overlap length, which is known to modulate conductance of CNT junctions\cite{tripathy2016,xu2013}. These oscillations arise at fixed overlap as a function of energy, reflecting how electronic states in the two tubes interact. Their patterns vary across junctions, depending on chirality and metallicity of the constituent nanotubes, as well as on the junction heterogeneity, driven by the interplay between individual electronic structures and their interfacial coupling. Similar transmission suppression and interference-induced fragmentation have been observed in ReO$_{3}$ and TaSe$_{3}$, where hybridisation gaps and subband coupling yield oscillatory conductance patterns\cite{chen2021, gatti2021}.

\textbf{With Magnetic Field}. A perpendicular magnetic field modifies transport by introducing Peierls phases in the inter-tube coupling, thereby tuning the relative phase between interfering electronic paths (Fig.~\ref{Fig3}a). At specific energies, MR ($\Delta$R/R$_\mathrm{0\,T}$, Fig.~\ref{Fig3}b) can be either positive or negative. Local density of states (LDOS) analysis at zero field shows that positive MR arises from field-induced dephasing between initially well-coupled states, which suppresses transmission, whereas negative MR arises when the field improves coupling between initially mismatched states and thereby enhances transmission. Figure~\ref{Fig3}c shows MR--B curves for five CNT junctions spanning a range of chiralities and electronic characters: three homojunctions (metallic-metallic achiral (9,9)+(9,9), metallic-metallic chiral (12,6)+(12,6), and semiconducting-semiconducting chiral (12,8)+(12,8)) and two heterojunctions (semiconducting-metallic chiral (10,5)+(12,6) and metallic-metallic chiral (11,8)+(16,1)), at doping levels from 0\,eV (undoped) to -0.6\,eV (nitric-acid-treated fibres\cite{hayashi2020}). Most junctions show positive MR that increases with field--often quadratically at high B--though negative MR and nonmonotonic responses also occur, especially in chirality-mismatched cases. These trends arise from the interplay between magnetic-field-induced suppression of coherent tunnelling and enhancement of electronic state alignment across the junction.

Such variability of MR responses suggests that tuning the Fermi level--via doping or gating--could control both transmission and MR sign. Homojunctions typically show positive, often oscillatory MR, while heterojunctions (of CNTs of different chirality) display negative MR across most considered doping levels and lack periodic features (see Supplementary Fig.~\ref{FigS8}). The magnetic field enhances tunnelling between mismatched electronic states in heterojunctions but suppresses coherence in well-aligned homojunctions.

\begin{figure}[h!]
	\centering
	\includegraphics[width=0.95\textwidth]{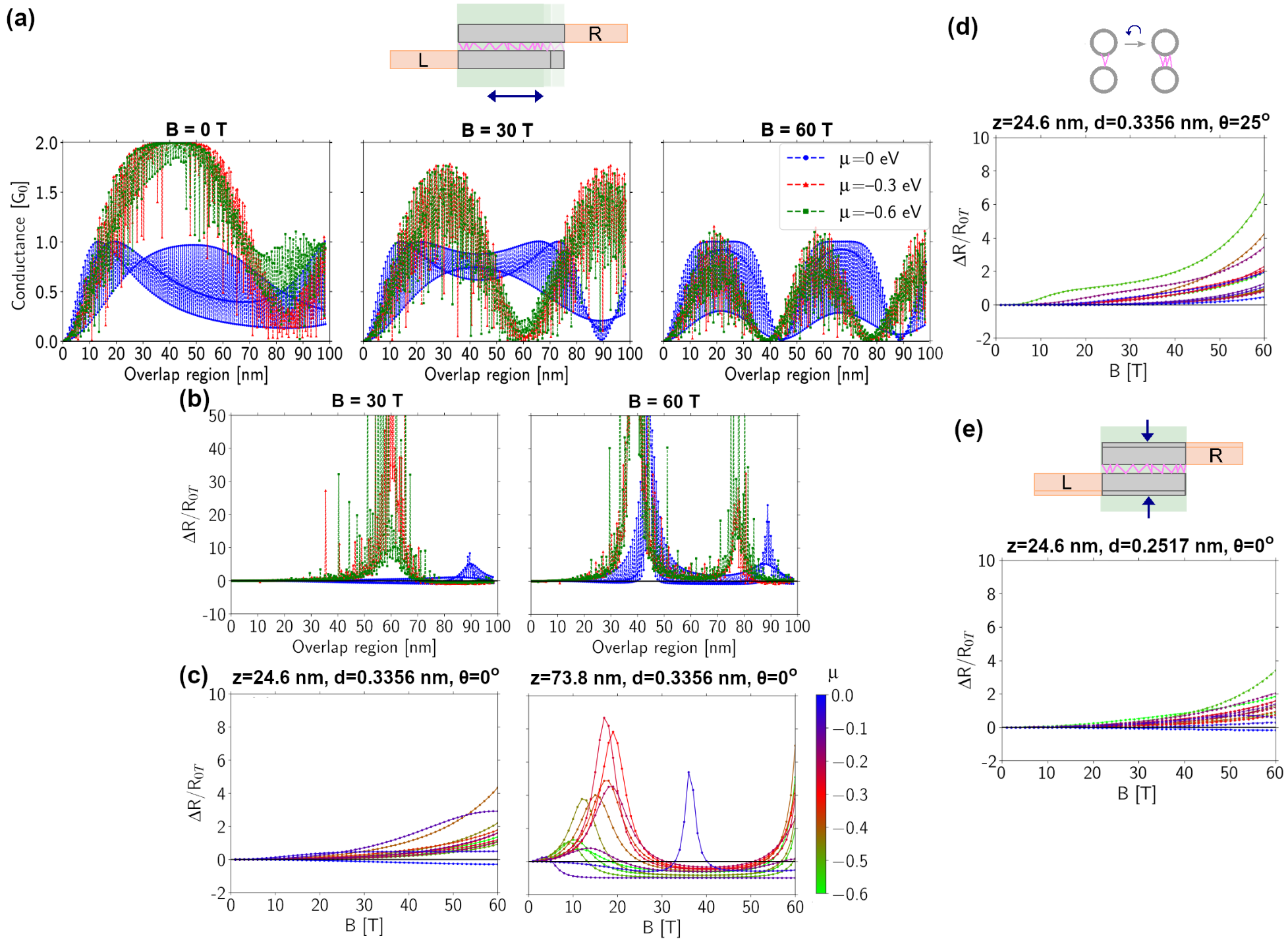}
	\caption{Magnetotransport properties of simple CNT junctions with varying overlap length, CNT stacking and intertube distance. a) The oscillations of conductance ($\upsigma$) and b) magnetoresistance  with CNT junction length of (9,9)+(9,9) metallic-metallic CNT junction under external perpendicular magnetic fields at the Fermi level ($\upmu$ = 0\,eV), nitric acid doping ($\upmu$ = -0.6 eV), and intermediate doping ($\upmu$ = -0.3\,eV). c)  Representative subsets of computed MR functions for different overlap (z) lengths: 24.6\,nm (left panel) and 73.8\,nm (right panel), plotted for doping levels in the range [-0.6, 0.0]\,eV, sampled at 0.001\,eV intervals. d) Representative subsets of computed MR functions for a CNT junction with an overlap length of 24.6\,nm but different relative angular positions of the upper nanotube ($\uptheta$ = 25\,$^o$). The upper nanotube is rotated around the z-axis, modifying atomic connections between the sublattices of tube 1 and tube 2 via hopping interactions, which form the junction, as illustrated in the scheme. For $\uptheta$ = 0\,$^o$ (panel c), left), only two connections are present, where atoms from sublattice A1 exclusively connect to atoms from sublattice A2 (Supplementary Fig.~\ref{FigS3}, Extended Data Fig.~\ref{FigE4}). For $\uptheta$ = 25\,$^o$ (panel d)), three connections appear: two A1-B2 connections and one B1-A2 connection (Extended Data Fig.~\ref{FigE4}). e) Representative subset of computed MR functions for a reduced separation distance between nanotubes. As illustrated in the scheme, the nanotubes are brought closer to a separation of d = 0.2517\,nm, effectively increasing the intertube coupling in the junction. In contrast, the junction with the same overlap length but a larger separation distance of d = 0.3356\,nm is shown in panel (c), left.}
	\label{Fig4}
\end{figure}

\textbf{Structural Factors}. We next examine how overlap length, tube rotation (stacking), and intertube separation influence conductance and magnetoresistance in junctions (Figs.~\ref{Fig4}a--e; Supplementary Sec.~\ref{secToverlap}-\ref{secTd}).

\textbf{Changing Overlap}. To isolate the effect of overlap length, we computed conductance and magnetoresistance for four junctions -- three homojunctions and one mixed-metallicity heterojunction - over overlap lengths up to $\sim$100\,nm. Metallic junctions are analysed at electrochemical potential $\upmu$\,=\,0\,eV, -0.3\,eV, and -0.6\,eV, whereas semiconducting and mixed-metallicity junctions are considered only at doped levels due to their band gap.
All junctions show overlap-dependent zero-field conductance oscillations at both undoped and doped levels (Fig.~\ref{Fig4}a, Supplementary Fig.~\ref{FigS10}), consistent with prior reports\cite{tripathy2016,xu2013}. Oscillation amplitude and period vary with junction chirality and doping. Metallic homojunctions behave similarly at -0.3\,eV and -0.6\,eV, while conductance is notably suppressed at -0.6\,eV compared to -0.3\,eV in semiconducting and mixed junctions.
Oscillations arise from interference between multiple conduction paths across the overlap region\cite{tripathy2016}. Depending on the stacking configuration, constructive or destructive interference occurs when the overlap matches an integer multiple of half the Fermi wavelength. Their frequency and amplitude depend on the tubes' electronic structure, with heterojunctions showing more complex and less regular oscillations. As overlap increases, oscillation frequency in transmission versus energy rises due to a larger number of coherent paths contributing to intertube tunnelling (Supplementary Fig.~\ref{FigS12}a,b). The magnetic field further modifies these patterns, suppressing overall conductance and increasing the oscillation frequency. MR vs. overlap length (Fig.~\ref{Fig4}b, Supplementary Fig.~\ref{FigS11}) also exhibits strong oscillations spanning two to three orders of magnitude and changing sign. By tuning the junction length alone, one can dramatically modify both the magnitude and sign of the MR. To illustrate this point, Fig.~\ref{Fig4}c and Supplementary Fig.~\ref{FigS12}c,d show that changing the overlap length of a single junction qualitatively alters its MR--B response--from predominantly positive and quadratic at high fields to strongly non-monotonic with sign changes--highlighting overlap length as a powerful tuning parameter for both conductance and MR.

\textbf{Rotations}. Rotating one nanotube around its axis at fixed overlap length (Fig.~\ref{Fig4}d) changes the set of intertube atomic pairs within the coupling range ($\le$3.6 \AA) and therefore the available hopping pathways between the tubes. Because electronic coupling in our model occurs only for atomic pairs within this range (Supplementary Sec.~\ref{secM}), rotation discretely alters intertube connectivity, leading to bcorresponding variations in conductance and magnetoresistance (Supplementary Fig.~\ref{FigS13}). The rotation effect is weaker than that of changing overlap (Fig.~\ref{Fig4}d) length but remains visible in the transmission and MR spectra (Supplementary Fig.~\ref{FigS14}). Although stacking effects are known to influence transport in aligned CNT junctions and carbon nanodevices\cite{nakanishi2001,ostovan2024}, they are likely unresolvable in macroscopic fibres lacking rotational order.

\textbf{Intertube Distance}. To assess the role of intertube coupling strength, we varied CNTs spacing in the junction, altering the number of atom pairs within the 3.6\,\AA$ $ cutoff instead of scaling Hamiltonian terms\cite{xu2013}. This approach better reflects experimental conditions, where compression can modify intertube contact. Field-induced forces at moderate magnetic fields have been predicted to distort molecular structures\cite{irons2021}, and while CNTs are rigid, the van der Waals gap between tubes may still respond. As shown in Supplementary Fig.~\ref{FigS15}, reducing spacing enhances intertube coupling. At the equilibrium distance (d$_1$), conductance behaves similarly to the weak-coupling regime predictions\cite{tripathy2016}, while at shorter spacing (d$_2$) it is strongly suppressed, consistent with ref.\cite{xu2013}. A similar conductance suppression can occur at d$_1$ for sufficiently long overlaps, where the junction approaches the bundle limit with gap-opening and quenched transmission\cite{bulmer2026}. For doped junctions ($\upmu$\,=\,-0.3 or -0.6\,eV), we observe oscillatory conductance patterns with no monotonic decay, consistent with previous findings\cite{xu2013} that doping can preserve finite conductance in strongly coupled systems by shifting the Fermi level out of the gap. Accordingly, modest doping can maintain transmission in compressed or strongly coupled regions of CNT networks. Although overlap length-dependent MR becomes more doping-sensitive at reduced spacing (Supplementary Fig.~\ref{FigS15}), the overall shape of MR(B) (Fig.~\ref{Fig4}e) remains similar. Thus, intertube distance affects MR oscillations and conductance amplitude, while overlap length is the dominant control parameter for MR polarity and magnitude.

\textbf{Spin}. Simulation of three metallic-metallic junctions with a Zeeman term (Supplementary Section~\ref{secTspin}, Supplementary Fig.~\ref{FigS17}) shows that spin effects do not alter MR(B) shape or doping trends, though they modestly enhance positive MR magnitude at high fields and are negligible for negative MR.

\begin{figure}[h!]
	\centering
	\includegraphics[width=0.95\textwidth]{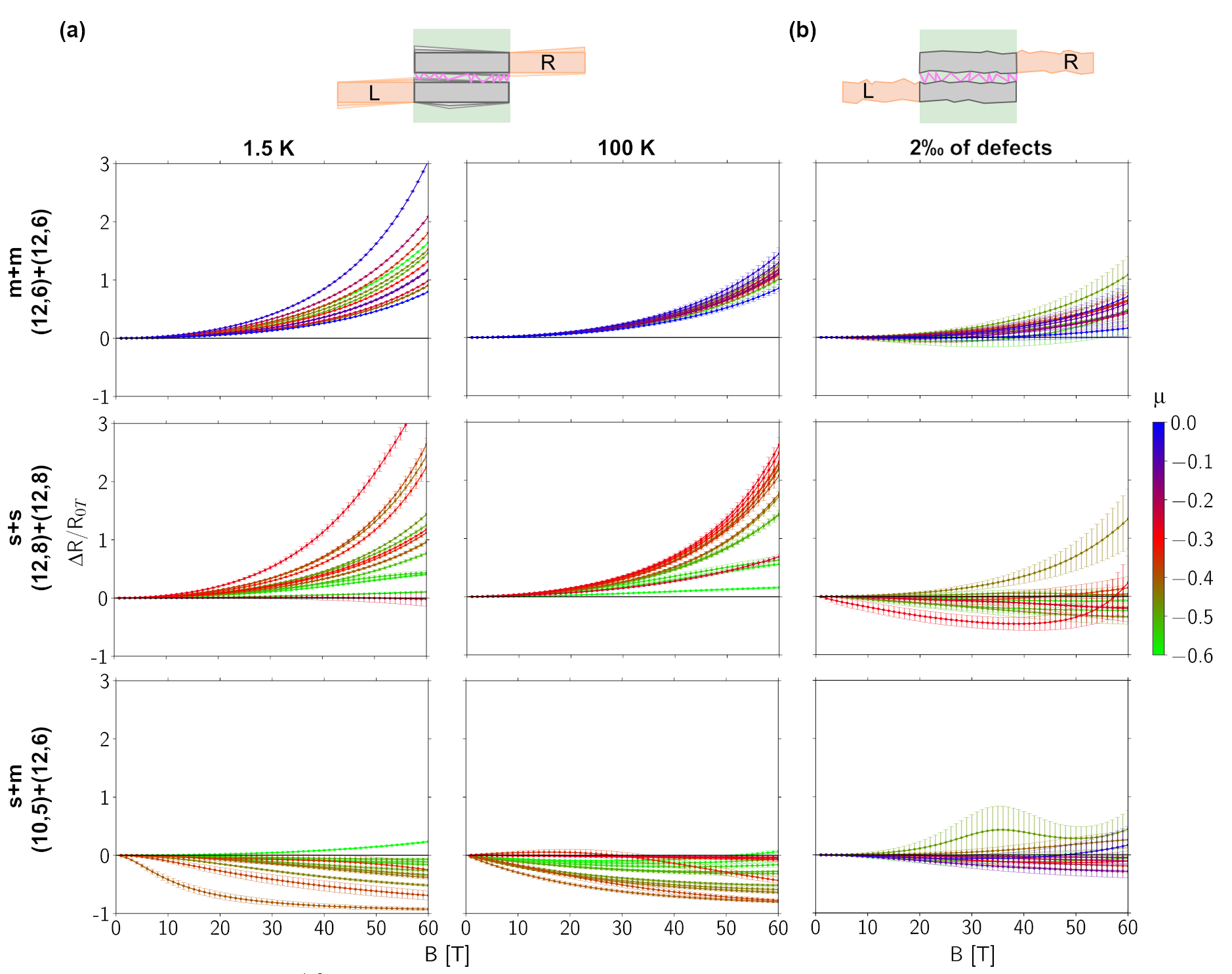}
	\caption{Magnetoresistance of simple CNT junctions under temperature and with structural defects a) The computed magnetoresistance as functions of the external perpendicular magnetic field (B) of metallic-metallic (z$_\mathrm{initial}$ = 22.6\,nm), semiconducting-semiconducting (z$_\mathrm{initial}$= 22.3\,nm) and semiconducting-metallic (z$_\mathrm{initial}$ = 22.6\,nm) CNT junctions at two temperatures and various doping levels. Representative MR functions are shown for doping levels in the range [-0.6, 0.0]\,eV, sampled at 0.001\,eV intervals, with subsets selected to reflect distinct trends. b) Computed magnetoresistance for the same CNT junction types containing 0.2\% Stone--Wales defects under external perpendicular magnetic fields and various doping levels.}
	\label{Fig5}
\end{figure}

\textbf{Disorder}. We now examine disorder from thermal fluctuations and structural imperfections, like Stone-Wales defects, both unavoidable in real systems.

\textbf{Temperature}. We first examine the effect of temperature on magnetotransport in three junctions (Figure~\ref{Fig5}a): metallic-metallic (12,6)+(12,6) and semiconducting-semiconducting (12,8)+(12,8) homojunctions, and semiconducting-metallic (10,5)+(12,6) heterojunction, the latter two requiring doping for conduction.
Temperature and phonon-assisted channels are known to modulate intertube tunnelling\cite{adinehloo2023}, and we observe a consistent trend. Rising temperature suppresses transmission, narrows existing band gaps, and dampens energy-dependent transmission oscillations (Supplementary Fig.~\ref{FigS18}, Supplementary Fig.~\ref{FigS19}). 
LDOS analysis shows that these trends originate from the gradual loss of phase coherence.
At 0\,K, well-aligned interference fringes appear in the positive-MR regime (Fig.~\ref{Fig3}b), but even 1.5\,K introduces phase shifts and mismatch between LDOS patterns on the two tubes (Extended Data Figs.~\ref{FigE5}a,b,d,e). At 100\,K, fringes broaden and lose alignment, reflecting thermal broadening.
Consequently, with increasing temperature, the MR spread across doping levels narrows and both positive and negative extrema decrease (Fig.~\ref{Fig5}a, Supplementary Fig.~\ref{FigS20}), consistent with experiment (Figs.~\ref{FigS2}a,e). In homojunctions, oscillatory MR--energy features are suppressed at higher temperature, reflecting thermal smearing of quantum interference at the junction (Supplementary Fig.~\ref{FigS21}), as reported for thermally disordered CNTs\cite{roche2001, yanagi2010}.

\textbf{Structural Defects}. We now turn to structural defects, inherent to real samples and essential for testing weak localisation theory, often indicated as a potential origin of CNT networks' magnetoresponse\cite{gao2021}. 
We therefore examine the effect of varying concentration of Stone-Wales defects (Fig.~\ref{Fig5}b and Supplementary Figs.~\ref{FigS22}--\ref{FigS25}) on five different junctions.
Even at 0.2\% concentration, defects strongly altered transmission (Supplementary Figs.~\ref{FigS22}--\ref{FigS23}): band-edge peaks  are suppressed, sharp features appear across the energy range, and overall conductance drops due to lattice distortion (Extended Data Fig.~\ref{FigE2}h). In gapped junctions, in-gap peaks arise from defect states\cite{roche2001}, confirmed by LDOS maps showing localised states (Extended Data Figs.~\ref{FigE5}c,f), which act as strong elastic scatterers disrupting coherent electron propagation along nanotubes (see Supplementary Sec.~\ref{secTd}-\ref{secTdtemp}).

Defects also alter magnetoresistance by introducing sign changes and irregularities in MR(B) across doping levels (Fig.~\ref{Fig5}b, Supplementary Fig.~\ref{FigS24}). These effects emerge in heterojunctions at low defect density and in all junctions by 1\%. However, their character differs between homo-and heterojunctions. In homojunctions, defects increase negative MR fraction and deepen MR minimum, while reducing the maximum positive MR at 0.2\% defect density (Supplementary Fig.~\ref{FigS25}), Supplementary Tab.~\ref{table}. At 1\%, positive MR rises again - surpassing the pristine case in (12,6)+(12,6) but not in (12,8)+(12,8). These trends align with weak localisation theory, where elastic scatterers enhance low-field negative MR via phase-breaking backscattering\cite{roche2001, salvato2012}. The effect is strongest in (12,6)+(12,6), with weaker signatures in (12,8)+(12,8), consistent with reports in single CNTs and bundles\cite{gao2021, nemec2007}, where coherent time-reversed paths can form due to orbital and band alignment between tubes, giving rise to interference effects that are sensitive to defect concentration and doping. For heterojunctions, the percentage of negative MR(B,$\upmu$) decreases with defect density and the MR minimum becomes less negative, diverging from weak localisation trends. Although maximum positive MR drops at 0.2\% and recovers at 1\%, the overall MR evolution is inconsistent with weak localisation expectations. Band-edge mismatch and orbital asymmetry likely suppress phase-coherent backscattering, rendering weak-localisation corrections less relevant. Instead, defect-induced states may modulate tunnelling via energy filtering or resonant effects, producing sign changes and irregular MR(B) curves\cite{adinehloo2023, nemec2007}. The periodic features in MR(B,$\upmu$) maps of pristine junctions vanish upon defect introduction (Supplementary Fig.~\ref{FigS25}). The loss of phase coherence and the emergence of localised states suppress the interference fringes characteristic of the pure system. In homojunctions, this behaviour is consistent with weak localisation\cite{roche2001, hikami1980}, whereas in heterojunctions it indicates a shift to incoherent, resonance-dominated transport\cite{roche2001, hikami1980}.

\textbf{Temperature \& Structural Defects}. 
Thermal fluctuations and structural defects coexist in real systems, making it important to examine their combined effect on magnetotransport.
For the metallic-metallic (9,9)+(9,9) homojunction, transmission declines more rapidly with temperature when 5--7 defects are present (Supplementary Figs.~\ref{FigS26}a--c). 
The effect intensifies with defect density, confirming the interplay between static disorder and thermal fluctuations. 
Comparison of the averaged conductance across the first transmission step (Supplementary Fig.~\ref{FigS5}d) for different chiral junctions further illustrates this trend: at 1.5\,K, the metallic-metallic (12,6)+(12,6) homojunction has the highest conductance, followed by the semiconducting-semiconducting (12,8)+(12,8) homojunction and the nearly order-of-magnitude lower semiconducting-metallic (10,5)+(12,6) heterojunction (Supplementary Fig.~\ref{FigS44}a). With increasing temperature, conductance decreases in all cases, most steeply for (12,6)+(12,6); by 100\,K, the two chiral homojunctions become comparable, while the heterojunction remains lower with a smaller absolute reduction. In the pristine achiral (9,9)+(9,9) homojunction, conductance briefly increases at 30\,K before decreasing, but this non-monotonicity vanishes with defects, where suppression strengthens with defect density.

Resistance--temperature curves (Supplementary Fig.~\ref{FigS26}(j--l)) show that in the pristine (9,9)+(9,9) homojunction, resistance increases with temperature, with doping-dependent low-temperature slopes and weak non-monotonic features (Supplementary Fig.~\ref{FigS26}j). In contrast, defective junctions show higher absolute resistance and, on log--log axes, flatter low-temperature behaviour followed by steep increases at higher temperature (Supplementary Fig.~\ref{FigS26}k,l). This behaviour reflects limited tunnelling in the initial geometry (A-A coupling only, Extended Data Fig.~\ref{FigE4}, Supplementary Fig.~\ref{FigS3}b), Supplementary Fig.~\ref{FigS14}). Modest heating can increase the number of intertube interactions, partially compensating thermal broadening. But further temperature increase eventually amplifies disorder and scattering, accelerating resistance growth.

At 100\,K, temperature-induced structural deformation exceeds that caused by 0.1--0.2\% defects but remains below that induced by 1\% defects. While averaged conductance generally decreases with increasing deformation parameter $\upalpha$, the trend depends on junction type. For (9,9)+(9,9) variants, conductance becomes similar at high temperature despite differing $\upalpha$, especially under magnetic field (Supplementary Fig.~\ref{FigS44}c-d). The junction quality parameter $\upgamma$, which counts the number of interacting intertube atom pairs, provides further insight. As $\upgamma$ increases, indicating additional tunnelling paths, conductance decreases, especially at higher temperatures (Supplementary Fig.~\ref{FigS44}). This reflects increased backscattering under structural deformation\cite{milowska2019} and/or lattice mismatch.

The MR response reflects the interplay of disorder and temperature (Supplementary Figs.~\ref{FigS28}--\ref{FigS29}, Extended Data Fig.~\ref{FigE6}, Supplementary Tab.~\ref{table}). In pristine metallic-metallic (9,9)+(9,9) homojunctions, the fraction of negative MR increases at 30 K with a deepening MR minimum, but decreases above 50\,K as the minimum shifts upward and negative MR vanishes by room temperature. Junctions with 0.1\% defects follow a broadly similar evolution but exhibit a consistently higher fraction of negative MR at all temperatures and deeper MR minima at low and high temperature, with non-monotonic MR(B) behaviour emerging by 100\,K. At 1\% defects, negative MR remains prevalent across the full temperature range, while MR(B) curves become irregular already above 50\,K. The MR sign evolution suggests competing effects of weak localisation and defect-assisted tunnelling\cite{adinehloo2023, nemec2007}. Parabolic temperature-dependence of minimum MR and rising maximum MR at high temperature, aligns with our experiments and highlights the central role of junctions (cf. Supplementary Fig.~\ref{FigS29} and Extended Data Fig.~\ref{FigE3}).

\section*{Transport mechanisms arising from CNT junction connectivity}\label{discussion}

\begin{figure}[h!]
	\centering
	\includegraphics[width=0.45\textwidth]{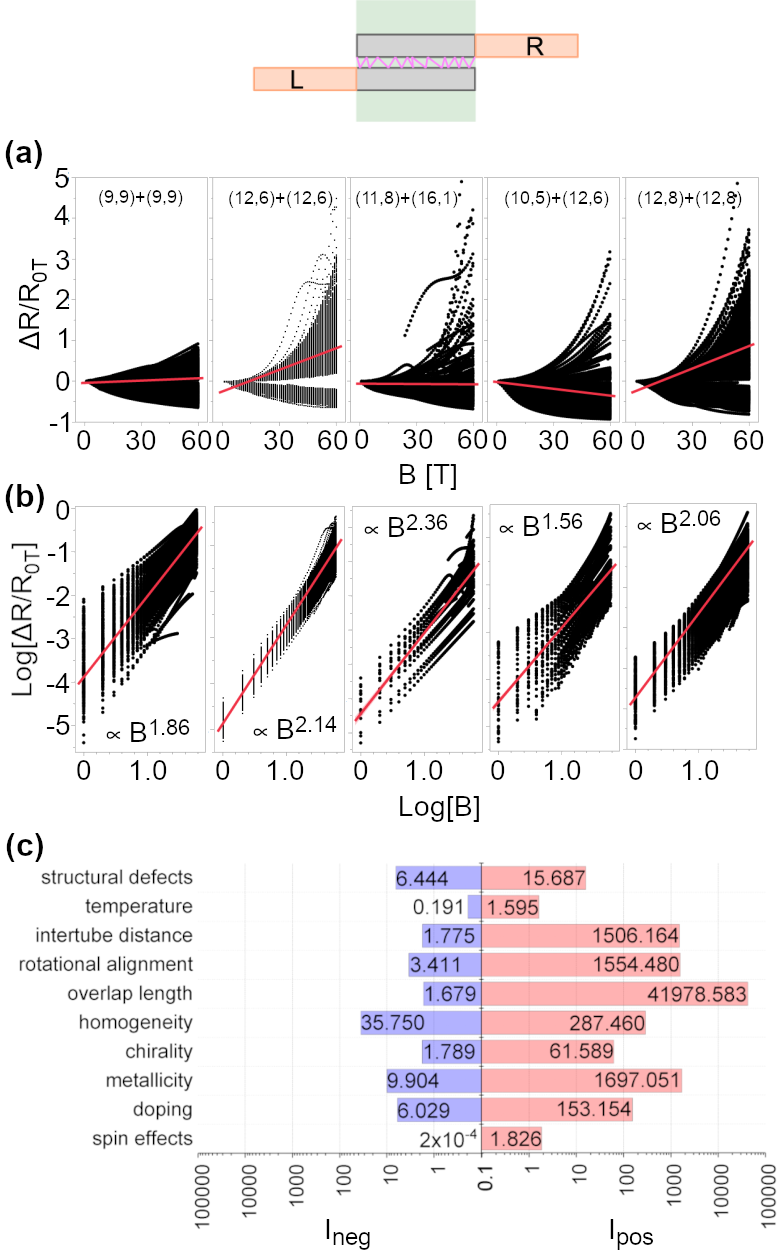}
	\caption{Statistical scaling and factor analysis of junction-controlled magnetotransport. 
	a) The calculated MR values vs magnetic field B across multiple datasets with red lines depicting the best fit. 
	b) Power-law analysis of positive MR for all considered junction chiralities. 
	The log plots of MR versus log B,  which exclude the influence of other input factors, indicate a power law dependence that approaches quadratic. The dependence becomes more clearly quadratic when the other input factors are incorporated in multivariate linear regression modeling (Supplementary Section~\ref{secStatistics2}).
    c) Key factors influencing magnetotransport in simple CNT junctions. Diverging bar chart showing the computed impact of ten factors on magnetoresistance. The dimensionless impact score (I) is defined as the product of the median absolute change in MR and the corresponding mean change in the proportion of the energy window exhibiting negative (I$_{\mathrm{neg}}$) or positive (I$_{\mathrm{pos}}$) MR. The evaluated factors include: doping, metallicity, chirality, homogeneity, overlap length, rotational alignment, intertube distance, spin effects, temperature, and structural defects. Details of the methodology used to calculate the impact scores are provided in the Supplementary Section~\ref{secTfactors}.}
	\label{Fig6}
\end{figure}

Statistical analysis of the large numerical dataset (734 000 datapoints, Supplementary Section ~\ref{secStatistics}) uncovered several key findings. Figure~\ref{Fig6}a plots MR against field for each junction species across doping, temperature, and defect density. In all cases, positive MR exhibits a robust quadratic dependence on field, consistent with our experimental results across multiple CNT fibre types. For homojunctions ((9,9)+(9,9), (12,8)+(12,8),(12,6)+(12,6)), this B$^2$ trend was evident in both bivariate logplots (Figure~\ref{Fig6}b) and multivariate linear regression models (Supplementary Section~\ref{secStatistics2}), with minor corrections from temperature, defects, and doping. In the heterojunctions ((10,5)+(12,6) and ((11,8)+(16,1)), this quadratic behaviour became more apparent only after incorporating doping in the linear regression model. 
Since single CNTs exhibit much weaker MR (Extended Data Fig.~\ref{FigE3}), the observed B$^2$ scaling must arise from inter-tube junctions rather than from intrinsic single-tube mechanisms. This aligns with impedance spectroscopy studies identifying junctions as bottlenecks in various 1D and 2D materials\cite{gabnett2024}. Similar quadratic MR trends have been reported in magnetic and altermagnetic compounds (Co$_3$Sn$_2$S$_2$, CrSb)\cite{liu2024,peng2025}, and semimetals (ReO$_3$, SiP$_2$)\cite{chen2021, zhou2020}, where transport is governed by multichannel or orbital scattering--suggesting B$^2$ scaling is a generic feature of systems with complex conduction pathways. Negative MR was consistently weaker than positive MR, with fitted field exponents ranging from 1 to 2 depending on chirality. Except for heterojunctions, doping, temperature, and defect density mainly affect MR amplitude rather than its field scaling.

To complement our analysis, we assessed how ten studied factors affect the occurrence of positive and negative MR across all modelled junctions (Fig.~\ref{Fig6}c). While covering key junction types, this sample remains limited, so conclusions should be interpreted with caution. We analysed not just extreme MR values but also the frequency of positive or negative MR across the -0.6 to 0\,eV energy range. Negative MR is primarily governed by junction homogeneity, with heterojunctions showing stronger responses, followed by metallicity, doping, and structural defects. In contrast, positive MR is overwhelmingly controlled by overlap length, with metallicity 25 times less influential. These trends suggest complementary design routes: tuning alignment and length to optimise positive MR, and tailoring chirality, doping, and defects to control negative MR.

\section*{Transport beyond single junctions: complex architectures} 

To address whether conclusions drawn from single-junction analysis extend to real CNT fibres, we examined more complex architectures--bundles, junction of bundles, loops, and multi-junctions (Extended Data Fig.~\ref{FigE7}, Supplementary Figs.~\ref{FigS42}--\ref{FigS43} and Supplementary Tab.~\ref{table}). Bundles and their junctions retain the core MR(B) trends seen in simple junctions. However, in (12,6)+(12,6) bundles, zero-field transmission resembles that of individual CNTs, with only broadened steps (Supplementary Fig.~\ref{FigS6}c). Because transmission through individual tubes dominates over intertube contributions by one to two orders of magnitude, (Supplementary Fig.~\ref{FigS6}b) bundle MR remains weak, with the maximum positive MR strongly reduced compared to the junction (Extended Data Fig.~\ref{FigE7}a). Weak negative MR, absent in pristine junctions, emerges in pristine bundles. Pristine junction of bundles (Extended Data Fig.~\ref{FigE7}b) and Supplementary Fig.~\ref{FigS42}d--f), whose transmission resembles that of a simple junction, show non-monotonic MR(B) behaviour, including sign changes, a feature that requires disorder in simpler junctions or their non-homogenity. Such behaviour parallels MR curves observed in experimental CNT films, though simple junctions remain better suited for isolating specific effects. In multi-junctions (Extended Data Fig.~\ref{FigE7}c), multiple hopping and internal reflections suppress transmission between sharp peaks, leading to enhanced MR modulations over a wider doping range (Supplementary Fig.~\ref{FigS43}a--c). Loops (Extended Data Fig.~\ref{FigE7}d) and Supplementary Fig.~\ref{FigS43}d--f) introduce phase-coherent pathways where perpendicular-field-induced Peierls phases modulate Fabry--P\'erot-like interference in the intermediate tubes. The wavefunction accumulates phase as it encircles the void, and destructive interference leads to near-zero conductance at specific fields, producing diverging MR. While bundles and their junctions amplify interference and broaden accessible transport regimes, only multi-junctions and loops introduce qualitatively new MR behaviour. Similar geometry-driven MR was seen in crystalline semimetals such as WP$_2$, MoO$_2$, and ZrSiS, where extended pathways or open Fermi surfaces produce angular MR modulations and orbital interference\cite{chen2021, chen2020, singha2017}.

\section*{From junction-scale physics to macroscopic transport}

 Our analysis clarifies why phenomenological models fall short in capturing CNT fibre magnetotransport. At the junction level, quantum interference, defect-induced scattering, and field-dependent tunnelling resonances produce signatures--sign-changing MR, power laws, Fabry--P\'erot oscillations--not accounted for in bulk fits. Some features, like enhanced negative MR in homojunctions, align with weak localisation, while others (e.g. band-edge filtering) are averaged out in macroscopic samples. Conversely, VRH-like behaviour seen in poorly conducting fibres does not emerge from coherent-junction simulations, suggesting it stems from large-scale disorder. Of the classical models, fluctuation-induced tunnelling aligns best with our results: temperature-driven conductance suppression, reduced MR spread, and disorder sensitivity resemble fluctuating-barrier transport, though our mechanism stems from thermal broadening and lattice-induced modulation of quantum interference. Overall, while CNT networks are often analysed using WL or VRH, their transport is fundamentally governed by junction-level quantum interference--partially masked by ensemble averaging. The parallels with crystalline semimetals and altermagnets suggest a shared interference-driven origin for MR across these diverse systems.

\section*{Outlook}\label{outlook}

By linking junction-scale quantum transport to macroscopic magnetoresistance, our framework provides a transferable route to interpret--and ultimately engineer--electronic transport in disordered networks of low-dimensional materials. A key next step is to quantify how processing routes shape the statistical distribution of junction mismatch modes (overlap length, chirality heterogeneity, and rotational disorder) and how this distribution sets an intrinsic transport baseline distinct from extrinsic defects and doping. Extending the present approach to experimentally measured network morphologies, and combining it with data-driven inference of junction statistics from transport fingerprints should enable predictive ''materials-by-design'' optimisation of fibres, films, and printed conductors across CNT, graphene, MXene, and TMD platforms.

\backmatter

\section*{Methods}\label{methods}

\subsection*{CNT fibre spinning}

The CNT fiber samples for testing were directly spun from an FC-CVD furnace\cite{lekawa2014}. In this process a hydrocarbon, ferrocene, sulphur precursor and hydrogen are introduced into the hot zone of the furnace. At temperatures above 1000\,$^o$C  the hydrocarbons and ferrocene crack releasing carbon atoms and iron particles, respectively. With the aid of sulphur and hydrogen, iron particles form small clusters which become catalysts. The carbon atoms arrange on the top of catalysts and the nanotubes grow suspended in the hot zone of the furnace. Further, the nanotubes are extracted from the furnace with the use of a metal rod, to which they stick trailing other nanotubes along. The extracted material is condensed via the spraying of acetone which wets the nanotubes and pulls together the individual CNTs and CNT bundles when evaporating. The evaporation process takes place within seconds and thus formed fibres are further stretched between a collecting and supporting spindle, so as to further improve the alignment of the CNTs and CNT bundles in the fibre. Additional details about tested samples are provided in Supplementary Section~\ref{secEr}.

\subsection*{Raman spectroscopy}

Raman spectroscopy was performed using excitation wavelengths of 514, 633 and 785 nm, with laser polarisation parallel and perpendicular to the fibre axis. Further details are provided in Supplementary Fig.~\ref{Fig1} and Supplementary Section~\ref{secRaman}.

\subsection*{Sample characterisation}
Sample morphology was characterised by scanning electron microscopy; imaging parameters are provided in Supplementary Fig.~\ref{Fig2} and Supplementary Section~\ref{secEsem}.

\subsection*{Electrical Transport Measurements}
Resistance--temperature (R--T) and magnetoresistance (MR=\,(R(B)-R(0))/R(0))  measurements were performed in pulsed magnetic fields up to 60\,T at the National High Magnetic Field Laboratory (NHMFL), Los Alamos, USA. Four-probe resistance was measured with lock-in detection, with current applied along the fiber axis and the magnetic field oriented perpendicular to the fiber (transverse MR). Sample preparation details, measurement protocols and methodology are detailed in the Supplementary Sections~\ref{secEsamples}-\ref{secERMR}.

\subsection*{Atomistic Modelling}
Magnetotransport properties were modelled using a novel multiscale framework that combines quantum-coherent tight-binding transport (TB--NEGF), Peierls substitution to account for the magnetic effects, and molecular dynamics (MD) methods. Simulations were performed for single and multi-tube CNT junctions, bundles, and loops, with systematic variation of chirality, overlap length, doping level, and defect density. Transmission spectra, magnetoresistance (MR), and density of states (DOS) were calculated. Full computational details, model geometries, and analysis procedures are provided in Supplementary Section~\ref{secM}.

\subsection*{Statistical Analysis}

The CNT-junction-CNT simulation calculated transmission and MR and as a function of magnetic field, with other input factors to include doping level, types of helicity, temperature, and defect density. Incorporation of temperature and defect density was a stochastic process and introduced error in the calculated output. Relative error greater than 50\,\% was generally excluded in the analysis. The (11,8)+(16,1) system had greater error and this restriction eliminated too many datapoints, so the level was brought to 100\,\% to uncover any possible trends.  Most trends with field could be captured with bivariate log plots of MR vs B (Supplementary Sec.~\ref{secStatistics2}) that excludes the impact from other input factors. Validated multivariate linear regression models in JMP (step-wise linear regression with a Bayesian information criteria stopping rule, Supplementary Sec.~\ref{secStatistics2}) were used to model MR as a 2nd order polynomial of all input factors and confirmed the results of the simple bivariate log plots. Validation was accomplished by an 80\,\%/20\,\% split of the data into training and validation sets before training. Supplementary Sec.~\ref{secStatistics} has implementation details. 

\section*{Data availability}

The study generated about 6\,TB of data, which are available from the
corresponding authors upon reasonable request.

\section*{Acknowledgements}

T.K. and K.Z.M. gratefully acknowledge the Interdisciplinary Centre for Mathematical and
Computational Modelling at the University of Warsaw, Poland (Grant No. G47-5) for providing
computer facilities and technical support. T.K. and K.Z.M. are grateful to the Agencia Estatal de
Investigacion, Ministerio de Ciencia e Innovacion, Spain for funding this research under
Proyectos de Generacion de Conocimiento 2022 program, PID2022-139776NB-C65. K.Z.M also
would like to thank the European Commission (Marie Sklodowska-Curie Cofund Programme;
grant no. H2020-MSCA-COFUND-2020-101034228-WOLFRAM2) for supporting this research.
T.K. acknowledges the 3rd edition of Microgrants in Action IV.4.1 - 'A complex programme of
support for UW PhD students', implemented as part of the 'Excellence Initiative - Research
University' (IDUB) Programme. A.L-R. acknowledges The Warsaw University of Technology Excellence Initiative IDUB Technologie Materialowe-3 ADVANCED POB 1820/359/Z01/POB5/2021 for supporting this research. A.L.-R. would also like to thank The Warsaw University of Technology Excellence Initiative (POSTDOC PW V edition PR-IDUB/374/Z01/Z10/2023. J.A.M. acknowledges the support from Centera2 project (FENG.02.02-IP.02.01-IP.05-T0004/23) funded with IRAFENG program of Foundation for Polish Science and cofinansed by the EUFANG Program. A portion of this work was performed at the National High Magnetic Field Laboratory, which is supported by National Science Foundation Cooperative Agreement No. DMR-2128556*, the State of Florida, and the U.S. Department of Energy.

\section*{Author contributions}

A.E.L.-R. performed high-field measurements with J.S.B., analysed the experimental data, initiated the collaboration on the theoretical framework, and wrote the experimental sections of the manuscript. 
J.S.B. also carried out SEM and Raman characterisation and performed statistical analysis. K.Z.M. developed the theoretical framework and supervised the modelling. T.K. performed the numerical simulations. M.M. and J.A.M. contributed to the interpretation of the theoretical results, with M.M. carrying out additional spin calculations. N.P. contributed to the development of the SIESTA implementation required for magnetic-field simulations. D.G.R. and F.F.B. supported the high-field experiments at Los Alamos. K.K., A.E.L.-R. and K.Z.M. secured funding. All authors discussed the results and contributed to the manuscript.

\section*{Competing Interests}

The authors declare no competing interests.

\section*{Additional information}
\textbf{Supplementary Information} The online version contains supplementary
material available at ~\ref{SI}

\FloatBarrier

\begin{appendices}
	
\section{Extended Data}\label{extended}
	
\begin{figure}[h!]
	\centering
	\includegraphics[width=0.95\textwidth]{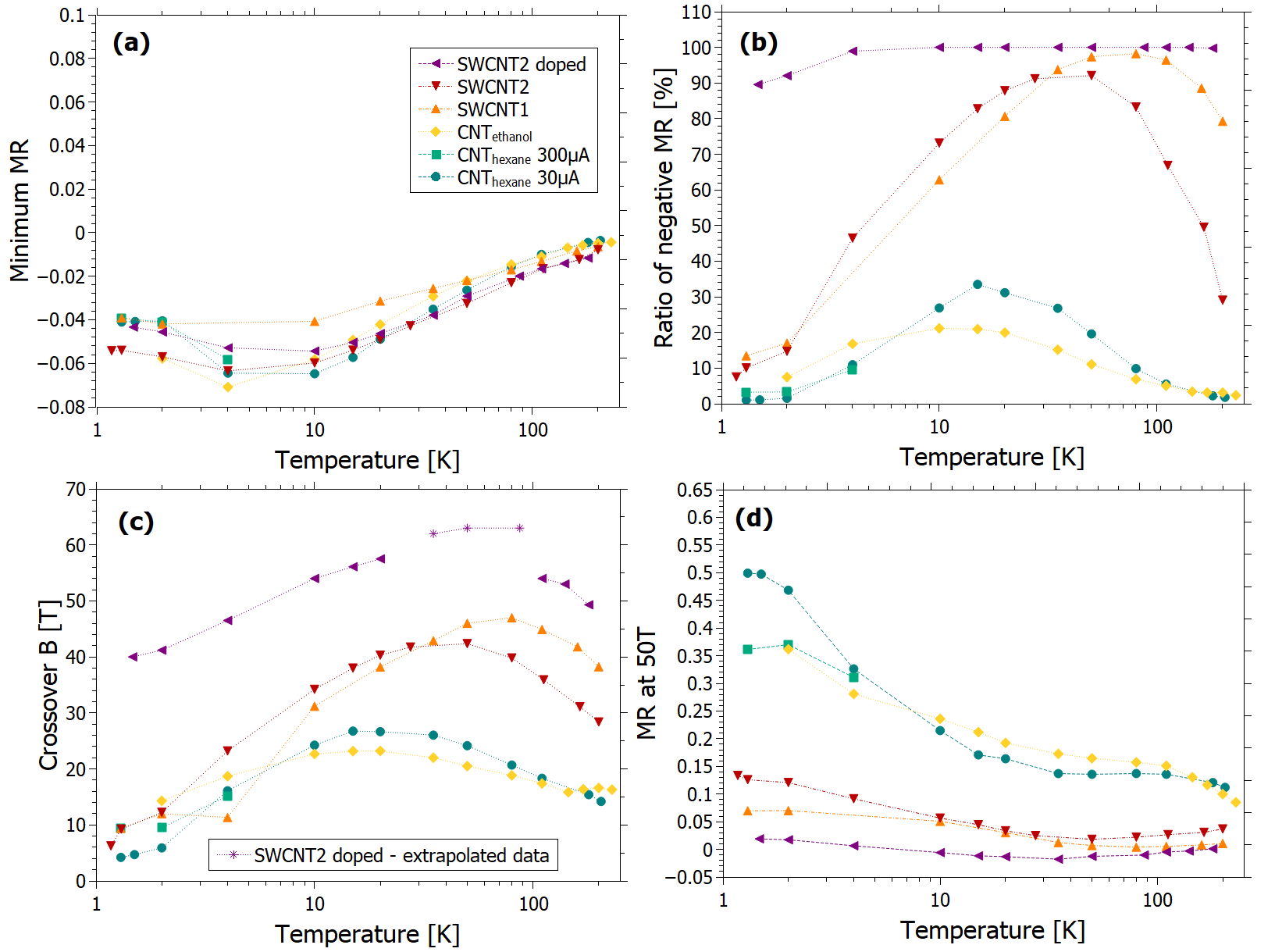}
	\caption{Analysis of the experimental MR data presented in Fig.~\ref{Fig2}. a) Maximum MR recorded at 50T against temperature for all samples. b) Percentage ratio of negative MR in relation to positive MR at specific temperatures for all tested samples. c) Minimum MR recorded for every temperature.  d) Magnetic field at which the MR of the samples changes from negative to positive versus temperature of the measurement. In the case of SWCNT doped sample the 3 points shown as stars are found by extrapolation of the MR curves above 60\,T.}
	\label{FigE1}
\end{figure}	
	
\begin{figure}[h!]
	\centering
	\includegraphics[width=0.95\textwidth]{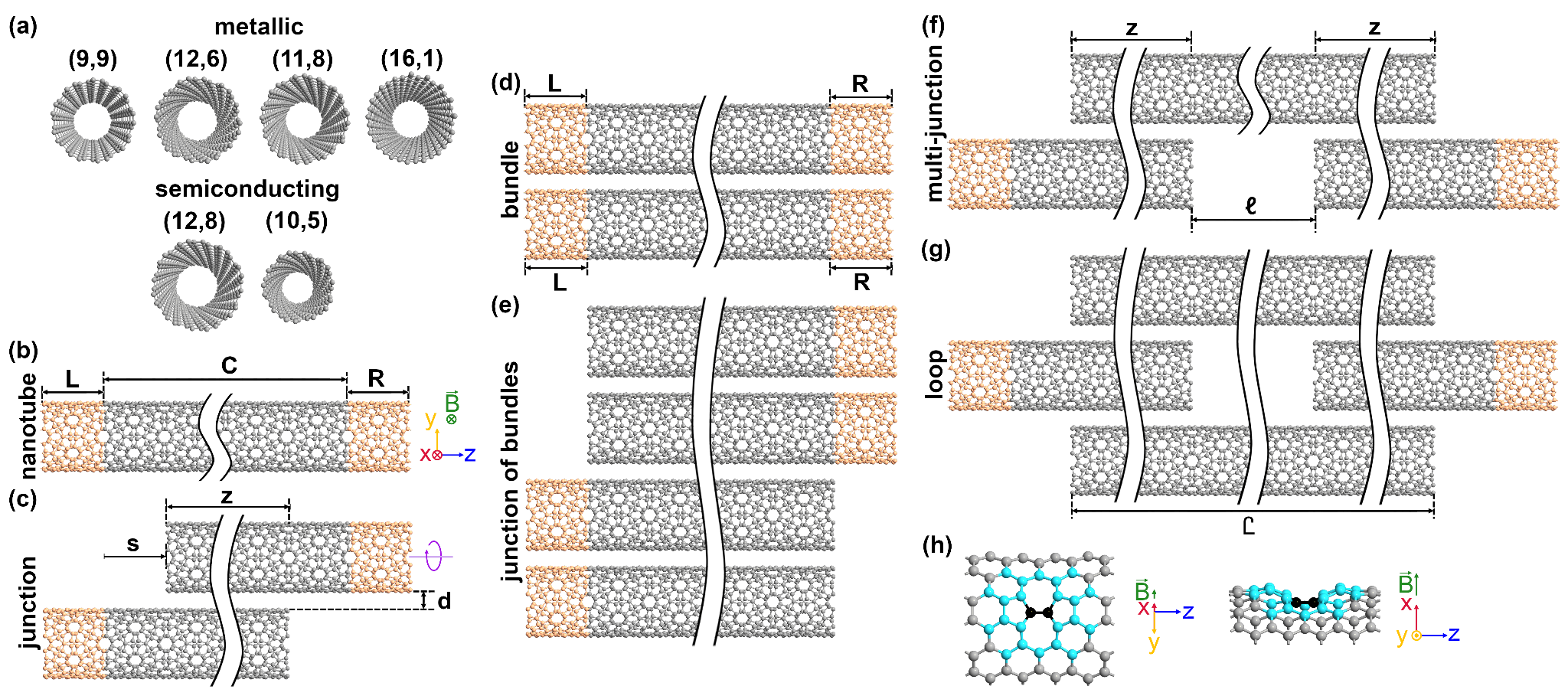}
	\caption{Atomistic views of models used for transport calculations. (a) Various types of SWCNTs considered: four metallic and two semiconducting nanotubes with similar diameters. The C-C bond length (d$_\mathrm{CC}$) was set to 1.42\,\AA. Examples of different device systems built from (12,6) nanotubes: (b) simple nanotube, (c) junction, (d) bundle, (e) junction of bundles, (f) multi-junction, and (g) loop coupled to pure SWCNT electrodes. The left (L) and right (R) electrodes, composed of the same type of nanotube as the connecting nanotube in the central region (C) of the device, are marked in gold in all panels. Each electrode in the bundle and junction of bundles systems contains two interacting nanotubes. Due to the large length of the actual models, the visualizations contain one or more breaks. Except for one case, nanotubes forming different types of junctions were placed at a distance (d) of 3.356\,\AA$ $ from each other. The length of the overlap region, z, was in some cases smaller than the length of the central region, as shown in (c), (f), and (g). In multi-junctions and loop systems, overlap regions were symmetrical. The mutual alignment of nanotubes in junctions and bundles was varied as indicated by the shift length, s, and the purple rotation axis displayed next to the upper junction's nanotube. In multi-junctions and loop systems, nanotubes sandwiched between electrodes were separated by a cavity with length $\ell$. The total length of each loop, $\mathrm{\mathcal{L}}$, is thus the sum of the lengths of two overlap regions (2z) and the cavity length ($\ell$). (h) Top and side views of a fragment of fully optimized (9,9) nanotube with Stone-Wales (SW) defect. Atoms forming the SW defect are marked in black, while atoms subject to geometry optimization are marked in light blue. Note that this defect not only transforms four hexagons into two pentagons and two heptagons but also induces significant deformation of the nanotube lattice.}
	\label{FigE2}
\end{figure}	

\begin{figure}[h!]
	\centering
	\includegraphics[width=0.95\textwidth]{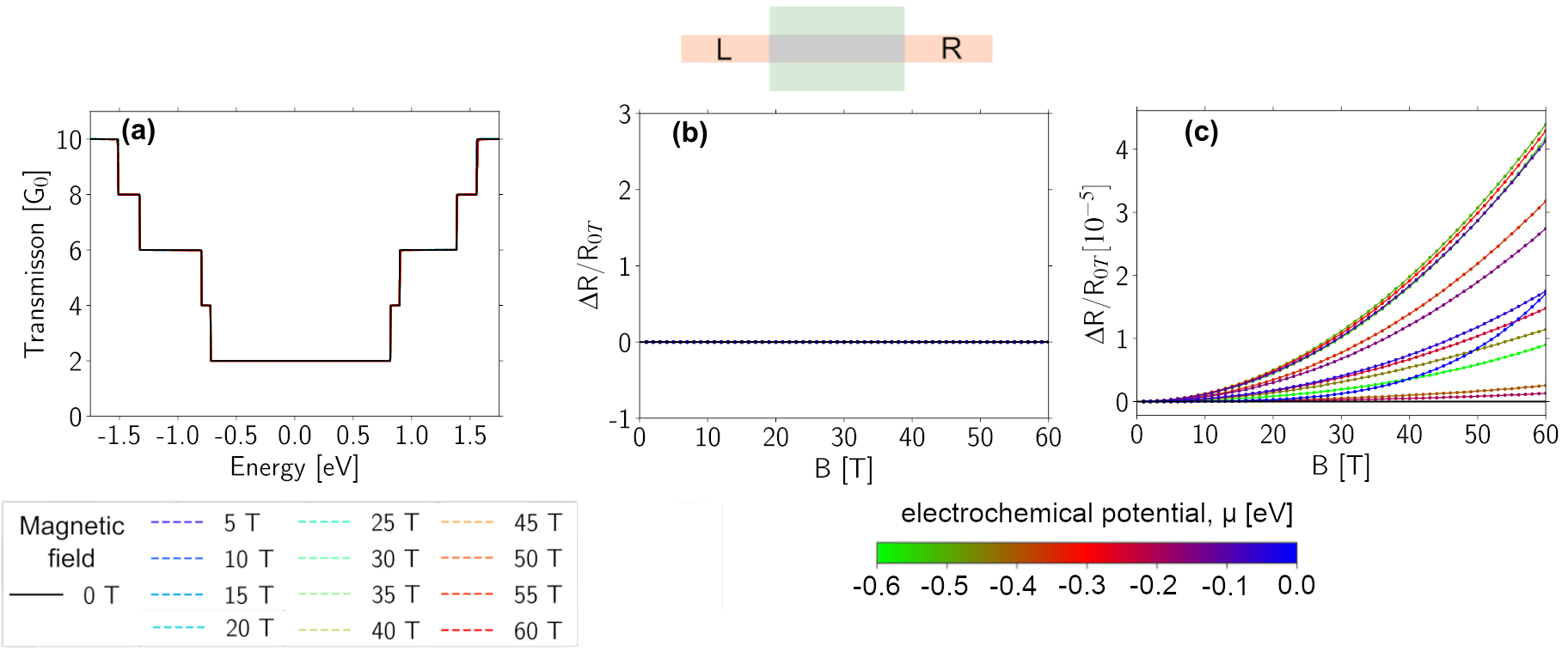}
	\caption{The computed zero-bias transmission spectrum and magnetoresistance of a single 22.6\,nm long (12,6) carbon nanotube under an external perpendicular magnetic field (B) at various doping levels. (a) Transmission spectrum showing the characteristic step-like quantisation of a pristine CNT. (b) Representative magnetoresistance ($\Delta$R/R$_\mathrm{0T}$) functions plotted for doping level (electrochemical potential $\upmu$) within the range [-0.6, 0.0]\,eV, sampled at 0.001\,eV intervals. MR functions were selected to ensure clarity and proportional representation of different trends, with the number of lines depending on the observed diversity and the need for visual clarity. (c) Zoomed view highlighting that, although the overall MR appears field-independent at this scale, a tiny positive MR response exists at the 10$^{-5}$ level. The scale in (b) matches that used later for junction MR(B) curves, illustrating that the apparent absence of MR in single tubes arises from its extremely small magnitude. As noted by Nemec and Cuniberti\cite{nemec2006}, perpendicular fields would need to reach thousands of tesla to induce appreciable magnetoelectronic effects in single CNTs.}
	\label{FigE3}
\end{figure}	

\begin{figure}[h!]
	\centering
	\includegraphics[width=0.95\textwidth]{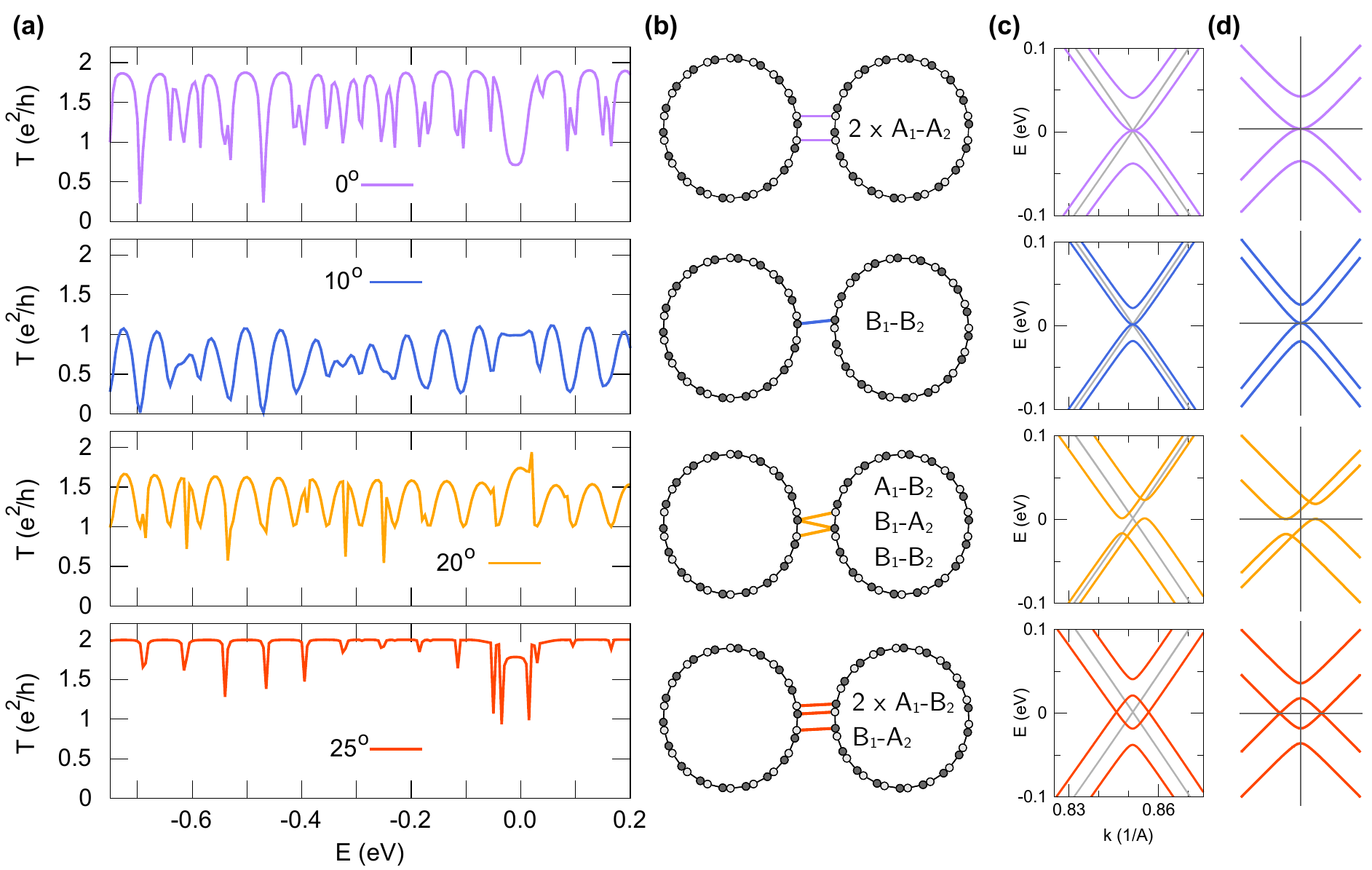}
	\caption{The transmission (a) through a (9,9)+(9,9) CNT junction with junction length 24.6\,nm, in four relative orientations whose cross-sections are shown in (b). The labels in (b) indicate how many atoms from which sublattices (A in light, B in dark grey) in tube 1 and 2 are connected by hopping, forming the junction. (c) The band structure of a (9,9)+(9,9) CNT bundle, with corresponding atomic configurations. Grey lines show the dispersion of the individual, uncoupled, CNTs. (d)  Band structures obtained with a minimal model of two coupled linear bands.}
	\label{FigE4}
\end{figure}

\begin{figure}[h!]
	\centering
	\includegraphics[width=0.95\textwidth]{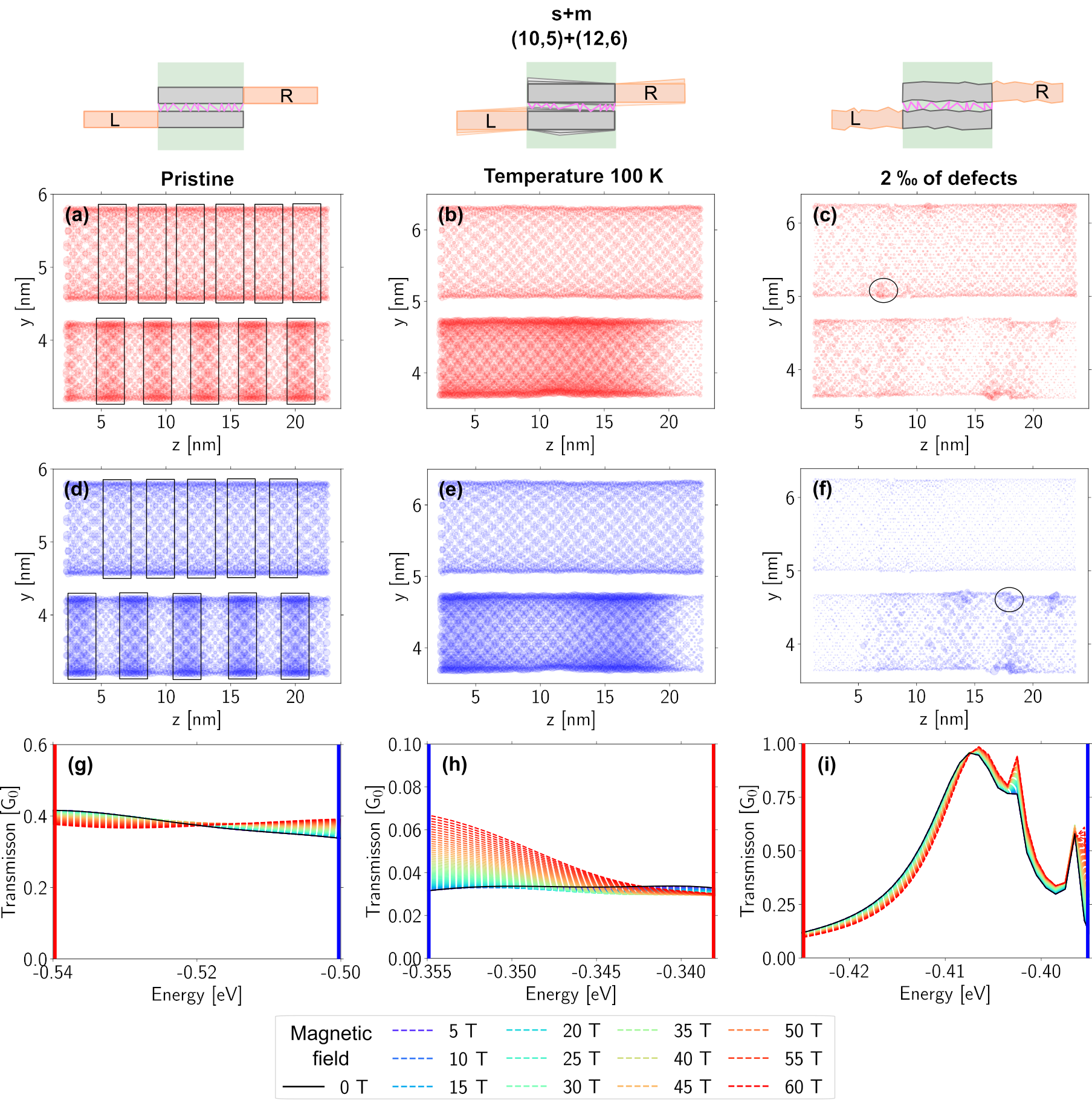}
	\caption{(a-f) Local density of states (LDOS) maps for the (10,5)+(12,6) semiconducting-metallic CNT junction (z$_\mathrm{initial}$ = 22.6\,nm) under different conditions: (left panel) pristine junction (no defects, no temperature effects), (middle panel) junction at 100\,K, and (right panel) junction with 0.2\,\% 5-7 defects (no temperature effects included). For each system, two LDOS maps are displayed: one at a doping level corresponding to negative magnetoresistance (MR, blue) and another at a nearby doping level corresponding to positive MR (red). These doping levels are chosen to be close in energy while maintaining similar transmission values, as illustrated in the transmission plots below (g-i). The exact doping levels are indicated by blue and red lines in the transmission plots: for the pristine system, LDOS maps correspond to -0.54\,eV (positive MR) and -0.50\,eV (negative MR); for the system at 100\,K, to -0.338\,eV (positive MR) and -0.355\,eV (negative MR); and for the system with 0.2\,\% defects, to -0.425\,eV (positive MR) and -0.395\,eV (negative MR). While transmission spectra are plotted for various external perpendicular magnetic field values, the LDOS maps are shown at B = 0\,T, representing the system's state before any magnetic field is applied.}
	\label{FigE5}
\end{figure}

\begin{figure}[h!]
	\centering
	\includegraphics[width=0.95\textwidth]{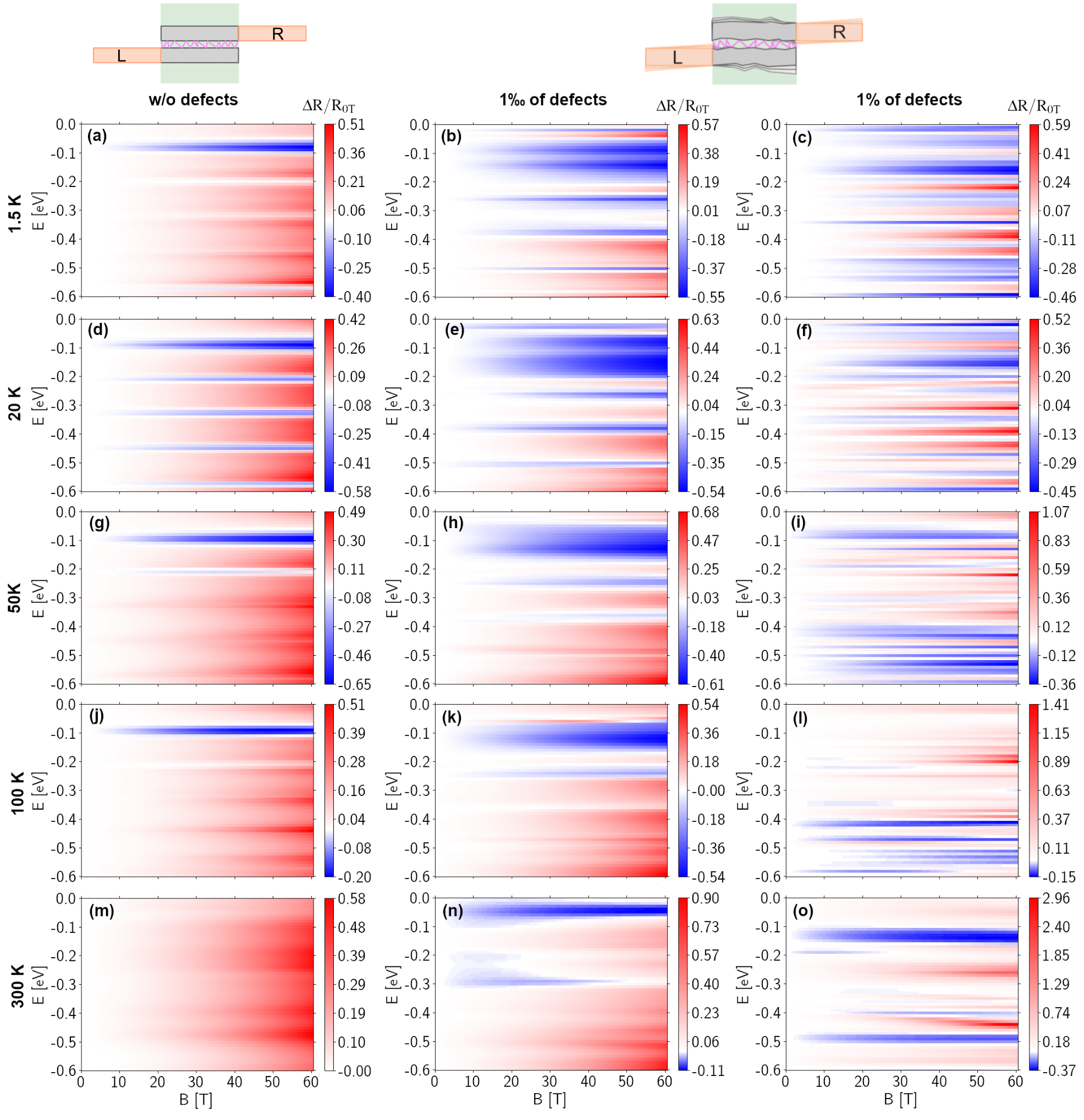}
	\caption{The computed magnetoresistance maps of (9,9)+(9,9) metallic-metallic (z$_\mathrm{initial}$ = 14.3\,nm) CNT junctions (a, d, g, j, m) without defects and with (b, e, h, k, n) 0.1\,\% of Stone-Wales defects (c, f, i, l, o) 1\,\% of Stone-Wales defects, at various temperatures, doping levels and under external perpendicular magnetic fields. Each panel displays a representative subset of MR functions plotted for doping levels in the range [-0.6, 0.0]\,eV, sampled at 0.001 eV intervals, with subsets selected proportionally to reflect different trends while ensuring clarity.}
	\label{FigE6}
\end{figure}

\begin{figure}[h!]
	\centering
	\includegraphics[width=0.75\textwidth]{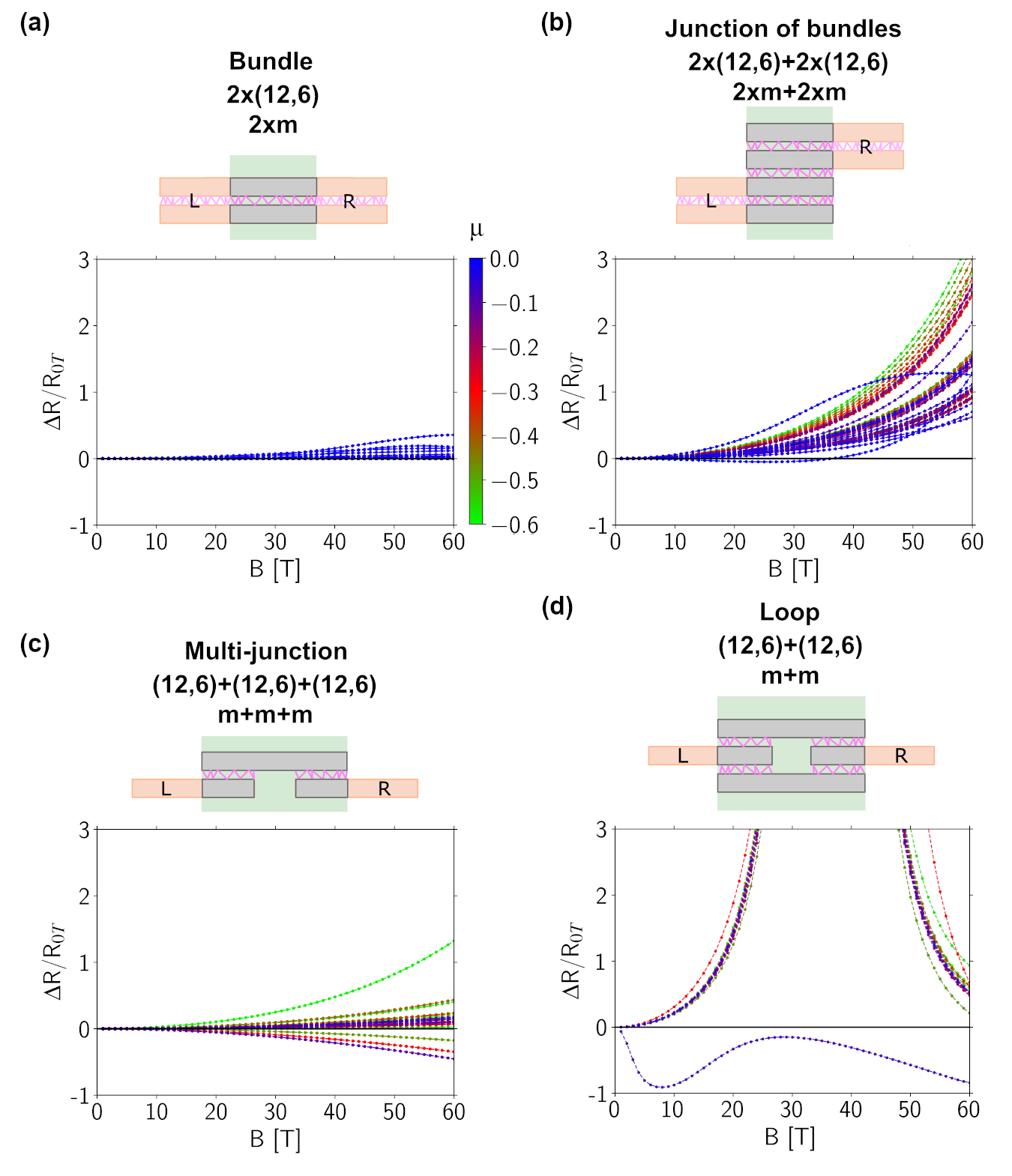}
	\caption{Magnetoresistance of complex connections. The computed magnetoresistance as a function of the external perpendicular magnetic field applied to metallic homogeneous systems: (a) a bundle, (b) a junction of bundles, (c) a multi-junction, and (d) a loop. Representative MR functions are shown for doping levels in the range [-0.6, 0.0]\,eV, sampled at 0.001\,eV intervals, with subsets selected to reflect distinct trends. The overlap region (z) is 22.6\,nm for systems (a) and (b), and 5.64\,nm for systems (c) and (d).}
	\label{FigE7}
\end{figure}

\FloatBarrier	
	
	\section{Supplementary Information}\label{SI}
	
	\subsection{Materials Characterization and  Additional Experimental Details}
	
	\subsubsection{Raman spectroscopy results}\label{secRaman}
	
Raman spectroscopy was performed using a silicon-calibrated Renishaw spectrometer with excitation wavelengths of 514, 633 and 785\,nm. Measurements were acquired with laser polarisation parallel and perpendicular to the CNT fibre axis. For each sample, spectra were collected at multiple spatial locations (at least 3 per sample) along the fibre to ensure reproducibility. 

\begin{figure}[h!]
	\centering
	\includegraphics[width=0.95\textwidth]{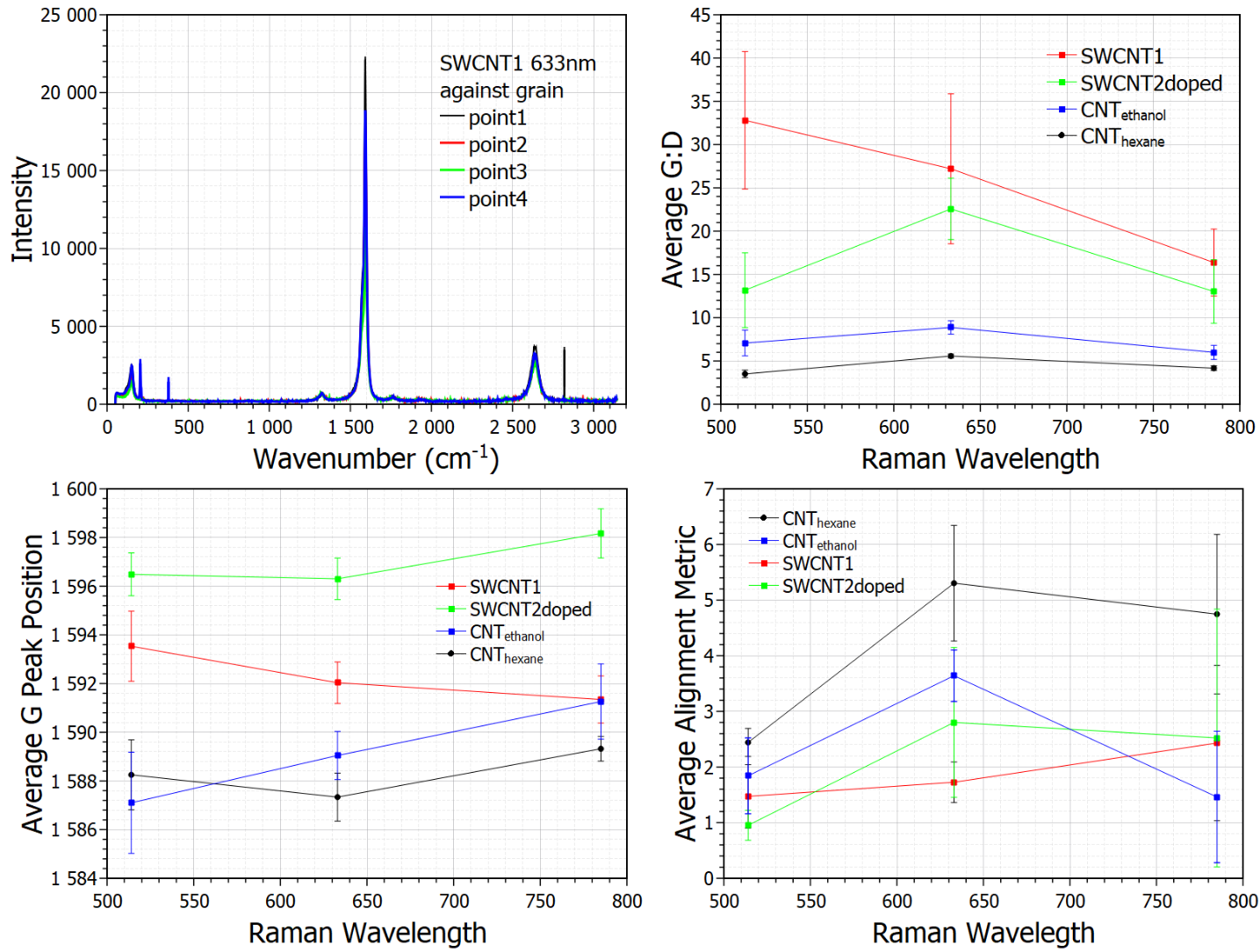}
	\caption{a) Representative Raman spectra measured at 4 spots of the SWCNT1 sample. b) Average G:D ratio, c) Average G peak position of measurements for all four samples. d) Average alignment metric calculated as the average magnitude of G peaks (from 4 spots) measured with grain and divided by the average values measured against grain. }
	\label{FigS1}
\end{figure}

Supplementary Figure~\ref{FigS1}a shows representative Raman spectra measured for the SWCNT\,1 sample. Every spectrum shows clear CNT features, i.e., G peak related to carbon-carbon bond vibrations within graphene plane; small D peak related to disorder of sp$^2$ bonds or presence of amorphous carbon; G' or 2D peak found in every graphite related material as well as radial breathing modes (RBMs) observed in the presence of SWCNTs and related to concentric stretching of their carbon-carbon bonds. RBM features were the most pronounced for the SWCNT\,1 sample. This confirmed that carbon disulfide fibre indeed contains mostly SWCNTs. The RBMs indicated both the presence of semiconducting and metallic chiralities. Taking into account that Raman picks up only some surface nanotubes, it is difficult to produce any conclusive answers about the share of specific chiralities in this sample. 

RBM features were much less visible in the CNT$_{\mathrm{hexane}}$ and CNT$_{\mathrm{ethanol}}$ samples, which may be explained by the fact that these fibres are composed of a mixture of nanotubes with respect to the number of walls. RBM visibility was also much poorer in the case of doped carbon disulfide fibre. This, however, should be better explained by the overall decrease in the intensity of Raman features in doped fibres\cite{lepak2018}. The same doping phenomenon may explain the lower G:D ratio of the SWCNT\,2 sample compared to the undoped SWCNT\,1 sample. The SWCNT\,1 sample showed the highest G:D ratio among all samples, indicating its highest purity and crystallinity (Supplementary Fig.~\ref{FigS1}b). As expected, due to the high impurity and MWCNT content, the lowest G:D ratio was observed in the CNT$_{\mathrm{hexane}}$ samples. While quite clean, CNTethanol samples containing single-, double-, and triple-wall CNTs had a G:D ratio higher than for CNT$_{\mathrm{hexane}}$ but lower than for the doped SWCNT\,2 sample. As shown in Supplementary Fig.~\ref{FigS1}c, the average G peak position of the SWCNT\,2-doped sample showed the highest up-shift of all samples, which clearly indicates p-doping. The slight up-shift of the G peak position of the SWCNT\,1 sample compared to the CNT$_{\mathrm{ethanol}}$ and CNT$_{\mathrm{hexane}}$ samples also indicates better conductivity of the clean, all SWCNT carbon disulfide sample. Finally, the average alignment plot (Supplementary Fig.~\ref{FigS1}d) shows that the carbon disulfide samples were not optimised for densification and alignment. It was, however, quite interesting that the best alignment was observed in hexane samples. 

\subsubsection{Room temperature resistance of the samples }\label{secEr}
	
	Before every MR measurement, the resistance of all samples was measured. The room-temperature DC resistance of the tested samples was 216\,$\Omega$, 81\,$\Omega$, 51\,$\Omega$, and 54\,$\Omega$ for CNT$_{\mathrm{hexane}}$, CNT$_{\mathrm{ethanol}}$, SWCNT\,1, and SWCNT\,2, respectively. The resistance of the SWCNT\,2 sample after doping dropped to 21\,$\Omega$. It was expected that the resistance of the CNT$_{\mathrm{hexane}}$ sample would be the highest, given the MWCNT structure of the fibres. The small difference in resistance between CNT$_{\mathrm{ethanol}}$ and SWCNT fibres may be attributed to poor densification and lower linear density of SWCNT fibres. 
	
\subsubsection{Scanning electron microscope analysis}\label{secEsem}

SEM imaging was performed using a Zeiss Gemini microscope with accelerating voltages of 2.00\,kV and working distances of 4.7--5\,mm, adjusted for each sample to optimise surface contrast and resolution. Detector configuration and pixel averaging were varied accordingly.

\begin{figure}[h!]
	\centering
	\includegraphics[width=0.95\textwidth]{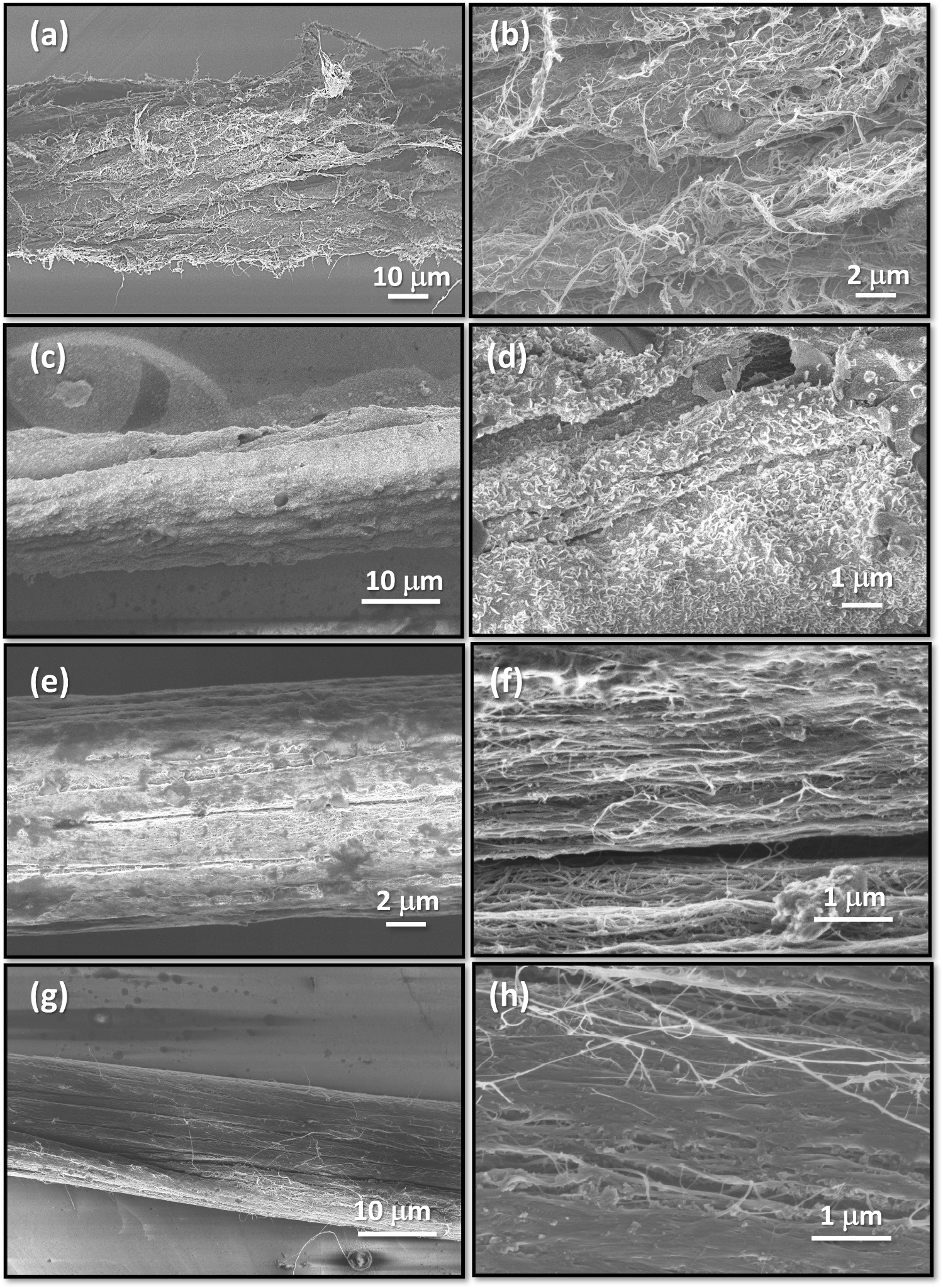}
	\caption{Scanning electron microscope images of a) and b) SWCNT\,1 sample, c) and d) SWCNT\,2 sample doped with nitric acid, e) and f) CNT$_{\mathrm{ethanol}}$ sample, g) and h) CNT$_{\mathrm{hexane}}$ sample.}
	\label{FigS2}
\end{figure}

SEM analysis reveals systematic differences in fibre surface morphology between the four CNT samples tested in the experiment (Supplementary Fig.~\ref{FigS2}). SWCNT\,1 shows a "fuzzy" surface with numerous CNTs/bundles protruding from the fibre body, often at large angles relative to the fibre axis, together with visible inter-bundle voids (Supplementary Figs.~\ref{FigS2}a,b). After nitric acid treatment, these voids appear partially filled, consistent with acid infiltration into the fibre structure as visible in the images of the doped SWCNT\,2 (Fig.~\ref{Fig1}c,d). CNT$_{\mathrm{ethanol}}$ fibres show a high degree of CNT alignment and densification (Supplementary Figs.~\ref{FigS2}e,f), whereas CNT$_{\mathrm{hexane}}$ samples display residual impurity clusters which fill in the voids between CNT bundles. These qualitative morphological differences are consistent with the variations in electronic transport discussed in the main text.

	\subsubsection{Preparation of samples}\label{secEsamples}

	Individual CNT fibres were mounted on sapphire substrates patterned with gold contact pads. Electrical measurements were performed in a four-probe configuration. The separation between the inner voltage contacts was 3\,mm, while the outer current contacts were spaced 5\,mm apart. This geometry ensured that the measured resistance reflected the collective transport of the fibre rather than the length of individual nanotubes. 
	The fibres were attached to the gold pads using conductive silver paint. To reduce mechanical motion during measurements, the fibre was fixed between the sapphire substrate and a second sapphire slab. The substrate assembly was mounted onto a plastic plug equipped with metal contact pins. Thin copper wires were soldered to the pins and connected to the gold pads using silver paint, providing electrical connection between the sample and the measurement probe.

\subsubsection{	Magnetotransport measurments}\label{secERMR}

Magnetotransport measurements were carried out at the National High Magnetic Field Laboratory (NHMFL), Los Alamos, using a short-pulse magnet capable of fields up to 65\,T. The magnetic field was applied perpendicular to the fibre axis, while the measurement current was applied along the fibre length.
The sample, a calibrated pick-up coil for field determination, and a temperature sensor were mounted together inside a removable multi-chamber cryostat positioned in the bore of the magnet. The cryostat was inserted into a liquid-helium refrigerator, enabling measurements over a wide temperature range.
Temperature-dependent resistance measurements at zero magnetic field were first recorded during cooling using standard instrumentation. For magnetoresistance measurements, the pulsed magnet generated magnetic-field pulses up to 60\,T, with a rise time of approximately 9\,ms and a total pulse duration of about 30\,ms.
To minimise electromagnetic noise associated with pulsed-field operation, resistance measurements were performed using an AC lock-in detection technique. A 28\,kHz sinusoidal excitation current was applied to the sample through a high-impedance circuit incorporating an isolation transformer. Depending on the measurement conditions, the excitation current ranged between 30 and 310\,$\upmu$A (RMS). Signal amplification and post-processing filtering were used to extract the magnetoresistance signal from the pulsed-field background noise.
Measurements were performed across a broad temperature range. Initial measurements above $\approx$110\,K were carried out using nitrogen cooling. At lower temperatures, the cryostat was cooled using helium, allowing measurements down to temperatures below 4\,K.

\subsubsection{Fitting of R--T data with VRH, WL and FIT models }\label{secFit}

Resistance--temperature data in Fig.~\ref{Fig1}b were analysed using the Mott three-dimensional variable range hopping (3D VRH) formalism\cite{yanagi2010}:
	\begin{equation}
R(T) = R_0 \exp\left[\left(\frac{T_0}{T}\right)^{1/4}\right],
\label{eqVRH}
\end{equation}
	where $R(T)$ is the resistance at temperature $T$; $R_{\mathrm{0}}$ is a prefactor, and is the characteristic Mott temperature. To enable comparison of multiple samples on the same plot, resistance was normalised by its value at 273.15\,K, and the data were plotted as:
\begin{equation}	
\ln\!\left(\frac{R}{R_{273.15\,\mathrm{K}}}\right)
\quad \text{vs.} \quad
T^{-1/4}.
	\label{eqVRH2}
\end{equation}

As shown in Fig.~\ref{Fig1}b, none of the samples exhibits linear behaviour over a broad temperature range, indicating that 3D VRH does not adequately describe the experimental data.

The temperature dependence of resistance in Fig.~\ref{Fig1}c was analysed using the fluctuation-induced tunnelling (FIT) model:
\begin{equation}	
R(T) = R_0 \exp\left[\frac{T_1}{T_0 + T}\right],
	\label{eqFIT}
\end{equation}
where R(T) and T denote the experimentally measured resistance and temperature, $R_{\mathrm{0}}$,$T_{\mathrm{0}}$ and $T_{\mathrm{1}}$ are fitting parameters of the FIT model\cite{xie2009}.

To compare temperature-dependent resistance across different samples on a single plot, resistance was normalised by its value at 273.15\,K and plotted as R/R$_{\mathrm{273.15\,K}}$ versus temperature (Fig.~\ref{Fig1}c). For visual clarity across the full measurement range, the temperature axis is shown on a logarithmic scale. The normalised data were then fitted using the FIT expression above. While the model reproduces only a limited range of temperatures, it does not provide a sufficient description of the experimental datasets for all samples.

The data in Fig.~\ref{Fig1}d were analysed using the two-dimensional weak localisation (2D WL) expression\cite{yanagi2010}:
\begin{equation}
	G(T) = G_0 + A \ln\!\left(\frac{T}{T_0}\right),
	\label{eqWL}
\end{equation}
where $G(T)$  is the conductance at temperature $T$, $G_\mathrm{0}$, $A$ and $T_\mathrm{0}$  are fitting parameters.

To compare different samples on a single plot, conductance was normalised by its value at 273.15\,K and plotted as  $G/G_{\mathrm{273.15\,K}}$ versus temperature. The temperature axis is shown on a natural logarithmic scale for visual clarity. Linear fits were applied to  $G/G_{\mathrm{273.15\,K}}$ as a function of $ln (T)$. While 2D weak localisation predicts linear behaviour in $ln (T)$, the experimental data do not exhibit consistent linearity of our data.

 \subsection{Modelling}\label{secM}
 
 To comprehensively model charge transport in CNT assemblies under conditions of high magnetic fields and finite temperatures, we have developed an integrated approach combining tight-binding (TB) calculations employing the non-equilibrium Green's function (NEGF) formalism, Peierls substitution (to accurately account for orbital effects of strong magnetic fields), and molecular dynamics (MD) simulations. This unified scheme enables simulation of magnetotransport through disordered CNT--CNT junctions, directly from atomistic configurations. Our approach builds upon the previously proposed MD-Landauer method\cite{markussen2017}, which couples classical MD simulations to electronic transport calculations via Landauer formalism. Unlike the Boltzmann transport equation (BTE), which typically relies on first-order perturbative electron-phonon coupling (via Fermi's golden rule), assumes linear screening of displacements, and requires harmonic phonon approximations and Bose--Einstein phonon populations, the MD-Landauer method solves the NEGF Green's function exactly for each atomistic configuration derived from MD simulations\cite{markussen2017}. In the Landauer setup used here, the contacts are treated as ideal reservoirs that inject and collect carriers, and the scattering inside the device is represented by elastic transmission through many thermally perturbed snapshots, thereby indirectly accounting for electron-phonon effects. This allows it to naturally include anharmonic and long-wavelength phonon scattering, structural disorder, and defect-induced scattering, without invoking assumptions of linear perturbation theory or equilibrium phonon distributions. If the MD region is sufficiently long, long-wavelength (Frohlich-type) scattering is also captured. In contrast, BTE-based methods can misestimate transport when higher-order phonon interactions (two-phonon processes) or deviations from harmonic behavior become significant\cite{hatanpaa2023}. While semiclassical Boltzmann transport (BTE) approaches remain highly successful in crystalline metals and semimetals--capturing Lorentz force magnetoresistance through Fermi-surface geometry and multicarrier compensation\cite{zhang2019}--they are less suited for non-periodic, junction-dominated systems such as CNT networks, where quantum-coherent tunnelling and interference govern charge transport. However, the MD-Landauer approach also has limitations: it does not include zero-point motion (because MD uses classical Boltzmann statistics rather than Bose--Einstein), nor does it incorporate phonon-assisted inelastic transitions involving energy exchange, such as phonon absorption events. Nonetheless, for disordered CNT--CNT junctions under ultrahigh magnetic fields and finite temperatures, where perturbative assumptions break down and junction disorder dominates transport, this approach provides more realistic and flexible transport predictions than BTE.
 
 At larger scales, complementary models have approached conduction in nanostructured assemblies using network or resistor-based formalisms that capture morphological connectivity and percolation effects\cite{yao2020}. However, such classical descriptions neglect the quantum-coherent and field-dependent transport that governs magnetoresistance in CNT junctions. Our TB--NEGF--MD--Peierls transport framework bridges this gap by resolving local interference and Peierls-phase effects within a network-like geometry, thereby linking atomistic and mesoscale descriptions of transport.
 
 \subsubsection{Models}
 
 As shown in Extended Data Fig.~\ref{FigE2}a), we focused on single metallic and semiconducting SWCNTs of similar diameters but different chiralities from which we constructed models for transport calculations. The systems that we investigated can be divided into the following groups: nanotube, junction, bundle, junction of bundles, multi-junction, loop, which are depicted in Extended Data Figs.~\ref{FigE2}b-g, respectively. 
 
  \subsubsection{Defects}
  
  The Stone-Wales (also known as 5-7) defects were homogeneously distributed in the selected systems, comprising approximately 0.2\% or 1\% of all bonds. Each rotated bond  (Extended Data Fig.~\ref{FigE2}h, black atoms) and surrounding carbon atoms up to nearest neighbors away from the defected bond (Extended Data Fig.~\ref{FigE2}h), light blue atoms) were relaxed using a semi-empirical interatomic potential for carbon (MEAM$\_$C$\_$2005)\cite{lee2005}, as implemented in the QuantumATK numerical package\cite{schneider2017, smidstrup2020, atk}. This potential, based on the modified embedded atom method formalism, includes second-nearest-neighbor interactions between carbon atoms and additional terms for many-body screening. It allows to reproduce reasonably well the physical properties of various nanocarbon systems, including defects and surface properties\cite{lee2005}. Geometry optimization was carried out until the maximum force acting on any atom was lower than 0.001\,eV/\AA.   For each defect concentration, ten different randomly chosen configurations of defects were prepared to generate a set of configurations for calculating the averaged transmission coefficient for the specific defected system.
  
  \subsubsection{Effect of temperature}
  
	To obtain perturbed non-ideal structures at non-zero temperatures, we performed a series of MD simulations of simple SWCNT junctions (Extended Data Fig.~\ref{FigE2}b) composed of metallic nanotubes with the same or different chiralities, semiconducting nanotubes, and mixed junctions containing both metallic and semiconducting nanotubes. The simulations were carried out in two steps:
	
\begin{enumerate}
	\item 	Electrodes simulation: NVT simulations were first applied to the electrodes. The final snapshot was used to construct the full device.
	\item Full Device simulation: We then performed NVT simulations on the full device, keeping the electrodes frozen. The final snapshot from each MD run was used for transport calculations.
\end{enumerate}

Both steps were repeated ten times for each temperature to generate a set of configurations for calculating the averaged transmission coefficient.

All simulations used the Langevin thermostat\cite{collins2017} with the MEAM$\_$C$\_$2005 potential\cite{lee2005}, implemented in the QuantumATK numerical package\cite{schneider2017, smidstrup2020, atk}. Random initial velocities were assigned according to the Maxwell-Boltzmann distribution, with the center of mass fixed. The simulations were carried out with a time-step of 0.1\,fs over a time period of 5\,ps (50\,000 steps) and a friction constant of 0.01\,fs$^{-1}$ at temperatures of 1.5\,K, 4\,K, and 100\,K. For the (9,9)+(9,9) junction, additional simulations were conducted at 2\,K, 10\,K, 20\,K, 30\,K, 50\,K, 80\,K, 200\,K, and 300\,K.
	
\subsubsection{Tight-binding calculations}

\begin{figure}[h!]
	\centering
	\includegraphics[width=0.65\textwidth]{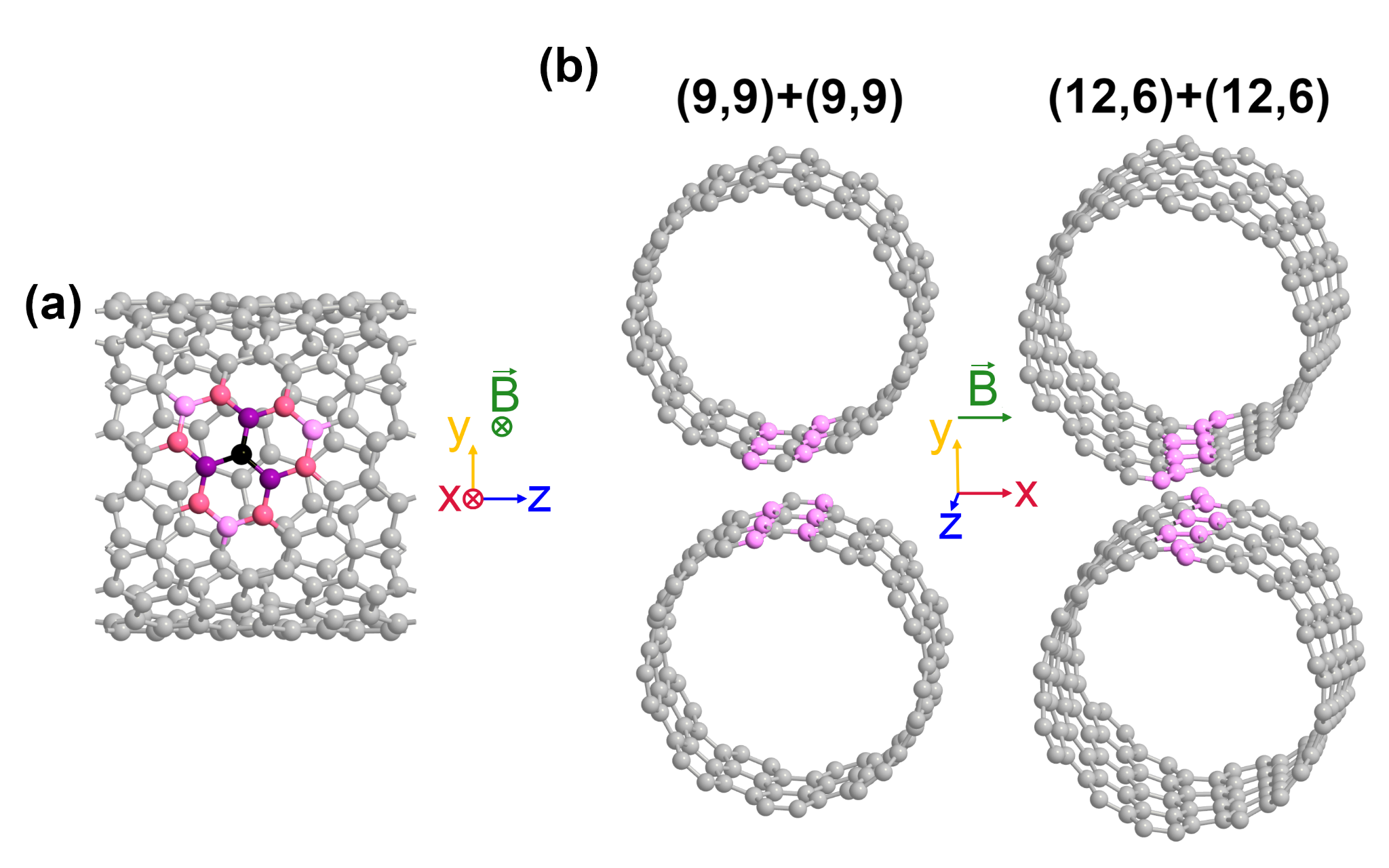}
	\caption{(a) Nearest neighbors (purple), next nearest neighbors (watermelon), and third nearest neighbors (pink) in the nanotube. The central atom is marked in black. For clarity, other carbon atoms are shown with a higher degree of transparency. (b) Atoms from different nanotubes that interact with each other are marked in pink, while other atoms are grey and shown with a higher degree of transparency. Interactions between nanotubes occur only as third nearest neighbors interactions and depend on their distance, chirality, and alignment. At d=3.356\,\AA, in a perfectly aligned (9,9)+(9,9) junction, only same-sublattice atoms from both nanotubes interact, while in a perfectly aligned (12,6)+(12,6) junction, atoms from both sublattices of each nanotube interact. }
	\label{FigS3}
\end{figure}	  
	  
We used sisl numerical package\cite{sisl} to define the tight-binding Hamiltonian with up to third nearest neighbors interactions\cite{hancock2010, saito1998}:
\begin{equation}
	  	H = - \sum_{i j \sigma} 
	  	\left( t_{ij}\, a^{\dagger}_{i\sigma} a_{j\sigma} + \mathrm{h.c.} \right) + U,
	  	\label{eqH}
\end{equation}
  which can be separated into two linear Hamiltonians $H_{\sigma=\uparrow}$ and  $H_{\sigma=\downarrow}$ for electrons with spin $\sigma = \uparrow$ and $\downarrow$. The first term of $H$ is hopping with $a^{\dagger}_{i\sigma}$   ($a_{i\sigma}$) being creation (annihilation) operators, and $t_{ij}$ 
  being a hopping integral from site $j$ to $i$.  In our model we take hopping parameters from Ref.~\cite{hancock2010} as: nearest neighbor (distance from the initial atom smaller than 1.5~\AA) 
  $t_1 = -2.97$~eV, next nearest neighbor  (distance between 1.5~\AA\ and 2.5~\AA) 
  $t_2 = -0.073$~eV, third nearest neighbor  (distance between 2.5~\AA\ and 3.6~\AA) 
  $t_3 = -0.33$~eV. In this model atoms do not interact when the distance between them is bigger than  3.6~\AA, and the interaction of atoms from different nanotubes is always of a third 
  nearest neighbor type (see Supplementary Fig.~\ref{FigS3}a). $U$ is a constant shift in energy that places the Fermi energy at 0\,eV. 
  Elements of an overlap matrix $S$ between basis states centered on different lattice 
  sites can be written in local basis states, $|i\rangle_{\sigma}$, as
 $S_{ij\sigma} = {}_{\sigma}\langle i | j \rangle_{\sigma}$, with values in our model taken from Ref.~\cite{hancock2010} as: nearest neighbor   $s_1 = 0.073$, next nearest neighbor $s_2 = 0.018$, third nearest neighbor $s_3 = 0.026$. The density matrix elements are:
 \begin{equation}
 	\rho_{ij\sigma} = \sum_{\alpha=1}^{\mathrm{occup.}} 
 	\; {}_{\sigma}\langle i | \Psi \rangle_{\sigma}^{\alpha}
 	\; {}_{\sigma}^{\alpha}\langle \Psi | j \rangle_{\sigma},
 	\label{eqRho}
\end{equation}
 where $|\Psi\rangle_{\sigma}^{\alpha}$ is the $\alpha^{\mathrm{th}}$ single-particle eigenstate.
 For spin-unpolarized calculations the Hamiltonian is reduced to:
 \begin{equation}
 	H = -2 \sum_{ij}
 	\left( t_{ij}\, a^{\dagger}_{i} a_{j} + \mathrm{h.c.} \right) + U.
 	\label{eqH2}
 \end{equation}
As the applied external magnetic field B is very high, it cannot be treated in a perturbative manner. Therefore, we add the external magnetic field to the Hamiltonian  by Peierls substitution~\cite{goldman2016,nemec2006, peierls1933} that changes the hopping  term by multiplying it by a phase factor:
\begin{equation}
	- \sum_{ij\sigma} t_{ij} a^{\dagger}_{i\sigma} a_{j\sigma}
	\;\rightarrow\;
	- \sum_{ij\sigma} t_{ij}
	\left(
	e^{\, i \frac{2\pi}{\Phi_0} \int_{\mathbf r_j}^{\mathbf r_i} \mathbf A \cdot d\mathbf r}
	\right)
	a^{\dagger}_{i\sigma} a_{j\sigma},
	\label{eqSubP}
\end{equation}

with $\mathbf r_i = (x_i, y_i, z_i)$ being the $i^{\mathrm{th}}$ site coordinates,
$\Phi_0$ the flux quantum, and $\mathbf A$ the magnetic vector potential
defined up to the gauge. In our model we choose
\[
\mathbf A = (-y B_z,\, 0,\, y B_x),
\]
and what follows:
\begin{equation}
	i \int_{\mathbf r_j}^{\mathbf r_i} \mathbf A \cdot d\mathbf r
	= \frac{i}{2} (-B_z \Delta x + B_x \Delta z)(y_i + y_j).
	\label{eqAz}
\end{equation}
Taking into account that in our experiments the magnetic field was perpendicular to the sample's plane, we mostly focus on the case $B_z = 0$:
\begin{equation}
	i \int_{\mathbf r_j}^{\mathbf r_i} \mathbf A \cdot d\mathbf r
	= \frac{i}{2} B_x (z_i - z_j)(y_i + y_j).
		\label{eqAz2}
\end{equation}
  
	  \subsubsection{Spin effects}\label{secTspin}
	  
	  For spin-polarized calculation, we need to account also for the Zeeman effect in the usual way:
	  \begin{equation}
	  	H = - \sum_{ij\sigma}
	  	\left( t_{ij}(B)\, a^{\dagger}_{i\sigma} a_{j\sigma} + \mathrm{h.c.} \right)
	  	+ \mu_B B \sum_{i\sigma} \sigma\, a^{\dagger}_{i\sigma} a_{i\sigma}.
	  	\label{eqHZeeman}
	  \end{equation}
	  Since we disregard the spin-orbit effects, the spin quantization axis is parallel to the applied
	  magnetic field.
 \subsubsection{Transport Calculations}
	  
To perform transport calculations we used TBtrans\cite{papior2017}, the tight-binding transport code implementing non-equilibrium Green function formalism. A unique feature of TBtrans is its  ability to allow any kind of user intervention in the Hamiltonian of the central part of the device. However, this code does not support the inclusion of the Peierls substitution in electrodes, so the scattering region must be long enough to compensate for this limitation. Obtained physical quantity is transmission of an electron of energy E between two different electrodes L and R, which is computed as:
\begin{equation}
	  	T_{L \to R}(E)
	  	= \mathrm{Tr}
	  	\left[
	  	\Gamma_R(E)\, G(E)\, \Gamma_L(E)\, G^{\dagger}(E)
	  	\right],
	  	\label{eqT}
 \end{equation}
 where $\Gamma_L$ is a scattering matrix from L, and $G$ the Green function that can be expressed in terms of an overlap matrix $S$, Hamiltonian $H$, and self-energy $\Sigma$ as:
  \begin{equation}
  	G^{-1}(E)
  	= S(E + i\eta) - H - \delta H
  	- \sum_i \Sigma_i - \delta \Sigma.
  	\label{egG}
  \end{equation}
with the limit $\eta \to 0^{+}$. The differences in the Hamiltonian, and what follows in the self-energy, coming from the Peierls substitution when the magnetic field is applied are, respectively, $\delta H$ and
$\delta \Sigma$.

\begin{figure}[h!]
	\centering
	\includegraphics[width=0.95\textwidth]{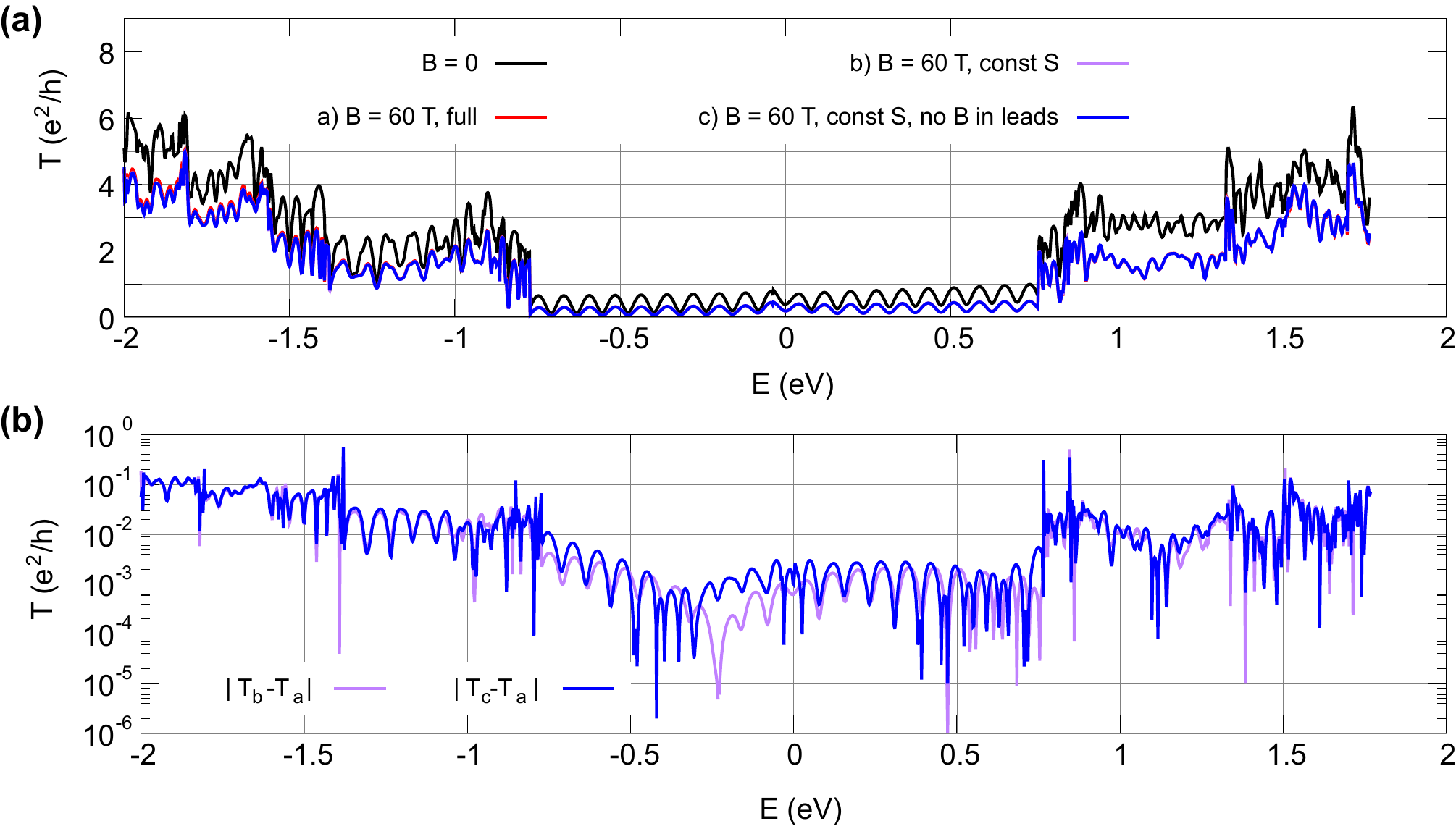}
	\caption{The consequences of including the Peierls phase in the overlap matrix S and in the Hamiltonian of the leads on the example of transmission through a (12,6)+(12,6) junction with length 24.6\,nm. (a) Transmission at B=0 and  B=60\,T with the different approaches. (b) The difference between the transmission calculated with the overlap matrix of the junction unmodified by the Peierls phase (T$_\mathrm{b}$), that with none of the magnetic field effects included in the leads (T$_\mathrm{c}$) and the one which includes all effects (T$_\mathrm{a}$). In this field, in the range from -0.6\,eV to 0\,eV, the error of neglecting the Peierls phase in the two approximations is below 1\,\%.}
	\label{FigS4}
\end{figure}

Note that TBtrans takes into account both the Hamiltonian $H$ and the   overlap matrix $S$ during transport calculations. The code uses   non-equilibrium Green's function method (NEGF) formalism, where the   Green's function is computed by combining both $H$ and $S$   (Eq.~\ref{eqT}), ensuring that the non-orthogonal nature of the basis is   fully accounted for during the simulation. In systems with a   non-orthogonal atomic orbital basis, the overlap matrix $S$ is not the   identity matrix, and it explicitly enters into the transport   calculations through the Green's function formulation. The overlap   matrix modifies how the energy dependence of the Green's function is   calculated, as well as how electrons propagate through the system.  In the presence of the external magnetic field, both the Hamiltonian   $H$ and the overlap matrix $S$ should acquire the Peierls phase, to   ensure results valid for a general (also non-orthonormal) basis~\cite{cresti2021}.   However, in the current version of the TBtrans code, when a Peierls   phase is introduced, it only alters the Hamiltonian hopping elements,   while the overlap matrix retains its original values (Eq.~\ref{egG}). As shown in Figure~\ref{FigS4}, specifically, the omission of the Peierls phase   in $S$ results in transmission values that are slightly lower than   expected, with differences on the order of $2 \times 10^{-3}\, \mathrm{G}_\mathrm{0}$   at the highest applied $B$ values. $\mathrm{G}_\mathrm{0}$ is the quantum conductance   equal to 2e$^2$/h, where e is elementary charge and h is Planck's   constant.
  
While running the TBtrans code, apart from taking the above-mentioned Hamiltonian, one has to specify other parameters of the calculations. As the structure is not periodic, a k-point sampling is set to 1$\times$ 1$\times$ 1. For inputs without temperature, smearing is set to the default value (300\,K), while for temperature computations at 1.5\,K, 2\, K, 4\,K, 10\,K, 20\,K, 30\,K, 50\,K, 80\,K, 100\,K, 200\,K, or 300\,K all the electronic temperature is set to 1.5\,K, 2\,K, 4\,K, 10\,K, 20\,K, 30\,K, 50\,K, 80\,K, 100\,K, 200\,K, or 300\,K, respectively. The energy grid was calculated with a step equal 0.001\,eV.
	  
For each non-ideal system (at non-zero temperature or with defects), we repeated the simulations 10 times and then averaged the transmission arithmetically (see Supplementary Fig.~\ref{FigS5}a). The Kolmogorov-Smirnov test was used to evaluate the normality of the transmission values for each energy level. The results indicate that at a 5\% significance level, no significant deviations from normality were detected. Therefore, we assume error bars to be the standard error of the mean value.
	  
Local density of states (LDOS) on orbital $\nu$ for an energy $E$
was calculated as the following matrix element:
	  \begin{equation}
	  	\mathrm{LDOS}_{\nu}(E)
	  	= -\frac{1}{\pi}
	  	\mathrm{Im}\!\left[ G(E) S \right]_{\nu\nu},
	  	\label{eqLDOS}
	  \end{equation}
	  where $G$ is the Green function and $S$ an overlap matrix.
	  
	  In the post-processing of the data one can determine the zero-bias  conductance
	  \begin{equation}
	  	\sigma(E,B) = G_0 \cdot T_{L \to R}(E),
	  	\label{eqGG}
	  \end{equation}
	  the resistance
	  \begin{equation}
	  	R(E,B) = \frac{1}{\sigma(E,B)},
	  	\label{eqR}
	  \end{equation}
	  and the magnetoresistance
	  \begin{equation}
	  	\mathrm{\Delta R/ R_\mathrm{0\,T}}
	  	\equiv \frac{R(E,B) - R(E,B=0)}{R(E,B=0)}
	  	= \frac{\sigma(E,B=0) - \sigma(E,B)}{\sigma(E,B)} \equiv\mathrm{MR}(E,B),
	  	\label{eqMR}
	  \end{equation}
	  where	 $E$ corresponds to the chosen doping level ($\mu$).
	  Within the Landauer formalism, conductance is determined at the electrochemical potential. Varying the Fermi level therefore mimics electrostatic gating or chemical doping, rendering conductance, resistance, and MR explicitly energy dependent.
	  For non-ideal systems, $R$ and $\mathrm{MR}$ were calculated using Monte Carlo estimation with standard error of the mean
	  (see Figure~\ref{FigS5}b).

\begin{figure}[h!]
	\centering
	\includegraphics[width=0.75\textwidth]{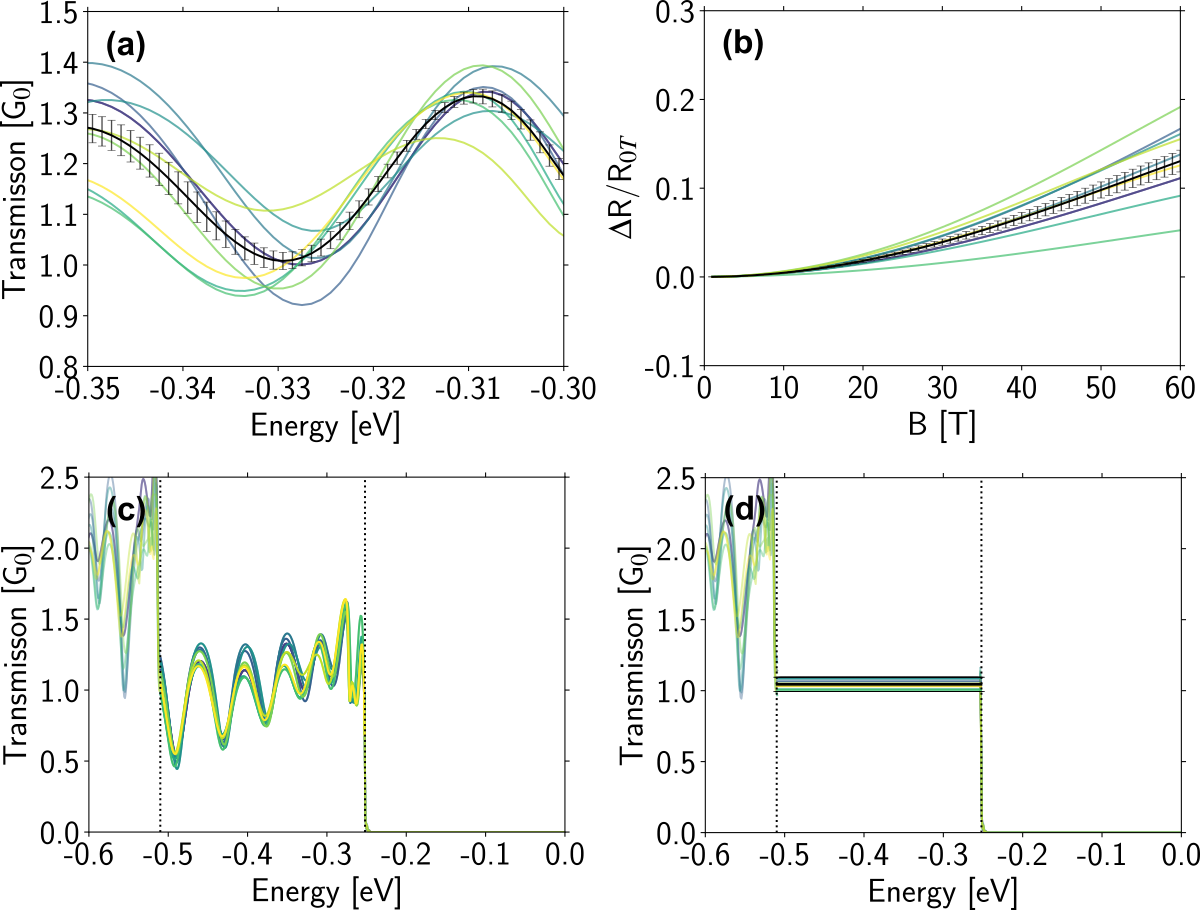}
	\caption{(a) An example of averaging transmission spectra over 10 different atomic configurations (coloured lines) obtained from MD simulations. The average transmission is shown as a black line, with error bars representing the standard error of the mean. (b) An example of magnetoresistance calculation from 10 different atomic configurations (coloured lines) obtained from MD simulations. For each temperature or defect concentration, the averaged MR (black line) was calculated using Monte Carlo Estimation with standard error of the mean. (c) An example of the first step of conductance calculation procedure -- averaging results for each of the 10 different atomic configurations (coloured lines) over the energy range corresponding to the first transmission step. The boundaries of the energy range are marked with vertical dotted lines. (d) An example of the second step of conductance calculation procedure -- averaging over 10 different atomic configurations (coloured lines). The boundaries of the energy range are marked with vertical dotted lines. The mean value is indicated by the central horizontal black line, while the two additional horizontal lines above and below represent error bars corresponding to the standard error of the mean.}
	\label{FigS5}
\end{figure}		  
	  
Computed zero-bias transmission spectra and magnetoresistance are typically presented over the energy range from 0\,eV to -0.6\,eV, covering both undoped and moderately p-doped regimes. In  experiments, untreated CNT fibres are generally slightly p-doped due to residual chemicals from synthesis and processing\cite{hayashi2020}. The exact doping level likely varies between samples, depending on the degree of acid infiltration during CNT fibre pretreatment. For nitric-acid-treated samples, prior estimates suggest Fermi level shifts of approximately -0.6\,eV\cite{hayashi2020}. As the precise doping level in our experimental devices was not measured, this energy window is taken as representative of the experimentally relevant conditions.
	  
	  \subsubsection{Deformation and junction quality parameters}
	  
In order to measure disorder in our system, we established deformation and junction quality parameters. The first one focuses on how far the researched nanotube is from the ideal one and was computed as an
average difference between positions of atoms of non-ideal $(x_i, y_i, z_i)$ and ideal
$(x_i^{\mathrm{ideal}}, y_i^{\mathrm{ideal}}, z_i^{\mathrm{ideal}})$ systems:

\begin{equation}
	\alpha
	=
	\frac{1}{N}
	\sum_{i=1}^{N}
	\sqrt{
		(x_i - x_i^{\mathrm{ideal}})^2
		+
		(y_i - y_i^{\mathrm{ideal}})^2
		+
		(z_i - z_i^{\mathrm{ideal}})^2
	},
	\label{eqAlpha}
\end{equation}
where $N$ is the total number of atoms in the device.

To gain insight into the quality of the junction, the parameter $\gamma$ was introduced to describe the number of atomic pairs $Y_{ij}$, atom $i$ from the first nanotube and $j$ from the second,
that interact with each other:
\begin{equation}
	\gamma = \frac{Y_{ij}}{N_1 \cdot N_2},
	\label{eqGamma}
\end{equation}
where $N_1$ ($N_2$) is the total number of atoms in the first (second) CNT.
	
 \subsubsection{Quantification of property-specific impact on magnetoresistance}\label{secTfactors}	
      
To quantify the relative influence of various structural and electronic properties on magnetoresistance, a reference configuration was defined for each property and compared against a modified counterpart. MR values were extracted over a defined energy window within the doping range. For metallic systems, this window spanned [0,-0.6]\,eV relative to the Fermi level (E\,=\,0), while for semiconducting systems it began at the onset of non-zero transmission and extended to -0.6\,eV, thereby excluding band gap regions, as shown on magnetoresistance maps. The studied properties included: doping, metallicity, chirality, homogeneity, overlap length, rotational alignment, intertube distance, spin effects, temperature, and structural defects. Reference systems were selected based on physical relevance and availability of equivalent junction configurations.

For each case, the effect on magnetoresistance was quantified using key descriptors: the minimum and maximum MR values ($\mathrm{MR}_{\min}$, $\mathrm{MR}_{\max}$), and the percentage of the energy window corresponding to negative and positive MR ($P_{\mathrm{neg}}$, $P_{\mathrm{pos}}$). For a set of $N$ junctions or configurations indexed by $k$, the absolute change in extreme MR values was calculated as:
\begin{equation}
	\left< \Delta MR_{\min} \right>
	=
	\frac{1}{N}
	\sum_{k=1}^{N}
	\left(
	MR^{\mathrm{test}}_{\min,k}
	-
	MR^{\mathrm{ref}}_{\min,k}
	\right),
	\label{eqMRmin}
\end{equation}
\begin{equation}
	\left< \Delta MR_{\max} \right>
	=
	\frac{1}{N}
	\sum_{k=1}^{N}
	\left(
	MR^{\mathrm{test}}_{\max,k}
	-
	MR^{\mathrm{ref}}_{\max,k}
	\right),
\label{eqMRmax}
\end{equation}
where $MR^{\mathrm{ref}}_{\min,k}$ and $MR^{\mathrm{ref}}_{\max,k}$ are the minimum and maximum MR values,
respectively, for the reference configuration of junction $k$, while $MR^{\mathrm{test}}_{\min,k}$ and
$MR^{\mathrm{test}}_{\max,k}$ are the corresponding values for the test system. However, due to the large variability and sign changes observed across junctions, the final scoring was based on the median of the per-junction absolute MR changes, which provided a more robust and representative measure of typical behaviour.
          
The mean change in the percentage of negative and positive MR regions within the doping range was computed as:
\begin{equation}
      	\left< \Delta P_{\mathrm{neg}} \right>
      	=
      	\frac{1}{N}
      	\sum_{k=1}^{N}
      	\left(
      	P^{\mathrm{test}}_{\mathrm{neg},k}
      	-
      	P^{\mathrm{ref}}_{\mathrm{neg},k}
      	\right),
      	\label{eqPneg}
\end{equation}
\begin{equation}
      	\left< \Delta P_{\mathrm{pos}} \right>
      	=
      	\frac{1}{N}
      	\sum_{k=1}^{N}
      	\left(
      	P^{\mathrm{test}}_{\mathrm{pos},k}
      	-
      	P^{\mathrm{ref}}_{\mathrm{pos},k}
      	\right),
      \label{eqPpos}
\end{equation}
where $P_{\mathrm{neg},k}$ and $P_{\mathrm{pos},k}$ denote the percentage of the selected energy range for which MR is negative or positive, respectively, for junction $k$.

 To evaluate the combined strength of a given effect on magnetoresistance, a dimensionless impact score was defined as the product of absolute values of the median MR change and the corresponding MR percentage change:
\begin{equation}
      	I_{\mathrm{neg}}
      	=
      	\left| \mathrm{Median}\!\left(\Delta MR_{\min}\right) \right|
      	\,
      	\left| \left< \Delta P_{\mathrm{neg}} \right> \right|,
      	 \label{eqIneg}
\end{equation}
\begin{equation}
      	I_{\mathrm{pos}}
      	=
      	\left| \mathrm{Median}\!\left(\Delta MR_{\max}\right) \right|
      	\,
      	\left| \left< \Delta P_{\mathrm{pos}} \right> \right|.
       \label{eqIpos}
\end{equation}
This approach allowed for stable comparison of property-specific impact on magnetoresistance while retaining sensitivity to significant trends and minimising the influence of statistical outliers.     
These metrics were calculated independently for each property by comparing test systems (e.g., doped, rotated, defected) with their corresponding reference configurations, such as undoped junctions, ideal angular alignments, or pristine systems.
  
For doping, MR values at $E = 0$ served as the reference and were compared against the full doping range for each junction $(9,9)+(9,9)$, $(12,6)+(12,6)$, $(12,8)+(12,8)$, $(10,5)+(12,6)$, $(11,8)+(16,1)$. In the case of metallicity, each metallic junction $(9,9)+(9,9)$, $(12,6)+(12,6)$, $(11,8)+(16,1)$ was compared individually to a single semiconducting-semiconducting reference junction $(12,8)+(12,8)$, preserving junction-specific electronic behaviour. Chirality impact was evaluated by referencing the achiral armchair $(9,9)+(9,9)$ system (30$^\circ$ chiral angle) against the chiral  $(12,6)+(12,6)$ junction (19.1$^\circ$ chiral angle). Homogeneity was assessed by comparing mixed-tube junctions $(11,8)+(16,1)$, $(10,5)+(12,6)$ with same-tube systems  $(9,9)+(9,9)$, $(12,6)+(12,6)$, $(12,8)+(12,8)$. For this case, the MR descriptors ($\mathrm{MR}_{\min}$, $\mathrm{MR}_{\max}$, $P_{\mathrm{neg}}$, $P_{\mathrm{pos}}$) were first averaged over all doping energies and over all systems within each group (homogeneous or mixed). These averages were then used to define $MR^{\mathrm{test}}_{k}$ and $P^{\mathrm{test}}_{k}$ for the mixed-tube systems and     $MR^{\mathrm{ref}}_{k}$ and $P^{\mathrm{ref}}_{k}$ for the same-tube systems, thereby capturing the general effect of structural uniformity. For overlap length, each junction $(9,9)+(9,9)$, $(12,6)+(12,6)$, $(12,8)+(12,8)$, $(10,5)+(12,6)$ at $\sim$22\--25~nm overlap was used as a reference; MR data for other overlaps were compared within the same doping range. Rotational alignment and intertube distance were studied using the $(9,9)+(9,9)$ junction as a baseline, with fixed overlap and systematically varied angular or spatial configurations (see Supplementary Fig.~\ref{FigS13} and Supplementary Fig.\ref{FigS15}). Spin effects were quantified by comparing spin-polarised and spin-unpolarised calculations for selected junctions: $(9,9)+(9,9)$, $(12,6)+(12,6)$, $(11,8)+(16,1)$. Temperature effects were assessed by comparing results at 100~K (test) against those at 1.5~K (reference) for $(12,6)+(12,6)$, $(12,8)+(12,8)$, and $(10,5)+(12,6)$ junctions. Although the initial overlaps were identical for each junction, structural differences due to thermal motion were captured following molecular dynamics simulations.  The impact of structural defects was evaluated by introducing 1\% Stone--Wales defects in $(9,9)+(9,9)$, $(12,6)+(12,6)$, $(12,8)+(12,8)$, $(10,5)+(12,6)$, and $(11,8)+(16,1)$ junctions, and comparing their MR behaviour with the corresponding pristine counterparts. Final impact scores were normalised and presented as diverging bar charts (Fig.~\ref{Fig6}c) to visually compare the relative contributions of each property to both negative and positive MR behaviour.

 \subsubsection{Benchmarking Calculations}
 
\begin{figure}[h!]
	\centering
	\includegraphics[width=0.85\textwidth]{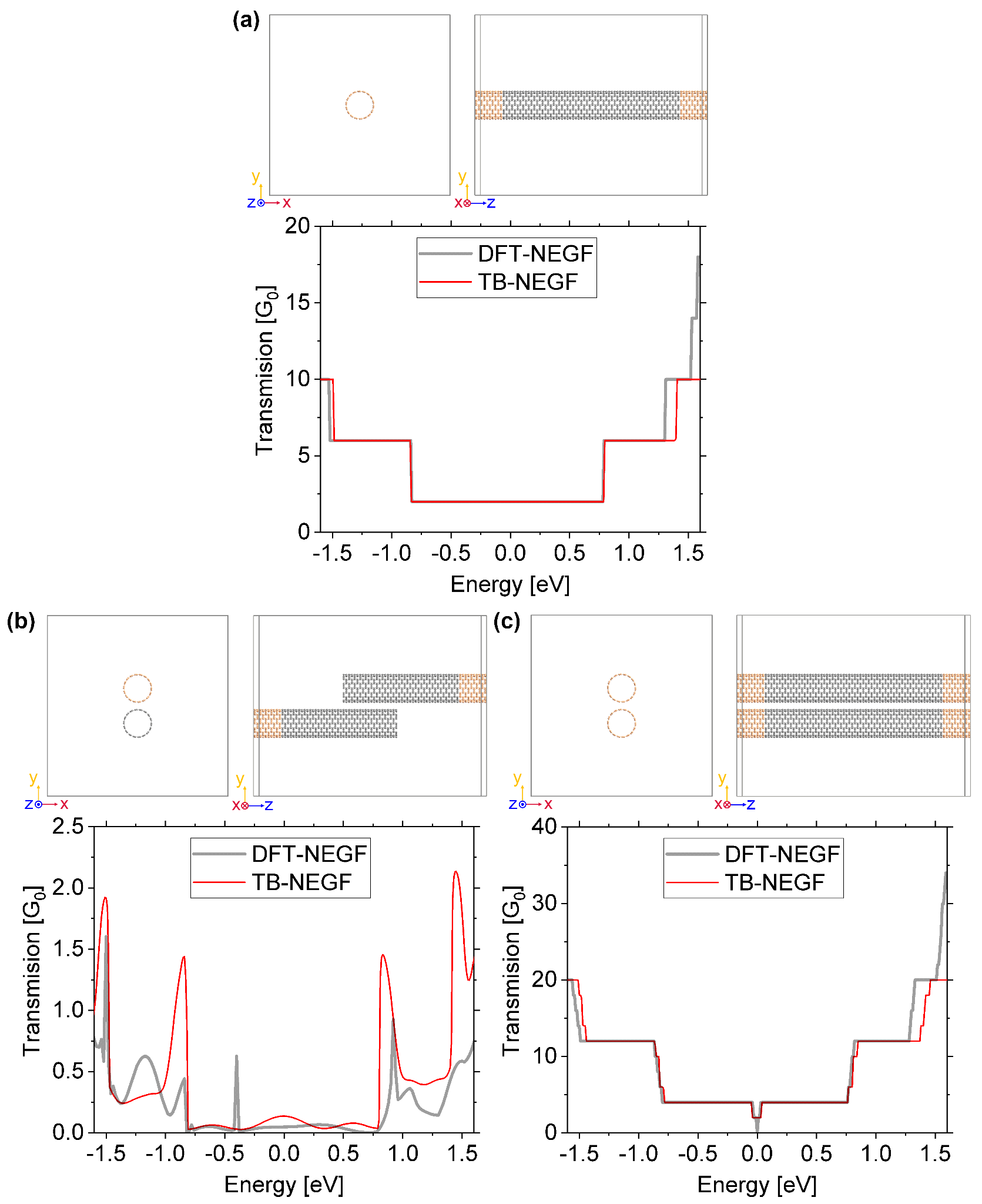}
	\caption{The accuracy and reproducibility of TB calculations. Transmission through a) (9,9) metallic nanotube, b) short (9,9)+(9,9) metallic-metallic CNT junction and c) (9,9)+(9,9) metallic-metallic bundle calculated at TB-NEGF (performed using TBtrans) and DFT/LDA(PZ)/SZP-NEGF (performed using QuantumATK) theory levels. Atomistic cross-sectional and side views of models used for transport calculations are presented at the top of each panel. In calculations performed using TBtrans, electrodes are marked in gold (5 units of (9,9) CNT). In calculations performed in QuantumATK, electrodes are outlined by a grey frame (1\,unit), while the remaining gold atoms represent the so-called electrode extensions (4 \,units). The length of the junction's overlap region is 2.5\,nm (10\,units). }
	\label{FigS6}
\end{figure} 
      
The DFT-NEGF calculations were performed employing density functional theory [T5, T6] in combination with the NEGF as implemented in QuantumATK\cite{atk, smidstrup2020, brandbyge2002}. The ab-initio calculations are based on the Kohn-Sham realization of the DFT with local density approximation (LDA) using the Perdew-Zunger (PZ)\cite{perdew1981} parameterization with a single-$\upzeta$ numerical orbital basis set of orbitals localized on atoms, including polarization functions (SZP). The kinetic cutoff for real-space integrals was set at 300\,Ry, whereas the Brillouin zone cut-off was set to 15\,\AA. The self-consistent field (SCF) cycle was iterated until the total energy changed by less than 10$^{-5}$\,eV/atom and the density matrix by less than 10$^{-5}$. SWCNTs comprising the central scattering regions of device configurations were coupled to two semi-infinite nanotube electrodes of the same type, as shown in Supplementary Fig.~\ref{FigS6} top panel. The Brillouin zone of the two-probe system was sampled using a 1$\times$1$\times$201 Monkhorst-Pack scheme. The transmission spectra were calculated in the [-1.6, 1.6]\,eV range within 601 points. The electronic temperature (smearing) was set to 300\,K.
	     
\subsection{Statistical data analysis of tight-binding simulation}\label{secStatistics}
      
The numerical simulations of the various CNT junction systems had wide-ranging magneto-resistance (MR) results as a function of multiple input factors, and particular statistical effort was required to usefully analyze the results and identify the most important trends. In this section, we describe how we statistically analyzed the results from the numerical tight-binding simulations, which is a distinct discussion from the analysis of experimental results.
      
Our simulation of magneto-resistance was a function of multiple input factors to include precisely-fixed parameters (magnetic field B, nanotube chirality, doping level) and stochastic parameters requiring averaging (defect density and temperature T); thus, our simulated MR values are an average with standard deviation. This wide-ranging and complex multivariate MR response was analyzed with the data exploration software JMP. The MR response was portioned according to positive and negative values, followed by taking the log of absolute value. This log transform was necessary for bivariate power-law analysis, where simple bivariate plots of Log(MR) against Log(B) identifies power-laws for both positive and negative MR. The log transform also reduced the spread in data and minimized the impact of outliers.
      
The simple bivariate power-law analysis however neglected the influence of other input factors that added noise to the bivariate response. We therefore also developed multivariate statistical models that using stepwise linear regression in JMP, with the Bayesian Information Criteria (BIC) as the stopping rule. This method statistically predicts the Log(MR) with an empirical expression composed of a best-fit sum of optimally-selected polynomials from the simulation input factors: linear terms represent primary effects, quadratic terms represent curvature, and cross terms between input factors interaction. These statistical multivariate models have greater predictive power than the bivariate power-law analysis and can uncover trends that may be obscured in complex multivariate data. In general, however, the polynomial expression generated is meant for predictive utility and does not necessarily match a governing physical law. Before generating the statistical multivariate model, the data was randomly portioned between a training set and validation set in an 80\,\%/20\,\% split. The training set data was used to find the optimum polynomials and associated fitted parameters without touching the validation data. Only after the statistical multivariate model was built, the validation data was used to calculate an unbiased validation R$^2$. A validation R$^2$ that closely matches the training R$^2$ indicates overfitting is not a concern. Note that, in some cases, the stochastic averaging resulted in significant error in the simulated MR response; therefore, MR values with relative error (=\,100\,\% standard deviation/average) greater than 50\,\% were excluded from analysis.
      
For the multivariate statistical models, we show: 1) plots of linear regression predicted value versus the actual simulated values, for training and validation datasets; 2) best linear regression      parameter fits for the selected polynomials, with error and statistical significance; 3) Fitting summary of the model with R$^2$, adjusted R$^2$, and number of observations; 4) JMP's ranking of parameter importance, distinguishing between main impact and total impact. 5) Slices of the   response surface--the fitted polynomial makes an n dimensional surface that is best visualized as taking slices of this surface; this is depicted by multiple traces where light red dashed lines indicate fixed parameters. In cases where parameter interaction complicates the response, different scenarios are selected showing the most interesting model features. Similar procedures with further detail may be found here\cite{bulmer2023, bulmer2020}.
      
\subsection{Additional Modelling Results: junctions}
      
\subsubsection{Theoretical modeling: CNT under external magnetic field}

\begin{figure}[h!]
	\centering
	\includegraphics[width=0.95\textwidth]{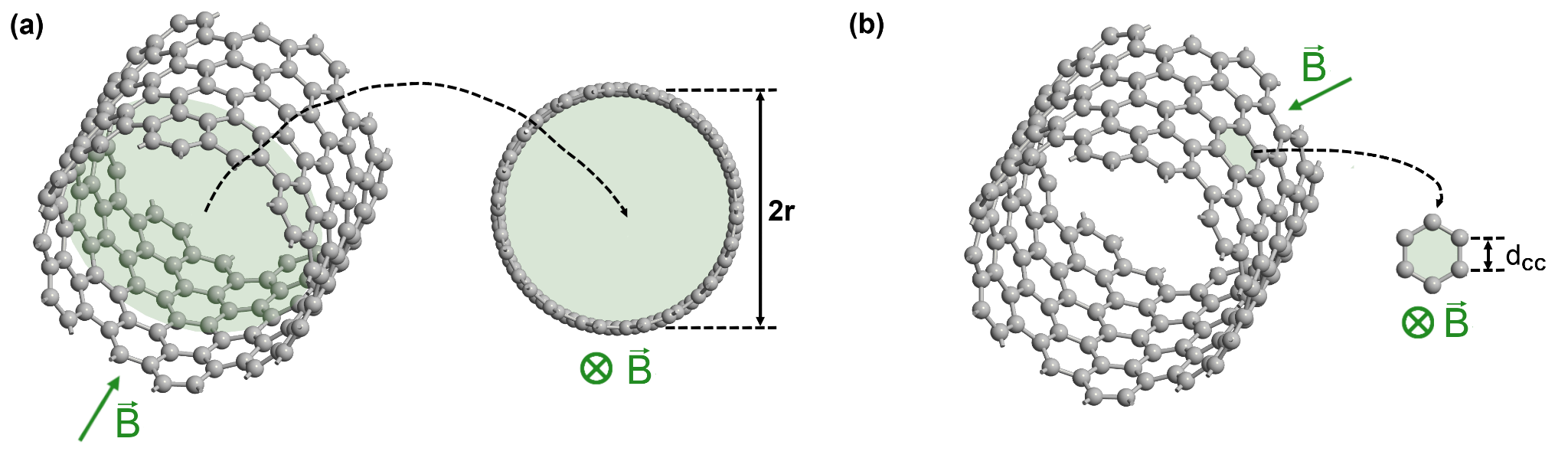}
	\caption{Magnetic field orientations relative to a carbon nanotube. (a) A magnetic field applied parallel to the nanotube axis (shown for one unit of a (12,6) CNT) threads flux through the tube cross-section of area $\uppi$r$^2$. (b) A perpendicular magnetic field penetrates the nanotube wall through its hexagonal plaquettes of area Ahex. Here, d$_\mathrm{CC}$ denotes the C--C bond length, and r is the nanotube radius. }
	\label{FigS7}
\end{figure}
      
In single carbon nanotubes, the response to a magnetic field depends on its orientation (Supplementary Fig.~\ref{FigS7}). A parallel field threads Aharonov-Bohm flux through the tube cross-section, shifts the circumferential boundary condition, and periodically modulates the bandgap with the flux  quantum $\Phi_{0}$. This drives metal--semiconductor transitions and produces magnetoconductance oscillations with period set by one flux quantum through the cross-section, (B$_{\parallel}$=$\Phi_{0}$/$\uppi$r$^2$ where r is nanotube radius). The required field to observe this behaviour scales as 1/r$^2$ and is therefore readily experimentally accessible in larger-diameter nanotubes.\cite{strunk2006, nemec2007, nanot2009}. Beyond gap tuning, in quantum-dot regimes, the parallel field deforms the longitudinal wavefunctions, altering tunnelling amplitudes to the contacts and thus the height of conductance resonances.\cite{marganska2019} By contrast, a perpendicular field threads the nanotube wall via its hexagonal plaquettes, setting an analogous flux scale B$\bot$=$\Phi_{0}$/A$_{\mathrm{hex}}$, with A$_{\mathrm{hex}}$=3$\sqrt{3/4}$ d$^{2}_{\mathrm{CC}}$\cite{saito1998}. The field couples to orbital motion around the circumference and mixes sub-bands. As the field increases, the magnetic length $\mathrm{\ell_{B}}$\,=\, $\sqrt{\hbar /|e|B}$ shrinks. Once $\mathrm{\ell_{B}}$ becomes comparable to nanotube radius r,  Landau-like quantisation develops. States localise where the normal component of the magnetic field on the tube wall is largest (near the 'top' and 'bottom' of the cylinder), while modes at azimuths where this normal component vanishes and changes sign remain extended along the tube axis. At sufficiently high fields the spectrum evolves towards a Hofstadter-like, pseudofractal structure approaching the graphene butterfly in large-diameter tubes, and transport becomes largely diameter- rather than chirality-controlled.\cite{nemec2006}  High-field interferometry on individual nanotube Fabry--Perot resonators reports the expected reorganisation of the conductance pattern and the onset of Landau levels under perpendicular magnetic field, consistent with these predictions\cite{raquet2008,nanot2009} These phenomena are well described within tight-binding approaches using Peierls substitution.\cite{goupalov2018, cresti2021} 
      
\subsubsection{Theoretical modeling: Effect of Chirality and Metallicity of CNTs on Their Simple Junctions}
 
\begin{figure}[h!]
	\centering
	\includegraphics[width=0.95\textwidth]{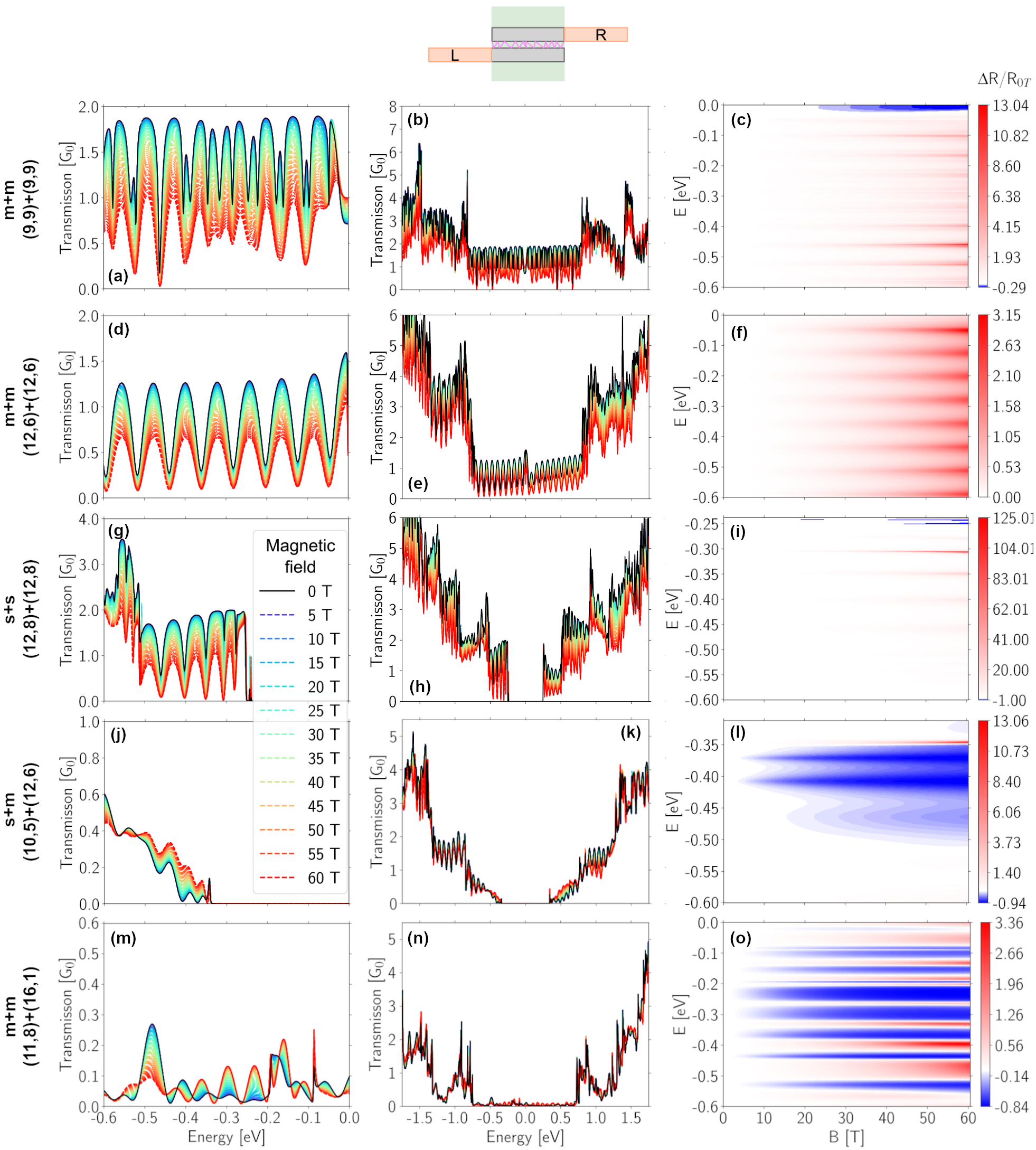}
	\caption{
		(a, d, g, j, m) The zero-bias transmission spectra, (b, e, h, k, n) transmission spectra in the wider energy range, and (c, f, i, l, o) magnetoresistance maps of different CNT junctions at various doping levels. The overlap regions (z) are 24.6\,nm, 22.6\,nm, 22.3\,nm, 22.6\,nm, and 23.5\,nm for the (9,9)+(9,9), (12,6)+(12,6), (12,8)+(12,8), (10,5)+(12,6), and (11,8)+(16,1) junctions, respectively. }
	\label{FigS8}
\end{figure} 
      
Figure~\ref{Fig3}c and Supplementary Figures~\ref{FigS8}c,f,i,l,o present MR as a function of perpendicular magnetic field for five distinct CNT junctions with varying chirality and electronic character, each probed  across a set of chemical potentials. Across all considered junctions, we observe two dominant MR trends. In the first, MR values remain strictly positive and increase monotonically with B, typically becoming quadratic at higher fields. This behaviour is most common in homo-junctions and reflects the gradual suppression of coherent tunnelling by the magnetic field. In the second,  MR remains strictly negative and decreases with B, often with a non-uniform slope that flattens at low and high B. This occurs more frequently in hetero-junctions and can be attributed to field-induced enhancement of intertube state overlap. In addition, we find nonmonotonic MR in certain cases, where resistance first increases with field, then decreases after reaching a maximum. Even more complex are sign-changing curves: in some junctions MR is initially negative and becomes positive at higher B, while in others the reverse trend occurs. These behaviours appear particularly in junctions with mixed metallicity or lattice mismatch and highlight the competing roles of orbital misalignment, tunnelling interference, and field-driven  state redistribution in determining transport.

\FloatBarrier      
\subsubsection{Theoretical modelling: Effect of Length of the Overlap Region on CNT Junctions}\label{secToverlap}

\begin{figure}[h!]
	\centering
	\includegraphics[width=0.95\textwidth]{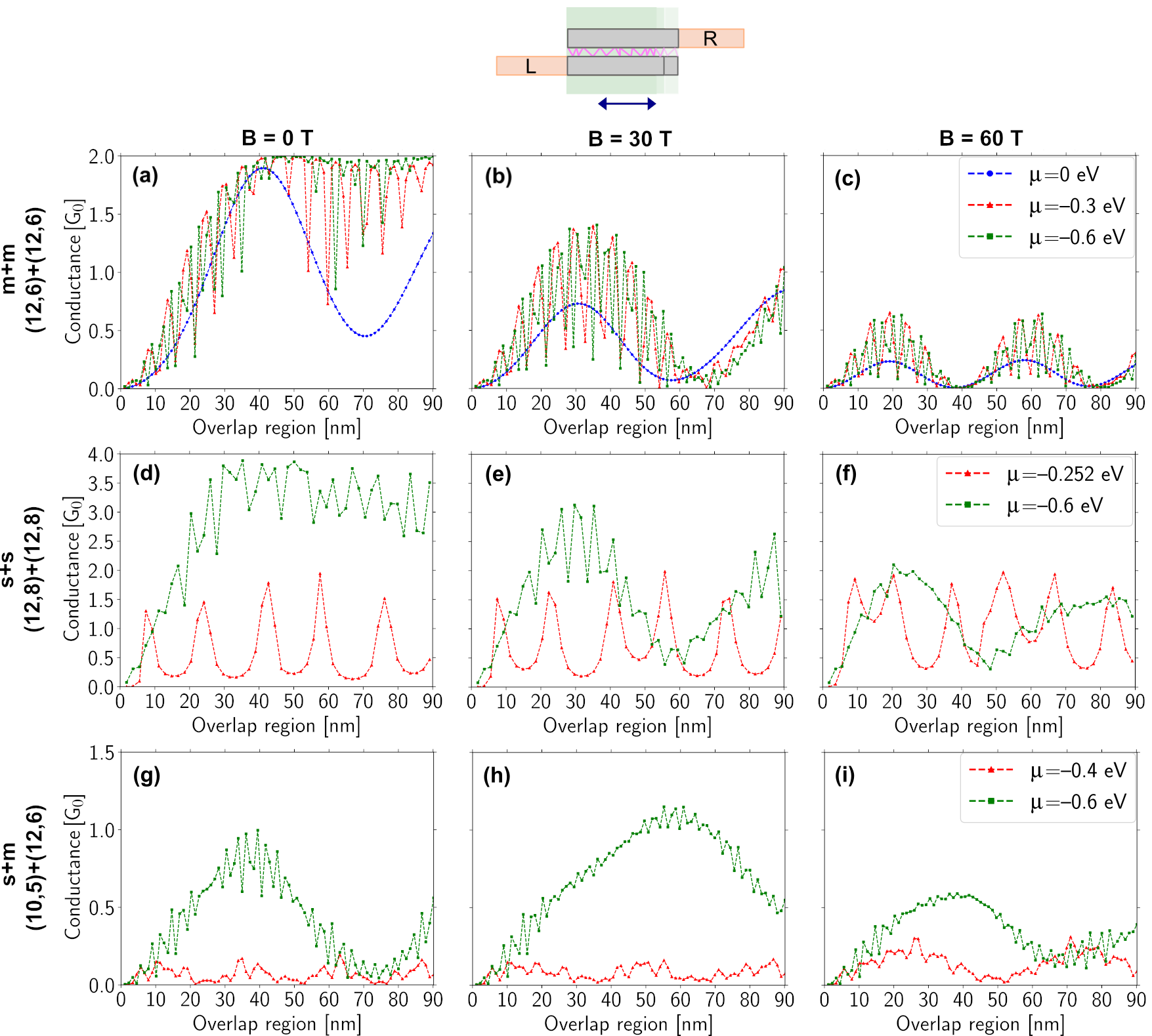}
	\caption{The oscillations of conductance ($\upsigma$) with CNT junction length of (a-c) (12,6)+(12,6) metallic-metallic, (d-f) (12,8)+(12,8) semiconducting-semiconducting, and (g-i) (10,5)+(12,6) semiconducting-metallic CNT junctions under external perpendicular magnetic fields at various doping levels. For metallic system, conductance was plotted for $\upmu$=0, -0.3 and -0.6\,eV. For (12,8)+(12,8) and (10,5)+(12,6) junctions, conductance at Fermi level is zero due to the presence of a band gap. Instead, conductance was plotted for the first $\upmu$ value yielding non-zero conductance, omitting -0.3\,eV if it was already close to this value.}
	\label{FigS10}
\end{figure}

\begin{figure}[h!]
	\centering
	\includegraphics[width=0.85\textwidth]{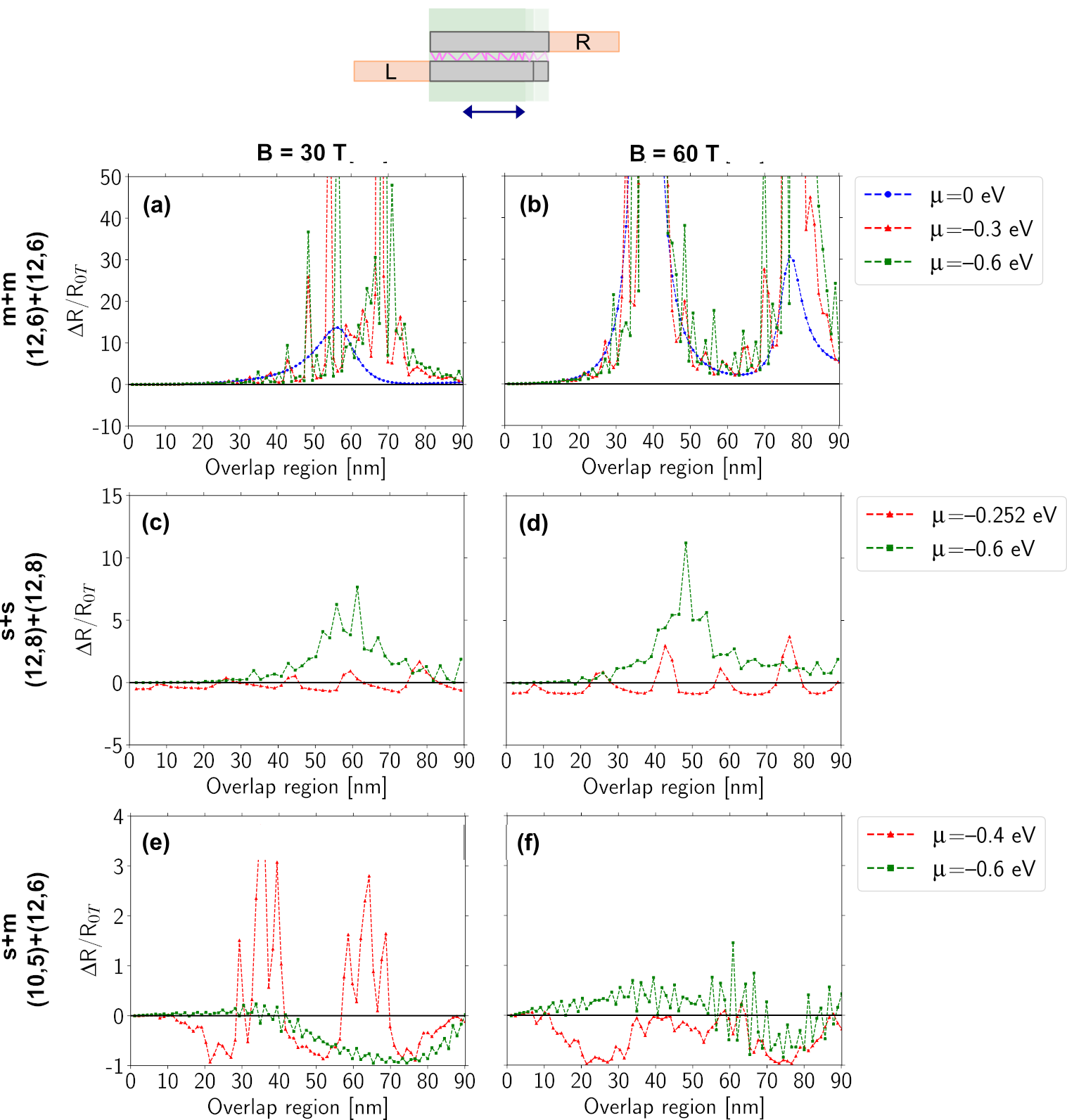}
	\caption{The computed oscillations of magnetoresistance with CNT junction length of (a-b) (12,6)+(12,6) metallic-metallic, (c-d) (12,8)+(12,8) semiconducting-semiconducting, and (e-f) (10,5)+(12,6) semiconducting-metallic CNT junctions under external perpendicular magnetic fields at various doping levels. For metallic system, magnetoresistance was plotted for $\upmu$=0 -0.3, and -0.6\,eV, while for (12,8)+(12,8) and (10,5)+(12,6) junctions, it was plotted for the first $\upmu$ value with non-zero conductance, omitting -0.3\,eV if similar, and for $\upmu$=-0.6\,eV.}
	\label{FigS11}
\end{figure}

\begin{figure}[h!]
	\centering
	\includegraphics[width=0.80\textwidth]{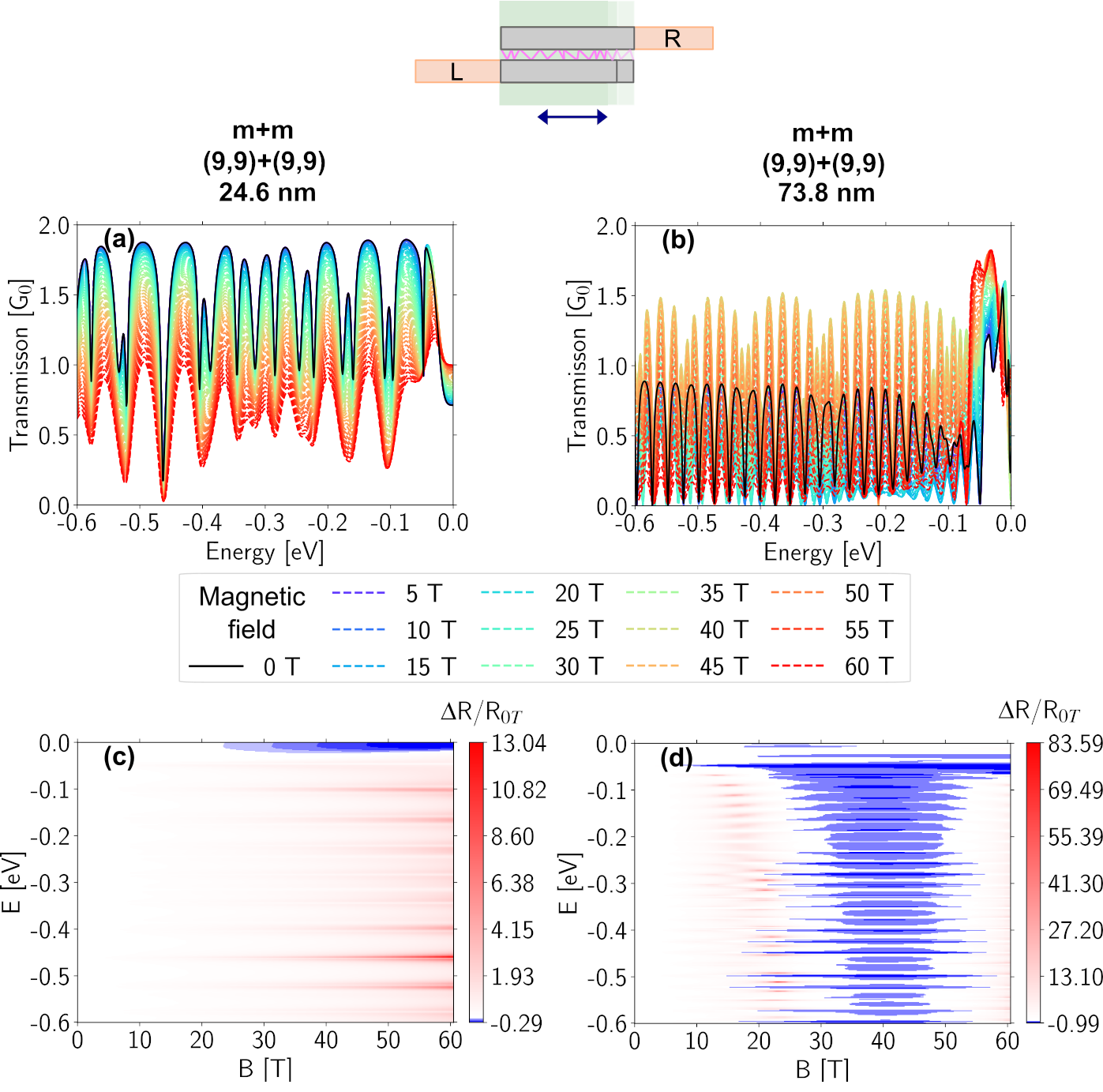}
	\caption{The computed (a, c) zero-bias transmission spectra and (b, d) magnetoresistance maps of (9,9)+(9,9) metallic-metallic CNT junctions with different overlap (z) lengths: 24.6\,nm -- upper panel, 73.8\,nm -- lower panel.}
	\label{FigS12}
\end{figure}

\FloatBarrier 
       
\subsubsection{Theoretical modeling: Effect of CNT rotation in CNT junctions}\label{secTrotation}
       
As shown in Supplementary Fig.~\ref{FigS13}, In our model, atoms from different nanotubes interact only if their separation distance falls within the range between the third-nearest neighbor distance in an individual nanotube (approximately 2.5\,\AA) and the 3.6\,\AA interaction cutoff. Since the initial separation distance between nanotubes is fixed at 3.356\,\AA (for 0\,$^\mathrm{o}$ rotation), rotating one nanotube around the z-axis changes the number of available atomic pairs contributing to intertube hopping, but this transition is not smooth. Instead, as rotation progresses, atomic connections appear or disappear abruptly, leading to the observed stepwise changes in transport properties.

\begin{figure}[h!]
	\centering
	\includegraphics[width=0.95\textwidth]{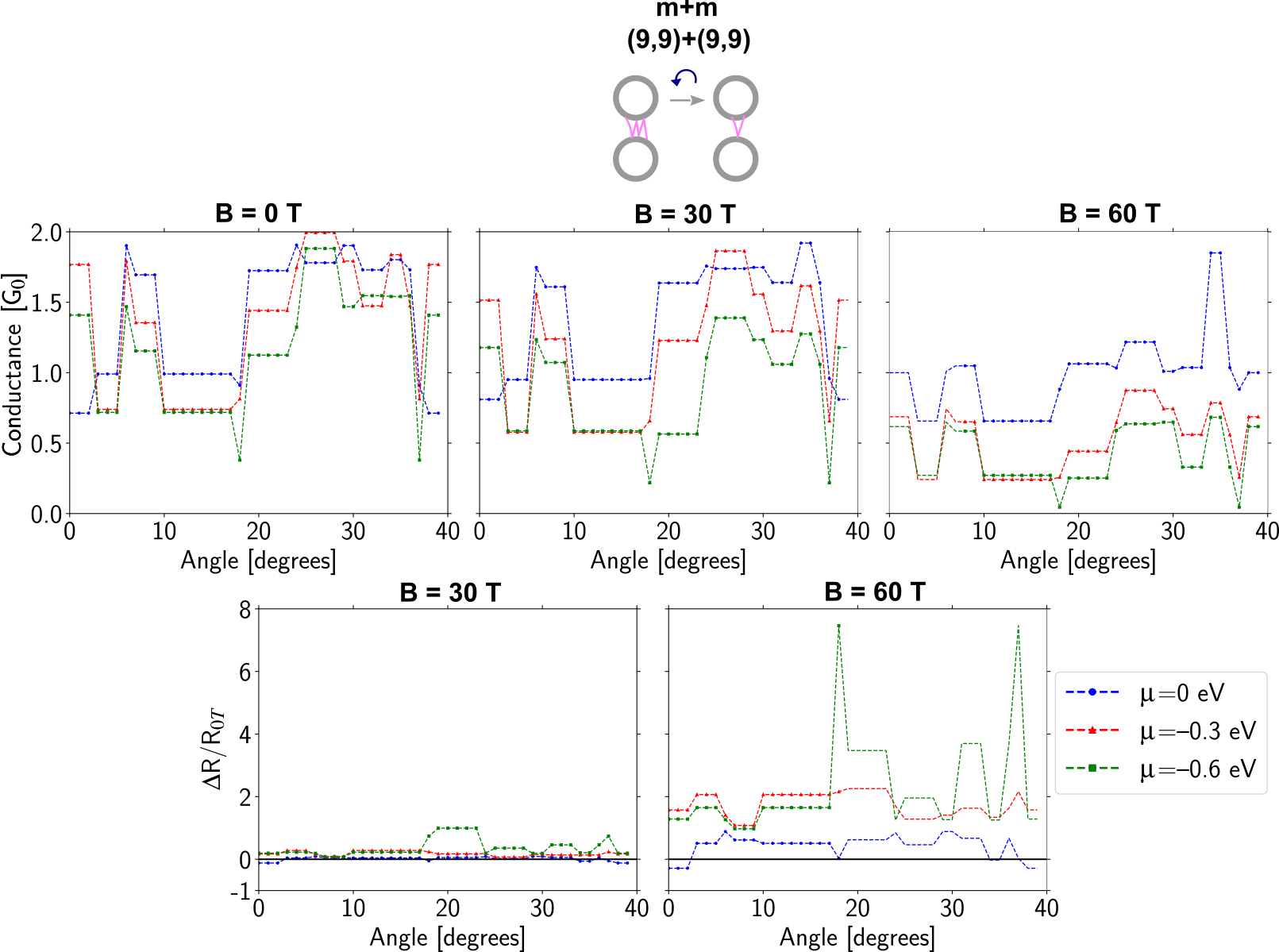}
	\caption{The computed oscillations of conductance ($\upsigma$) (top panel) and magnetoresistance (bottom panel) with varying angle between nanotubes in (9,9)+(9,9) metallic-metallic CNT junctions under external perpendicular magnetic fields at various doping levels. The overlap region (z) is 24.6\,nm, consisting of one hundred units of a (9,9) CNT. The observed step-like variations in conductance and magnetoresistance arise due to the discrete nature of atomic pair formations between the nanotubes, as illustrated in Extended Data Fig.~\ref{FigE4}}
	\label{FigS13}
\end{figure}

\begin{figure}[h!]
	\centering
	\includegraphics[width=0.8\textwidth]{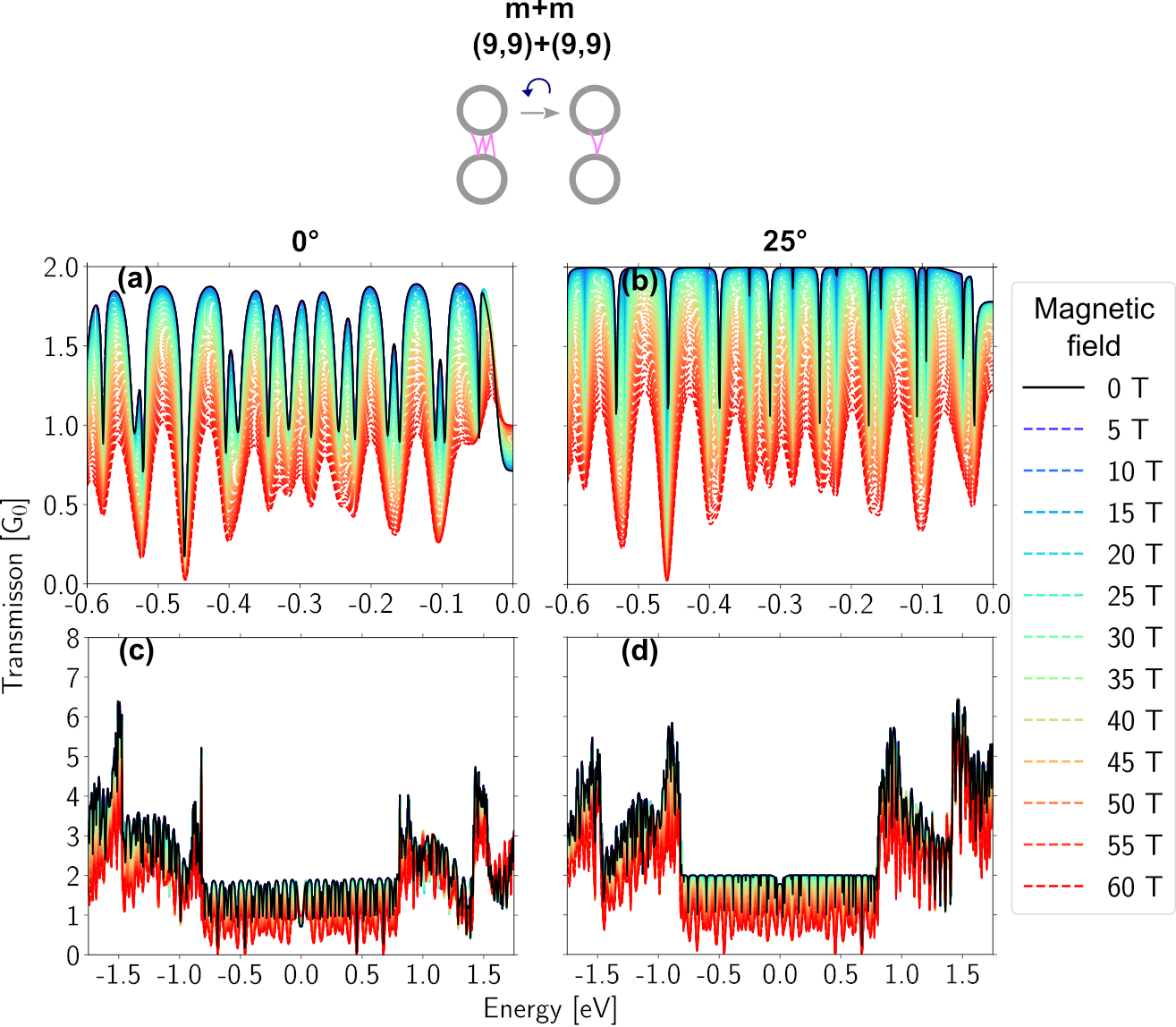}
	\caption{The computed zero-bias transmission spectra of (9,9)+(9,9) metallic-metallic CNT junctions with different relative angular positions (rotation around the tube's z-axis) under external perpendicular magnetic fields. The overlap region (z) is 24.6\,nm, consisting of one hundred units of a (9,9) CNT. Panels (a, c) show the transmission in the energy range of [-0.6, 0.0]\,eV, while panels (b, d) show the transmission over a wider energy range.}
	\label{FigS14}
\end{figure}

\FloatBarrier 

\subsubsection{Theoretical modeling: Effect of crossing angle between CNTs in CNT junctions}\label{secTangle}

We do not vary the crossing angle as it simultaneously affects overlap and stacking - both already analysed independently. This separation isolates their individual impacts on magnetotransport. Although experimental CNTs are not perfectly aligned (Supplementary Fig.~\ref{FigS2}), our simulated parallel configuration represents the high-conductance limit expected for small crossing angles\cite{khromov2017, larin2021, adinehloo2023}.

\FloatBarrier 
       
\subsubsection{Theoretical modelling: Effect of Separation Distance between CNTs on CNT Junctions}\label{secTd}

\begin{figure}[h!]
	\centering
	\includegraphics[width=0.95\textwidth]{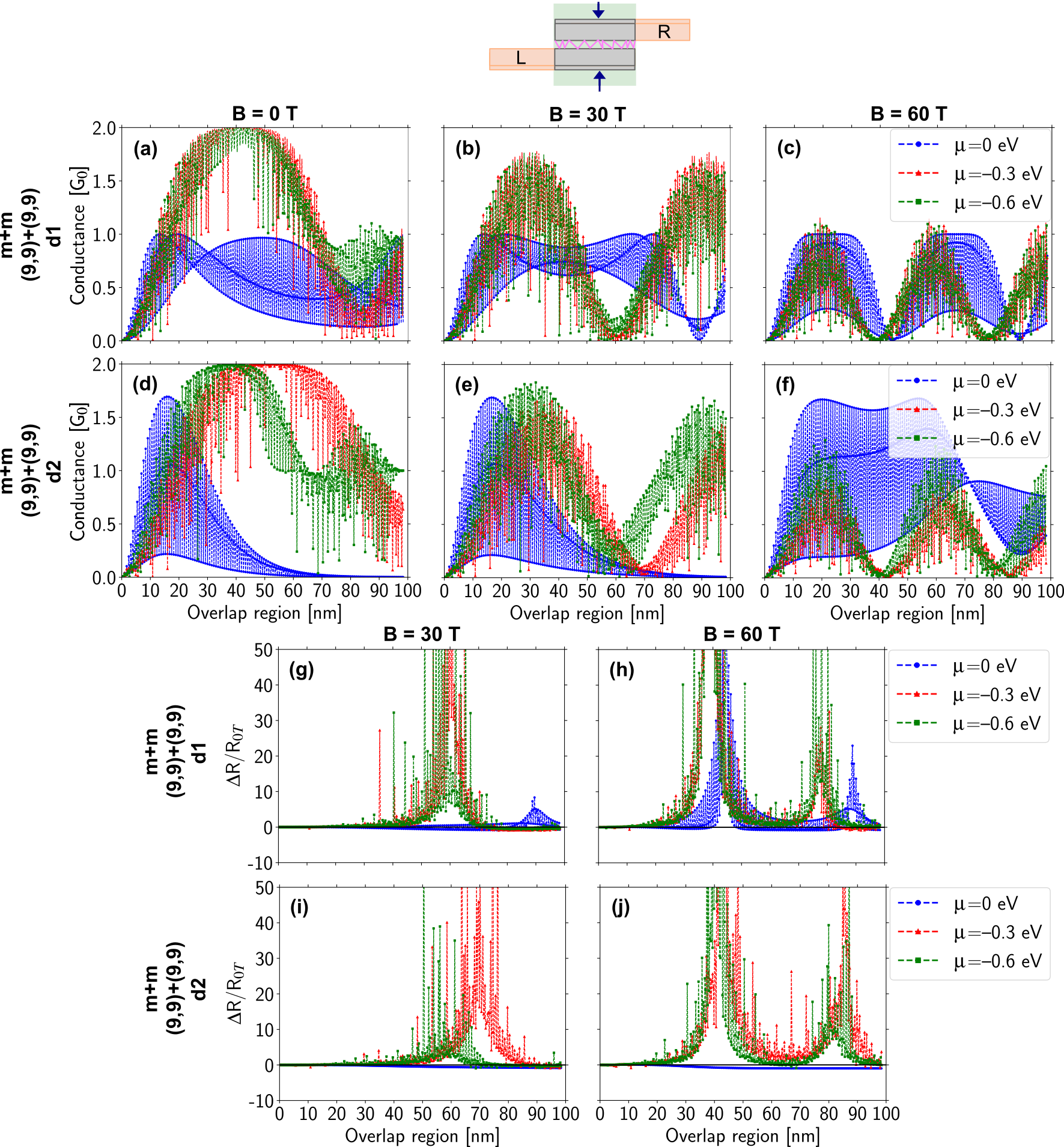}
	\caption{The computed oscillations of (a-f) conductance ($\upsigma$) and (g-j) magnetoresistance with CNT junction length of (9,9)+(9,9) metallic-metallic CNT junctions. The results are shown for different distances between nanotubes, d: (a-c, g, h) 0.3356\,nm and (d-f, i, j) 0.2517\,nm, under external perpendicular magnetic fields at various doping levels.}
	\label{FigS15}
\end{figure}

\textbf{Note on conductance suppression and intertube coupling mechanisms at B\,=\,0.}
	
Unlike the rapid and irreversible quenching of conductance predicted by Xu et al. using a
tight-binding model\cite{xu2013}, our results agree with the DFTB-based findings of Tripathy et al.~\cite{tripathy2016} and show persistent conductance oscillations as the overlap length increases. In both the (12,6)+(12,6) and (9,9)+(9,9) junctions at equilibrium separation (d$_1$), conductance remains oscillatory even up to 	90\,nm overlap (Supplementary Fig.~\ref{FigS10}). The absence of complete suppression likely reflects differences in stacking configuration and in the details of the tight-binding Hamiltonian. At equilibrium separation, the 3.6\,\AA intertube interaction cutoff includes only a limited number of atomic pairs, as visualised in Supplementary Fig.~\ref{FigS3}a), and can restrict coupling to a single sublattice depending on chirality and alignment. Reducing the separation to d$_2$ increases the number of interacting pairs, especially across sublattices, modifying the junction Hamiltonian and promoting behaviour more	consistent with Xu's predictions. As our stacking analysis shows (Extended Data Fig.~\ref{FigE4}), such 	interactions can lead to destructive interference depending on geometry, confirming that stronger 	coupling does not always enhance conductance. Therefore, while eventual conductance decay 	due to band gap opening is expected at sufficiently long overlaps--ultimately resembling the bundled CNT regime\cite{bulmer2026}--our data suggest that this transition may be delayed because is highly sensitive to stacking geometry and the coupling model.
	
\begin{figure}[h!]
	\centering
	\includegraphics[width=0.95\textwidth]{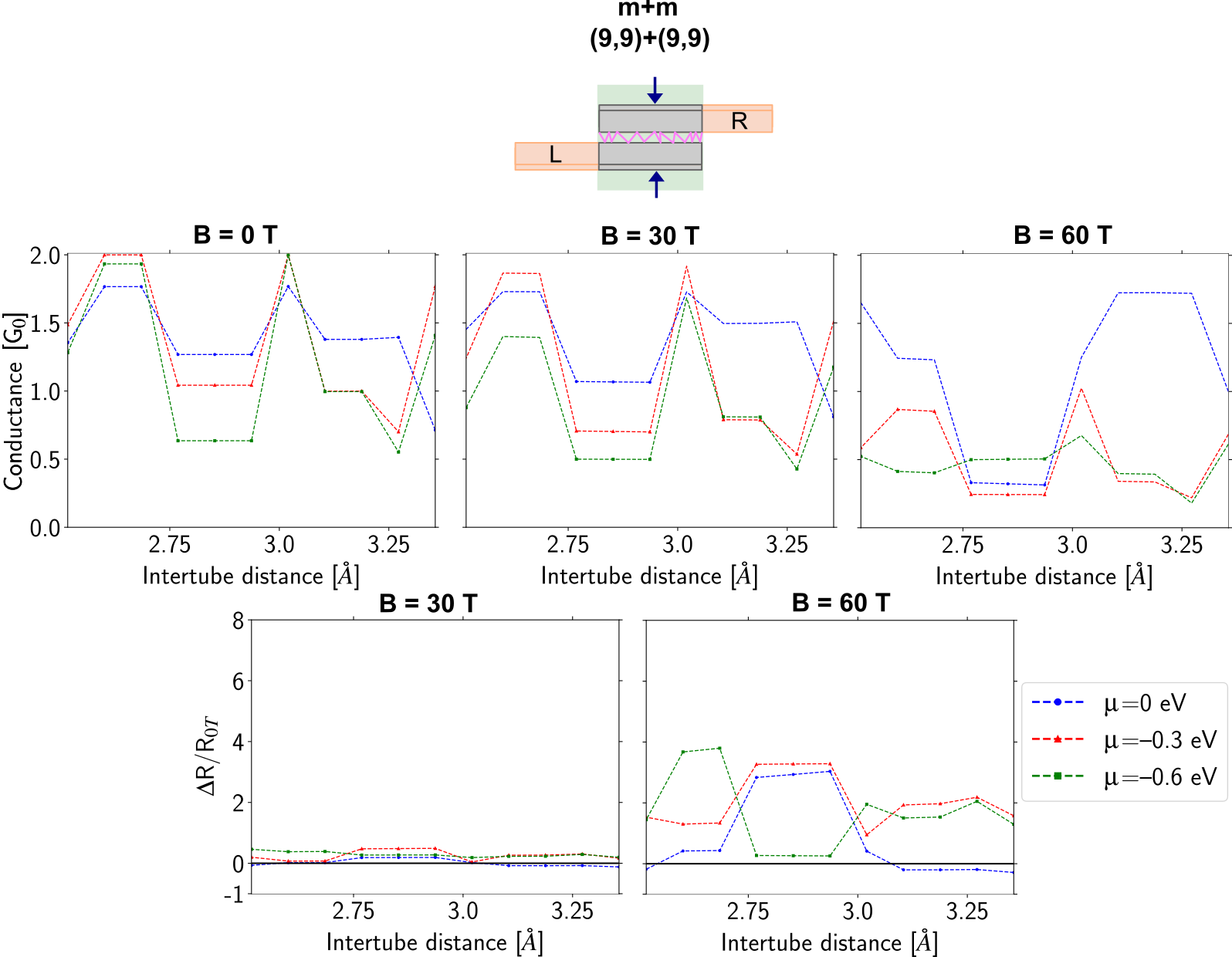}
	\caption{The computed oscillations of conductance ($\upsigma$) (top panel) and magnetoresistance (bottom panel) with varying distance between nanotubes in (9,9)+(9,9) metallic-metallic CNT junctions under external perpendicular magnetic fields at various doping levels. The overlap region (z) is 24.6\,nm, consisting of one hundred units of a (9,9) CNT.}
	\label{FigS16}
\end{figure}	
	
\textbf{Note on doping dependence of MR at reduced separation distances.}
	
As shown in Supplementary Figure~\ref{FigS15}, MR as a function of overlap length becomes more doping-sensitive at reduced separation distances: while for the equilibrium separation (d$_1$), the MR vs overlap curves for $\upmu$,=\,-0.3 eV and -0.6\,eV appear similar, at shorter distances (d$_2$), they begin to diverge 	significantly. This indicates that stronger intertube coupling not only alters transmission but also 	enhances the impact of doping on MR behaviour.
	
The observed step-like character of the oscillations in conductance and magnetoresistance with 	varying distance (Supplementary Fig.~\ref{FigS16}) arises directly from the adopted model, which assigns discrete hopping parameters based on interatomic distances rather than a smooth distance-dependent 	decay, as discussed in Ref.~\cite{trizon2004} and Ref.~\cite{lambin1994}. Specifically, hopping values are determined according to predefined neighbour classifications (t$_1$, t$_2$, t$_3$), with interactions cut off beyond 3.6\,\AA, leading to abrupt transitions in conductance as nanotube separation changes. By modifying the separation distance between nanotubes, we effectively alter the Hamiltonian of the system.

\FloatBarrier 
	
\subsubsection{Theoretical modelling: Spin Effects on CNT Junctions}

\begin{figure}[h!]
	\centering
	\includegraphics[width=0.95\textwidth]{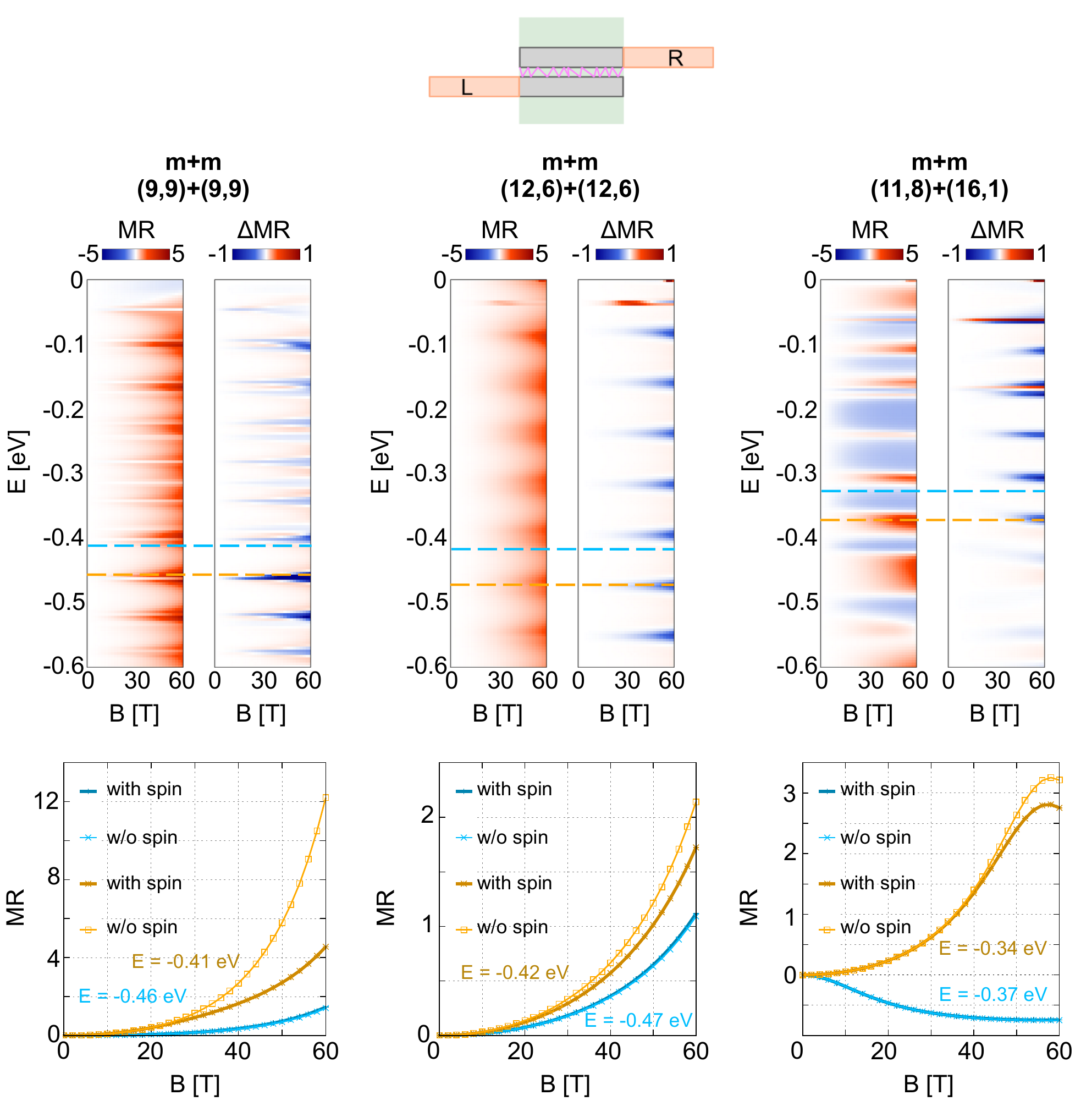}
	\caption{Spin contribution to the MR. The top row: MR(B,$\upmu$) maps computed with spin (Zeeman term included) for three metallic-metallic junctions: (9,9)+(9,9) (z=22.4\,nm), (12,6)+(12,6) (z=22.6\,nm), and (11,8)+(16,1) (z=22.6\,nm), shown alongside the corresponding $\Delta$MR(B,$\upmu$) maps, defined as the difference between MR computed with and without spin. For each junction, the MR and $\Delta$MR maps are displayed side by side. The bottom row: MR(B) line cuts at selected energy values ($\upmu$), comparing calculations with and without spin.}
	\label{FigS17}
\end{figure}	

\FloatBarrier 
	
\subsubsection{Theoretical modelling: Effect of Temperature on CNT Junctions}

\begin{figure}[h!]
	\centering
	\includegraphics[width=0.95\textwidth]{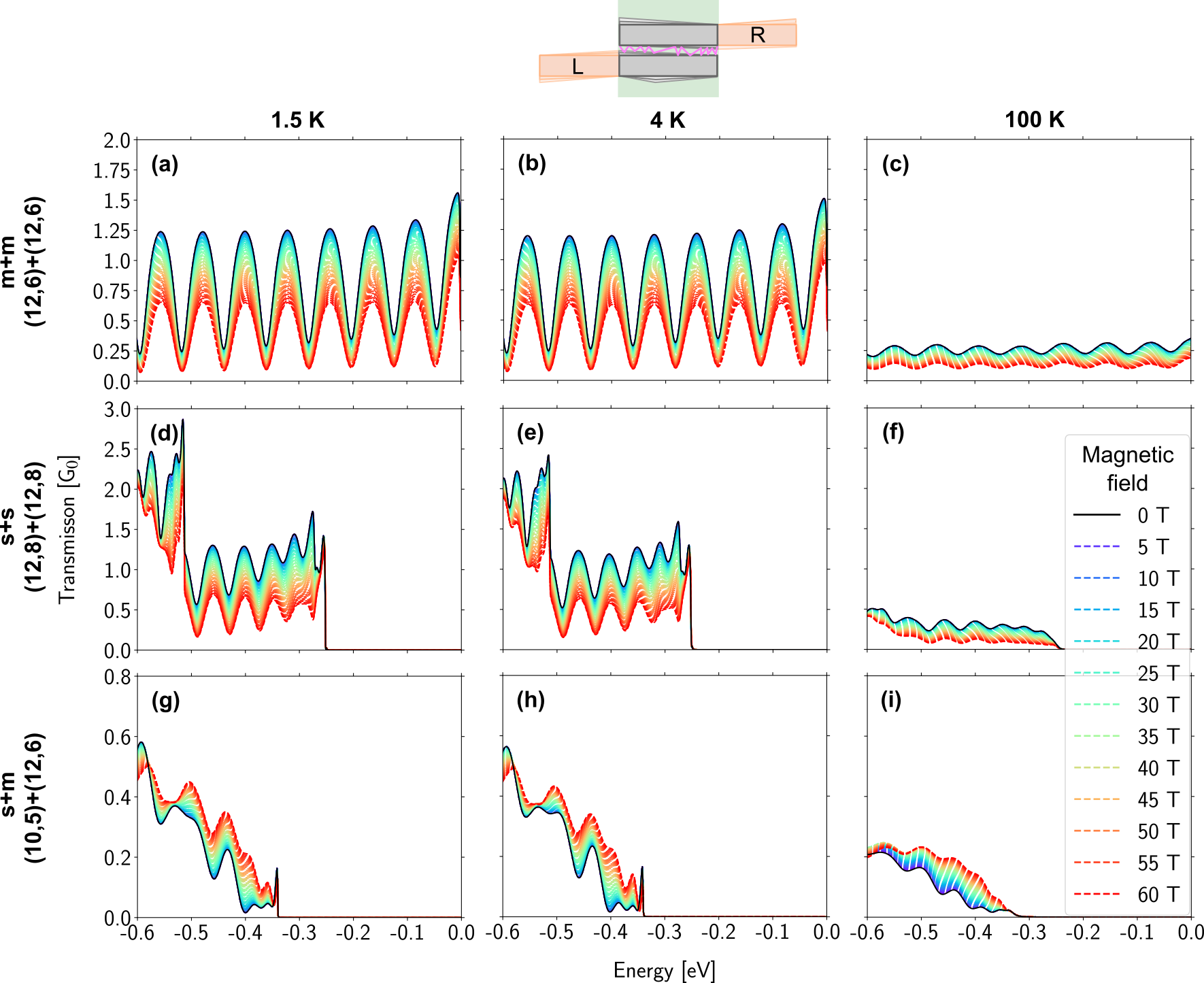}
	\caption{The computed zero-bias transmission spectra of metallic-metallic (z$_\mathrm{initial}$ = 22.6\,nm), semiconducting-semiconducting (z$_\mathrm{initial}$ = 22.3\,nm), and semiconducting-metallic (z$_\mathrm{initial}$ = 22.6\,nm) CNT junctions computed at chosen temperatures and under external perpendicular magnetic fields.}
	\label{FigS18}
\end{figure}

\begin{figure}[h!]
	\centering
	\includegraphics[width=0.95\textwidth]{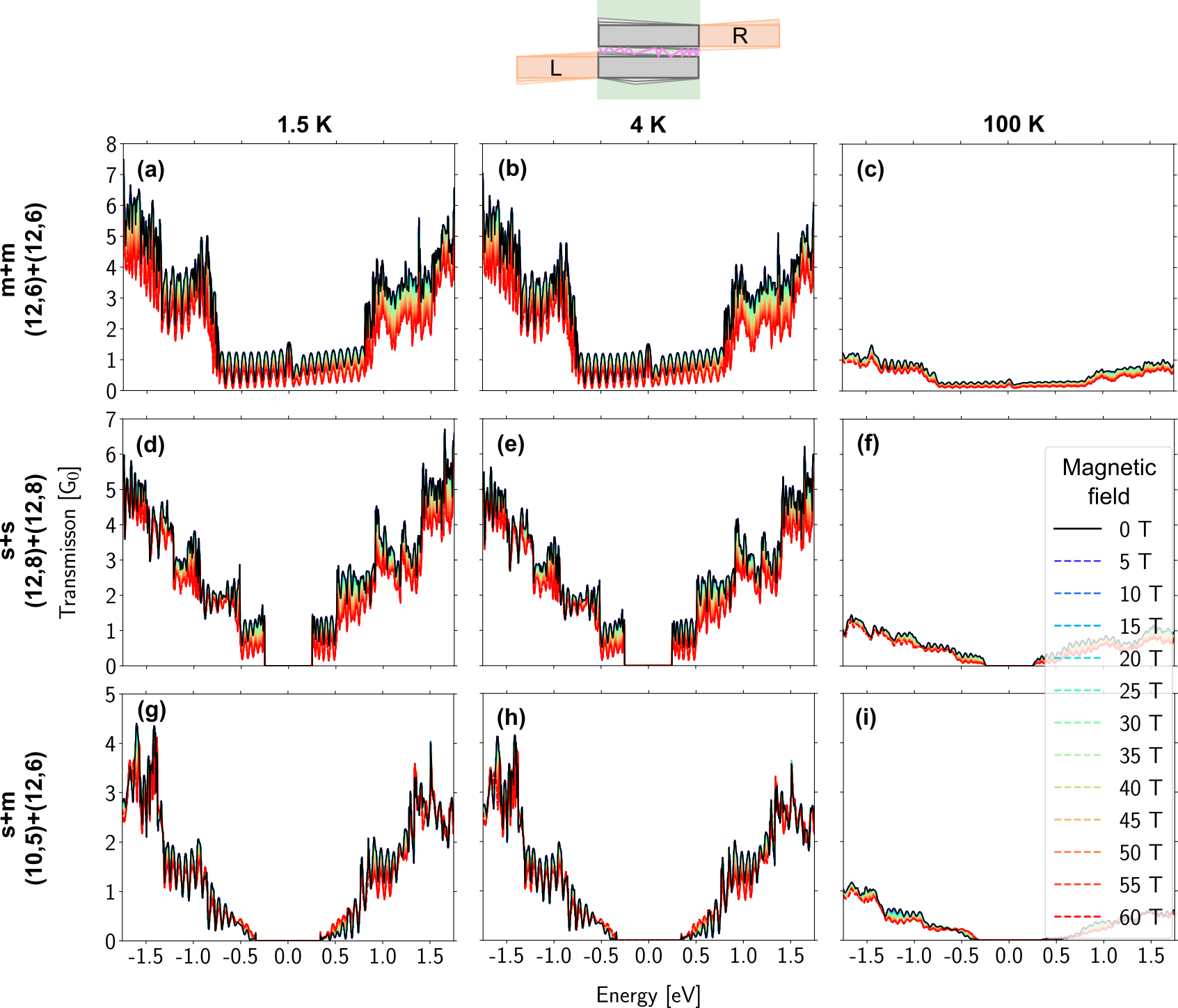}
	\caption{The computed zero-bias transmission spectra in the wider energy range of metallic-metallic (z$_\mathrm{initial}$ = 22.6\,nm), semiconducting-semiconducting (z$_\mathrm{initial}$ = 22.3\,nm), and semiconducting-metallic (z$_\mathrm{initial}$ = 22.6\,nm) CNT junctions computed at chosen temperatures and under external perpendicular magnetic fields.}
	\label{FigS19}
\end{figure}

\begin{figure}[h!]
	\centering
	\includegraphics[width=0.95\textwidth]{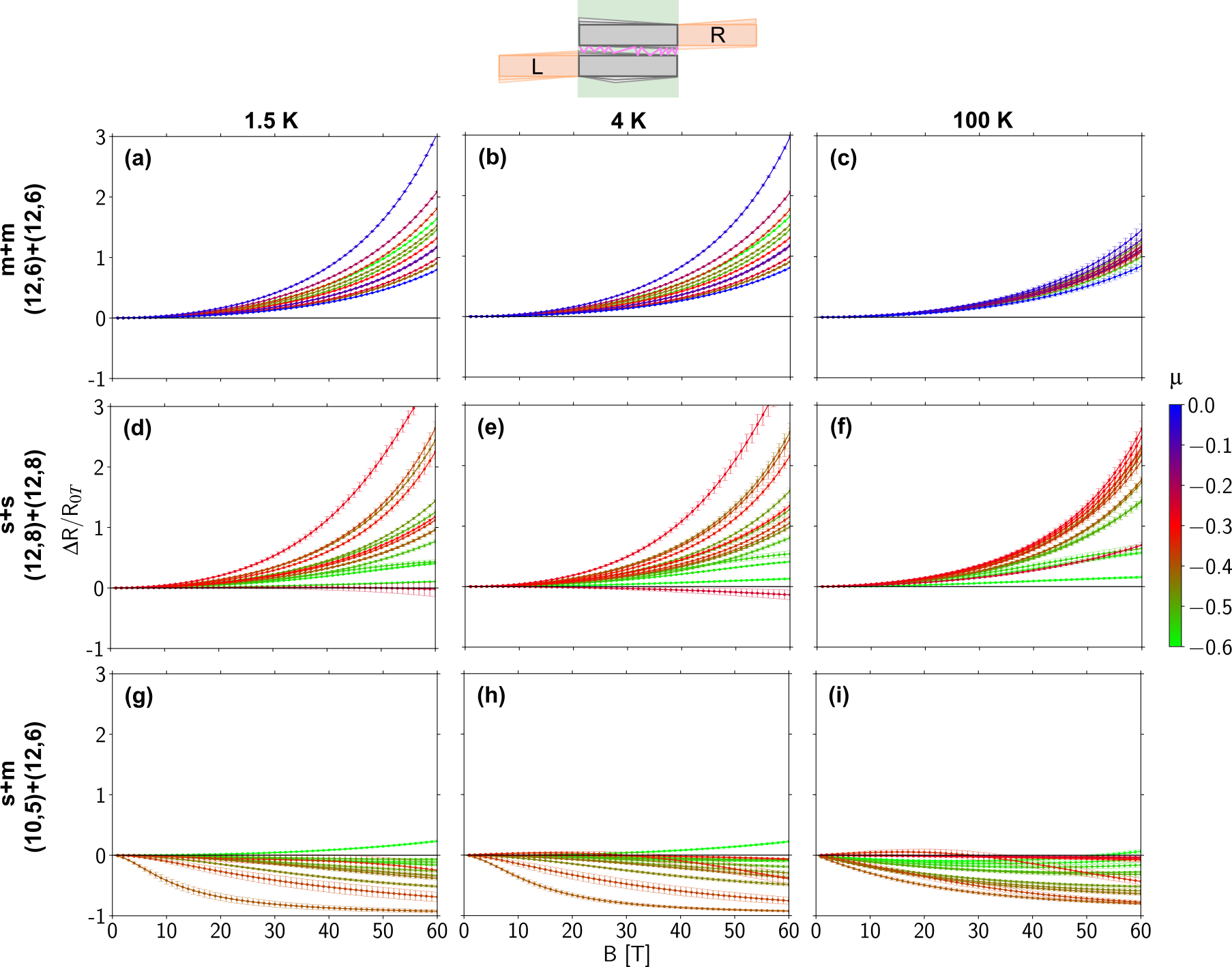}
	\caption{The computed magnetoresistance as functions of the external perpendicular magnetic field (B) of metallic-metallic (z$_\mathrm{initial}$ = 22.6\,nm), semiconducting-semiconducting (z$_\mathrm{initial}$ = 22.3\,nm) and semiconducting-metallic (z$_\mathrm{initial}$ = 22.6 nm) CNT junctions at different temperatures and doping levels. Each panel displays a representative subset of MR functions plotted for doping levels in the range [-0.6, 0.0]\,eV, sampled at 0.001\,eV intervals, with subsets selected proportionally to reflect different trends while ensuring clarity.}
	\label{FigS20}
\end{figure}

\FloatBarrier 

\begin{figure}[h!]
	\centering
	\includegraphics[width=0.95\textwidth]{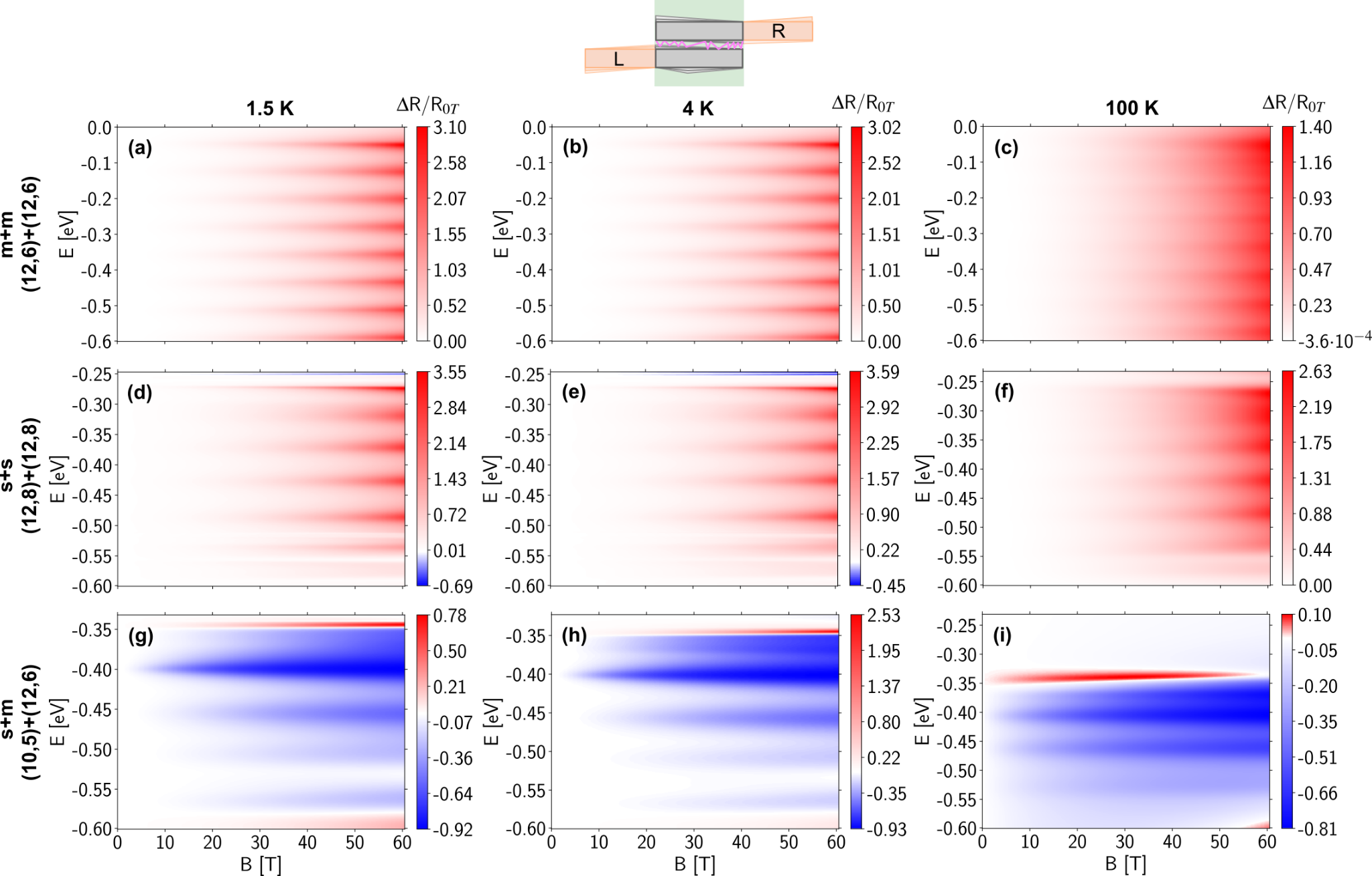}
	\caption{The computed magnetoresistance maps of metallic-metallic (z$_\mathrm{initial}$ = 22.6\,nm), semiconducting-semiconducting (z$_\mathrm{initial}$ = 22.3\,nm) and semiconducting-metallic (z$_\mathrm{initial}$ = 22.6\,nm) CNT junctions at different temperatures and doping levels.}
	\label{FigS21}
\end{figure}

While the MR(B) characteristics at 1.5\, K and 4\,K appear nearly identical in Supplementary Fig.~\ref{FigS20}, small but systematic differences become visible in the corresponding MR maps shown in Supplementary Fig.~\ref{FigS21}. As shown in Supplementary Tab.~\ref{table}, for the semiconducting-semiconducting (12,8)+(12,8) homo junction, increasing the temperature from 1.5\,K to 4\,K leads to a reduction in both the maximum positive MR and the minimum negative MR, accompanied by a slight increase in the percentage of 	negative MR cases. In contrast, for the semiconducting-metallic (10,5)+(12,6) heterojunction, the percentage of negative MR decreases, yet the minimal MR becomes more negative. These trends are notable, given that both systems require doping to conduct due to their intrinsic band 	gaps, yet they respond in opposite ways to low-temperature thermal fluctuations. At 100\,K, negative MR disappears entirely in the semiconducting-semiconducting case, while for the mixed junction, the percentage of negative MR increases, although the minimal value becomes less negative compared to 4\,K. This non-monotonic temperature dependence of both the 	percentage and magnitude of negative MR in the mixed junction mirrors the behaviour observed in experiment (Extended Data Figs.~\ref{FigE1}b,c), where the percentage of negative MR initially increases with 	temperature, reaches a maximum, and subsequently declines. The minimum MR exhibits a similar dome-like dependence, with the temperature corresponding to the extremum varying 	between samples.

\FloatBarrier 
	
\subsubsection{Theoretical modelling: Effect of Structural Defects on CNT Junctions}\label{secTd}

\begin{figure}[h!]
	\centering
	\includegraphics[width=0.85\textwidth]{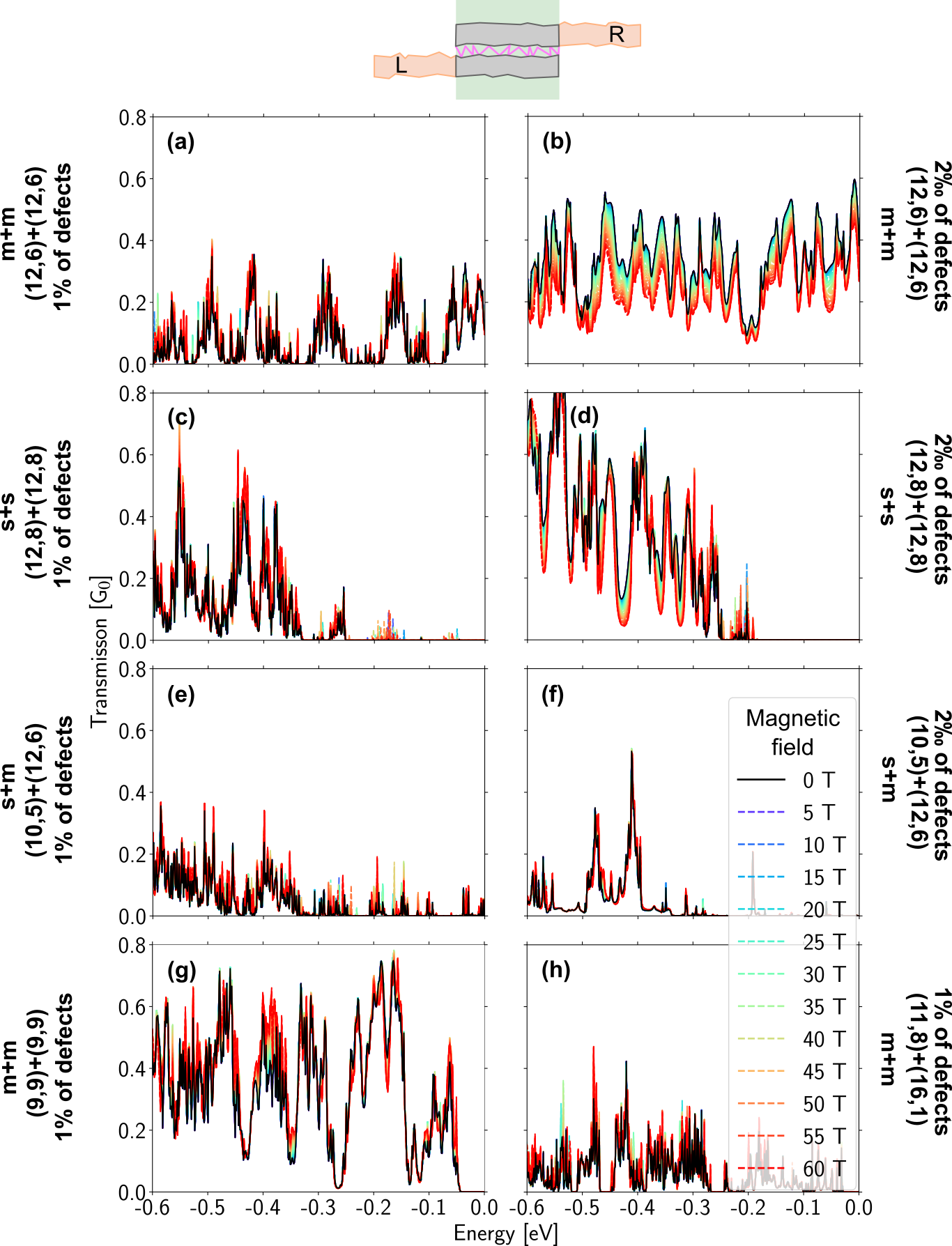}
	\caption{The zero-bias transmission spectra of a), b) (12,6)+(12,6) metallic-metallic (z$_\mathrm{initial}$ = 22.6\,nm), c), d) (12,8)+(12,8) semiconducting-semiconducting (z$_\mathrm{initial}$ = 22.3\,nm), e), f) (10,5)+(12,6) semiconducting-metallic (z$_\mathrm{initial}$ = 22.6\,nm) g) (9,9)+(9,9) metallic-metallic (z$_\mathrm{initial}$ = 24.6\,nm) h) (11,8)+(16,1) metallic-metallic (z$_\mathrm{initial}$ = 23.5\,nm) CNT junctions containing a),c),e),g),h) 1\,\% of Stone-Wales defects b),d),f) 0.2\,\% of Stone---Wales defects, computed under external perpendicular magnetic fields.}
	\label{FigS22}
\end{figure}

\begin{figure}[h!]
	\centering
	\includegraphics[width=0.85\textwidth]{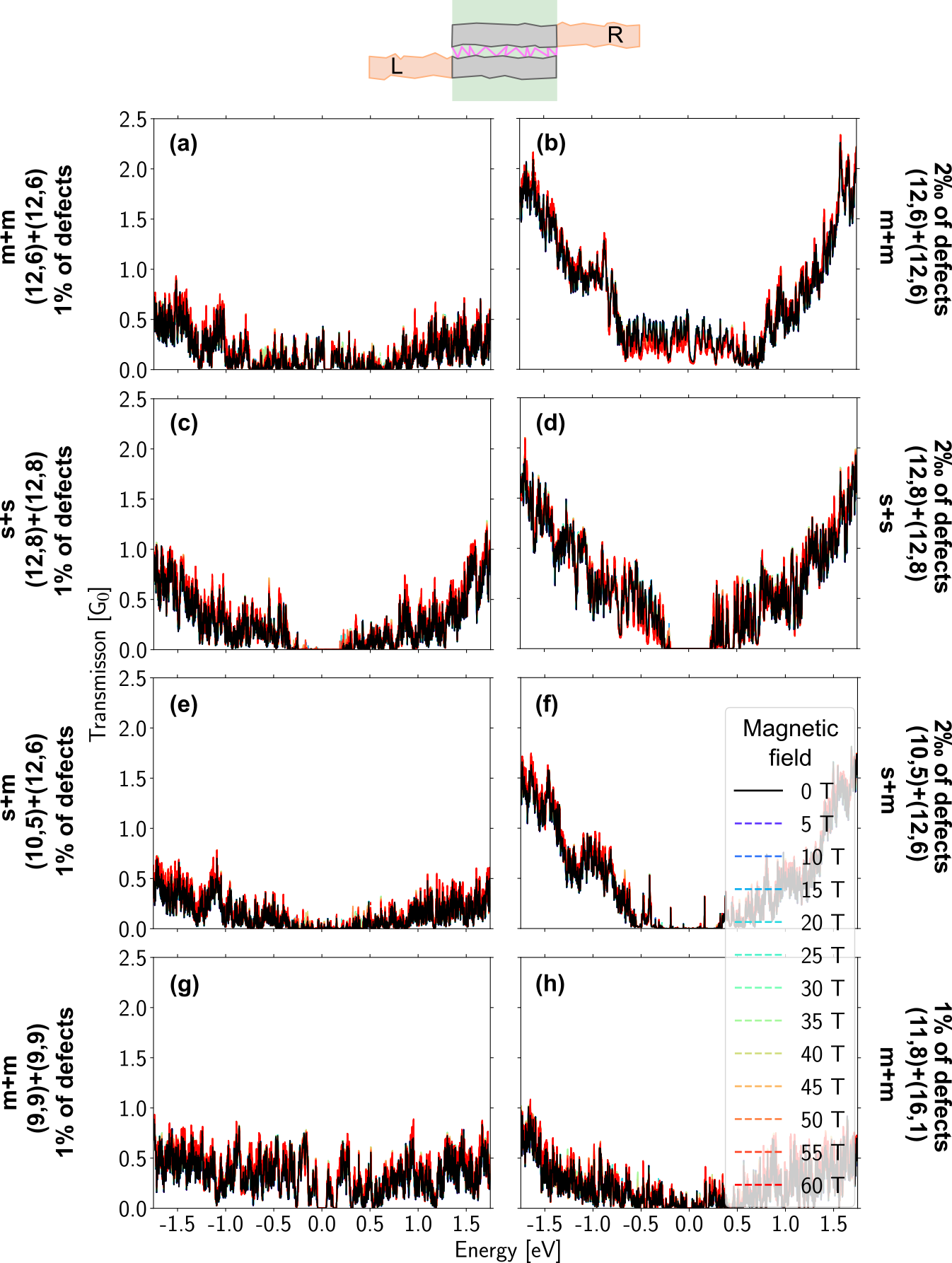}
	\caption{The zero-bias transmission spectra in the wider energy range of a), b) (12,6)+(12,6) metallic-metallic (z$_\mathrm{initial}$ = 22.6\,nm), c), d) (12,8)+(12,8) semiconducting-semiconducting (z$_\mathrm{initial}$ = 22.3\,nm), e), f) (10,5)+(12,6) semiconducting-metallic (zinitial = 22.6 nm) g) (9,9)+(9,9) metallic-metallic (z$_\mathrm{initial}$ = 24.6\,nm) h) (11,8)+(16,1) metallic-metallic (z$_\mathrm{initial}$ = 23.5\,nm) CNT junctions containing a),c),e),g),h) 1\,\% of Stone-Wales defects b),d),f) 0.2\,\% of Stone-Wales defects, computed under external perpendicular magnetic fields.}
	\label{FigS23}
\end{figure}

\begin{figure}[h!]
	\centering
	\includegraphics[width=0.85\textwidth]{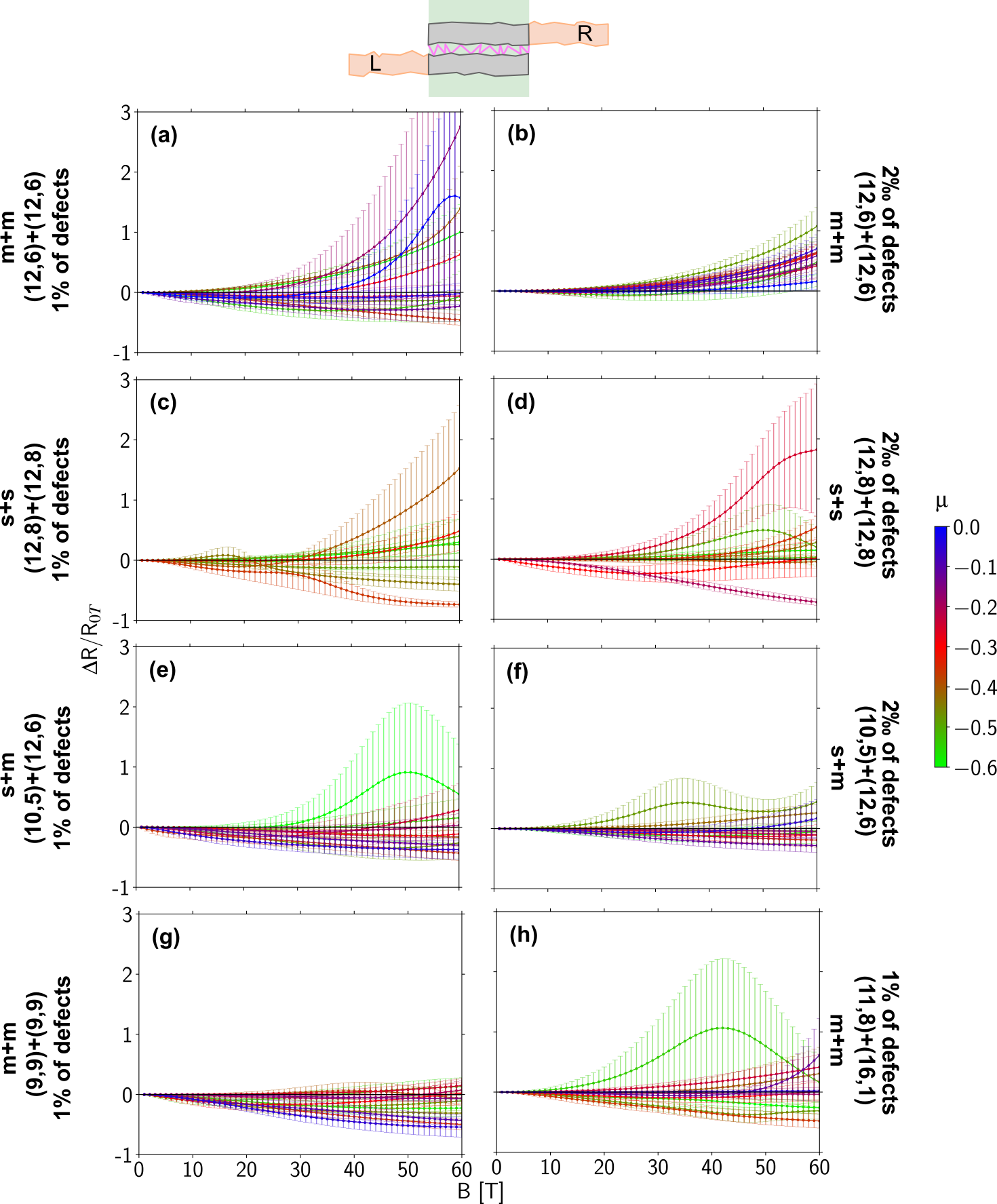}
	\caption{The computed magnetoresistance as functions of the external perpendicular magnetic field (B) of (a, b) (12,6)+(12,6) metallic-metallic (z$_\mathrm{initial}$= 22.6 nm), (c, d) (12,8)+(12,8) semiconducting-semiconducting (z$_\mathrm{initial}$ = 22.3\,nm), (e, f) (10,5)+(12,6) semiconducting-metallic (z$_\mathrm{initial}$= 22.6\,nm) g) (9,9)+(9,9) metallic-metallic (z$_\mathrm{initial}$= 24.6\,nm) h) (11,8)+(16,1) metallic-metallic (z$_\mathrm{initial}$= 23.5\,nm) CNT junctions containing  (a, c, e, g, h) 1\,\% of Stone-Wales defects (b, d, f) 0.2\,\% of Stone-Wales defects, under external perpendicular magnetic fields and at different doping levels. Each panel displays a representative subset of MR functions plotted for doping levels in the range [-0.6, 0.0]\,eV, sampled at 0.001\,eV intervals, with subsets selected proportionally to reflect different trends while ensuring clarity.}
	\label{FigS24}
\end{figure}

\begin{figure}[h!]
	\centering
	\includegraphics[width=0.95\textwidth]{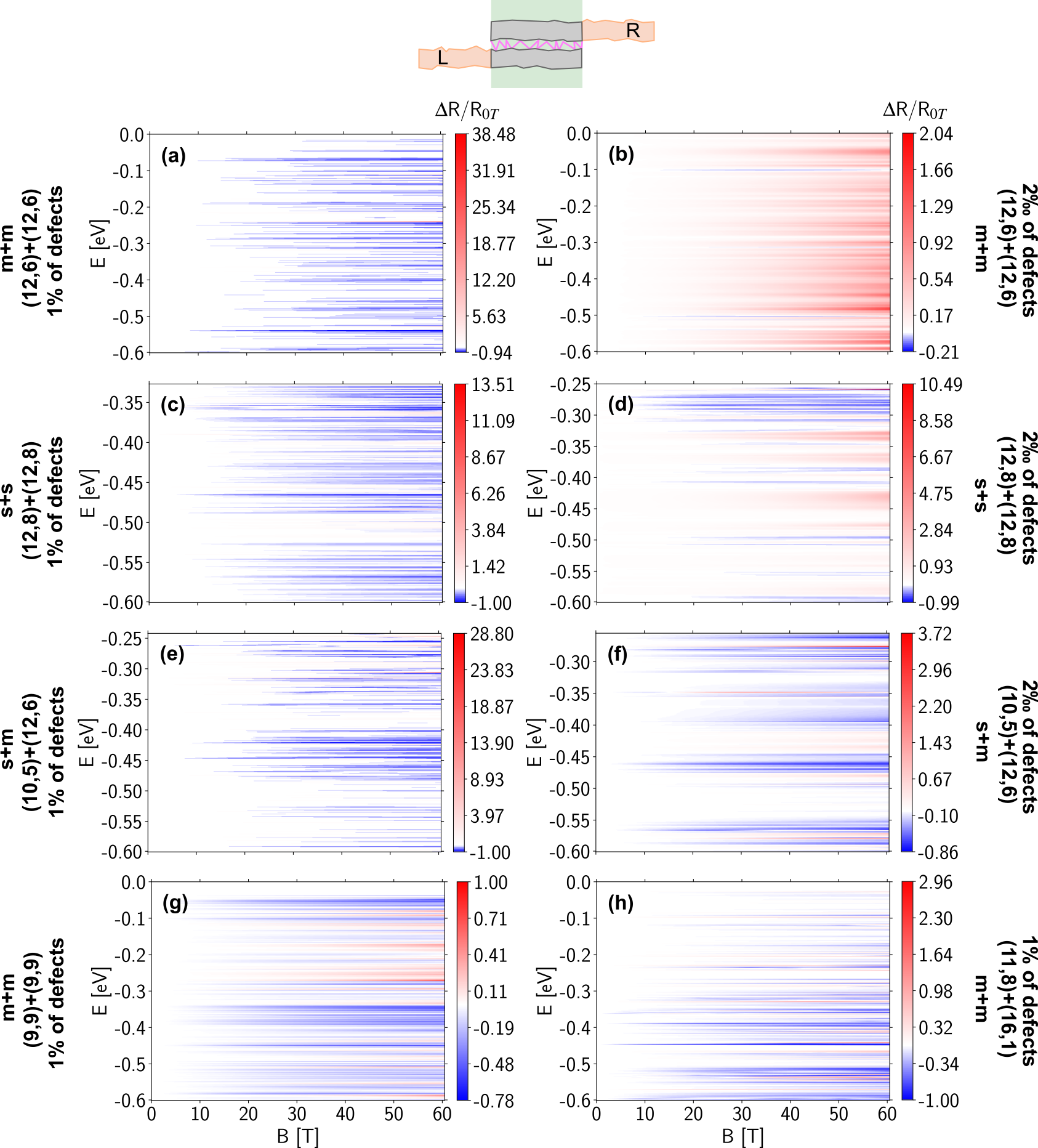}
	\caption{The computed magnetoresistance maps of (a, b) (12,6)+(12,6) metallic-metallic (z$_\mathrm{initial}$ = 22.6\,nm), (c, d) (12,8)+(12,8) semiconducting-semiconducting (z$_\mathrm{initial}$ = 22.3\,nm), (e, f) (10,5)+(12,6) semiconducting-metallic (z$_\mathrm{initial}$ = 22.6\,nm) g) (9,9)+(9,9) metallic-metallic (z$_\mathrm{initial}$ = 24.6\,nm) h) (11,8)+(16,1) metallic-metallic (z$_\mathrm{initial}$= 23.5\,nm) CNT junctions containing (a, c, e, g, h) 1\,\% of Stone-Wales defects (b, d, f) 0.2\,\% of Stone--Wales defects, under external perpendicular magnetic fields and at different doping levels.}
	\label{FigS25}
\end{figure}

\FloatBarrier 
	
At 0.2\,\% defect density, the transmission observed in pristine junctions become fragmented, and sharp features appear across the full energy window. In junctions with non-zero band gaps, localised peaks emerge within the gap region, corresponding to spatially confined defect-induced states, as visible in the LDOS maps (Extended Data Figs.~\ref{FigE5} c),f)). These states do not contribute to conduction but act as scattering centres. At 1\,\%, the number of localised states increases, and transmission is further suppressed. The lateral deformation of the CNT walls also becomes more pronounced (Extended Data Fig.~\ref{FigE2}h)).
	
Comparison of Supplementary Figs.~\ref{FigS8} and~\ref{FigS22} shows that the band gap evolves non-monotonically with increasing defect density. For the semiconducting-semiconducting junction, it narrows at 0.2\,\% defect concentration but broadens again at 1\,\%, while for the semiconducting-metallic junction it decreases progressively with increasing disorder.

\FloatBarrier 
	
\subsubsection{Theoretical modelling: Temperature vs. Structural Defects in CNT Junctions}\label{secTdtemp}
	
As shown in Extended Data Fig.~\ref{FigE5}, In the ideal case (left panel), black frames mark regions of increased state density on both nanotubes, highlighting interference effects. For LDOS maps corresponding to positive MR, the interference fringe patterns in tube 1 and tube 2 are closely spaced but do not 	match exactly, in contrast to the perfect alignment observed in Figure~\ref{Fig3}b. While in Figure~\ref{Fig3}b, the interference patterns from nanotube 1 and nanotube 2 for positive MR align exactly, in the present case, the fringe spacing differs between the two tubes, yet most of the maxima in 	nanotube 1 are positioned beneath maxima in nanotube 2. For LDOS maps corresponding to 	negative MR, the phase relationship between the fringes changes, and most maxima in nanotube 1 are instead positioned beneath minima in nanotube 2, indicating a phase shift between the 	electronic states of the two tubes.
	
When the junction is subjected to a temperature of 100 K (Extended Data Fig.~\ref{FigE5}, middle panel), the fringe spacing increases, and the distinct interference matching/mismatching patterns observed in the pristine case become significantly less pronounced, suggesting thermal broadening of electronic states.
	
In the presence of structural defects (Extended Data Fig.~\ref{FigE5}, right panel), local state localization is evident, as indicated by the oval frame. These defects introduce electronic states that act as scattering centers, disrupting coherent electron propagation along the nanotubes and further suppressing interference effects in the junction.
	
\subsubsection{Theoretical modelling: Effect of Temperature on Non-Ideal CNT Junctions}\label{secTdisorder}

\begin{figure}[h!]
	\centering
	\includegraphics[width=0.95\textwidth]{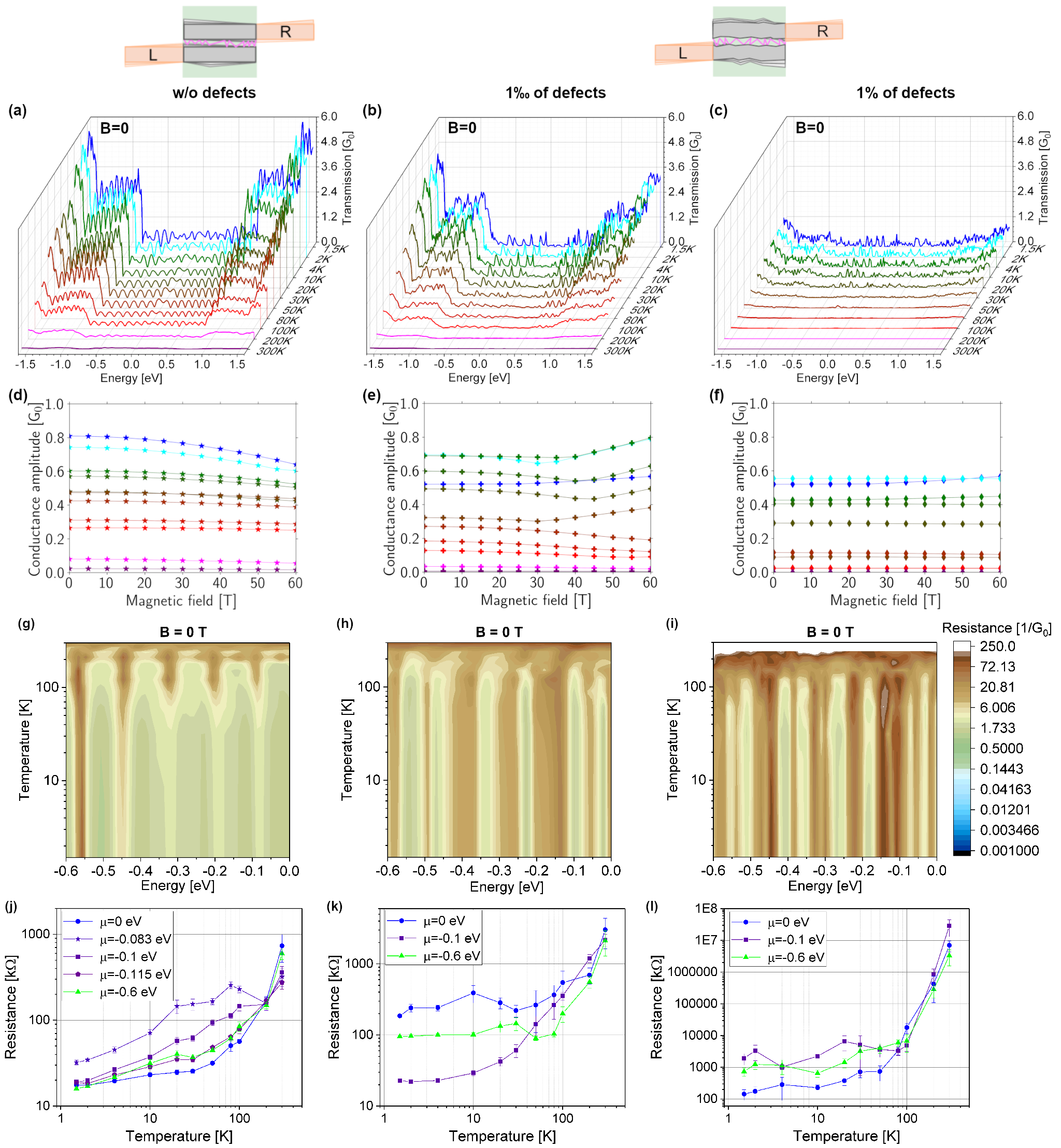}
	\caption{ (a-c) The computed transmission spectra at B=0\,T as functions of energy (effectively representing doping levels) and (d-f) conductance amplitude versus external perpendicular magnetic field B for pristine, 0.1\,\%, and 1\,\% Stone--Wales defected junctions (from left to right), at various temperatures. The initial overlap region (z$_\mathrm{initial}$) was 14.3\,nm (58\,units of a (9,9) CNT). The averaged conductance represents the mean transmission value over the energy range from -0.6\,eV (nitric acid doping\cite{hayashi2020}) to the Fermi level (0.0\,eV), as depicted in Supplementary Figs.~\ref{FigS5}c,d. The point colors in (d-f) correspond to the colours of the lines in (a-c), which represent different temperatures. (g-i) The computed resistance maps at B=0\,T for the same junctions, as functions of temperature and doping level. At low temperatures ($\le$200\,K), the pristine junction exhibits a strikingly regular pattern of alternating high- and low-resistance bands, progressively suppressed with increasing defect density. (j-l) Line cuts of resistance versus temperature taken at the energies indicated in panels (g-i): E = 0\,eV (undoped), E = -0.1\,eV (low hole doping), and E = -0.6\ eV (heavy hole-doping level typical of nitric acid treatment\cite{hayashi2020}. For the pristine junction (j), two additional profiles sample contrasting features of the map: E = -0.083\,eV (across a ridge of elevated resistance) and E = -0.12\,eV (across a valley of reduced resistance).}
	\label{FigS26}
\end{figure}

\begin{figure}[h!]
	\centering
	\includegraphics[width=0.95\textwidth]{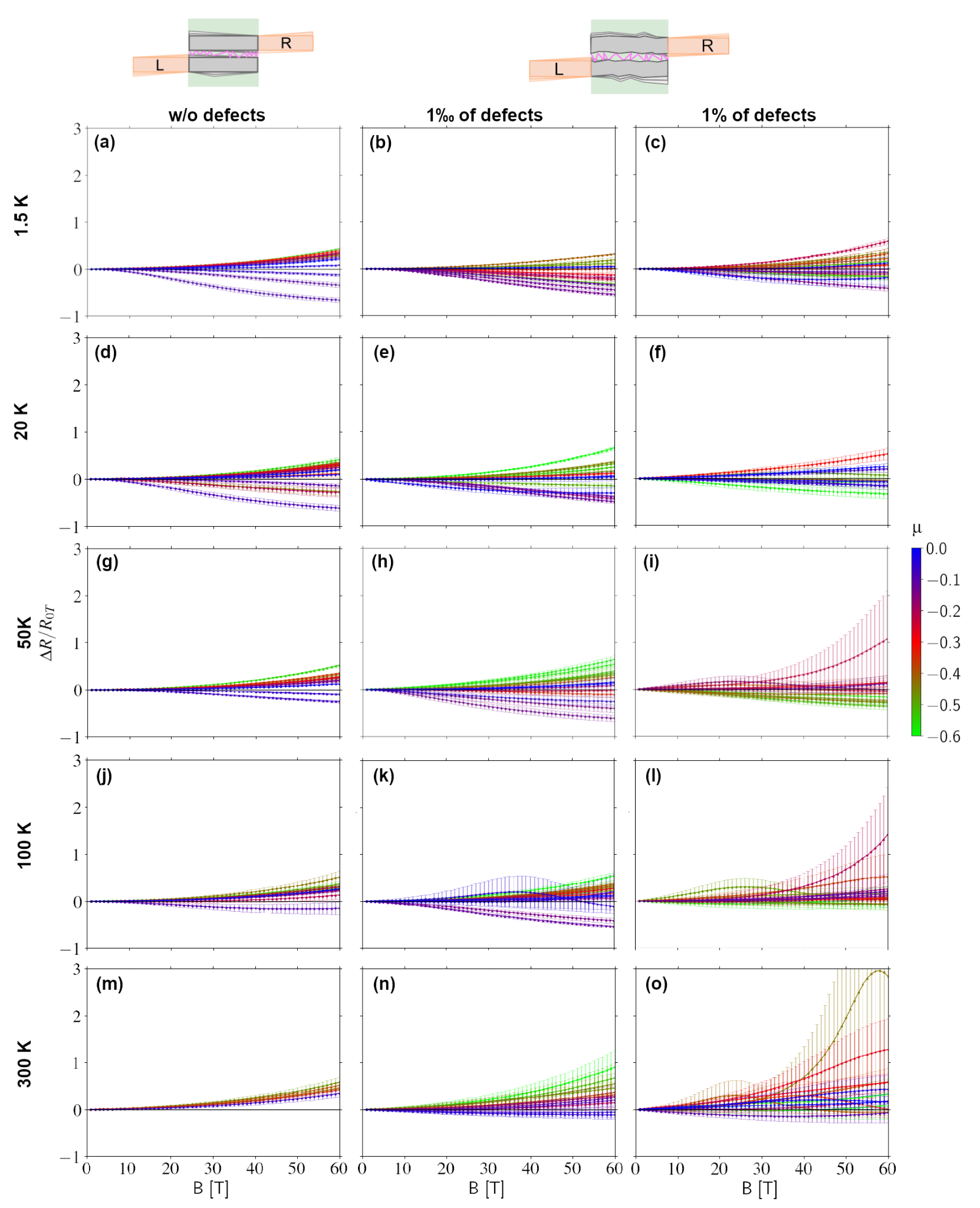}
	\caption{The computed magnetoresistance as functions of the external perpendicular magnetic field (B) of (9,9)+(9,9) metallic-metallic (z$_\mathrm{initial}$= 14.3\,nm) CNT junctions (a, d, g, j, m) without defects and with (b, e, h, k, n) 0.1\,\% of Stone-Wales defects (c, f, i, l, o) 1\,\% of Stone-Wales defects, at various temperatures, doping levels and under external perpendicular magnetic fields. Each panel displays a representative subset of MR functions plotted for doping levels in the range [-0.6, 0.0]\,eV, sampled at 0.001\,eV intervals, with subsets selected proportionally to reflect different trends while ensuring clarity.}
	\label{FigS28}
\end{figure}

\begin{figure}[h!]
	\centering
	\includegraphics[width=0.95\textwidth]{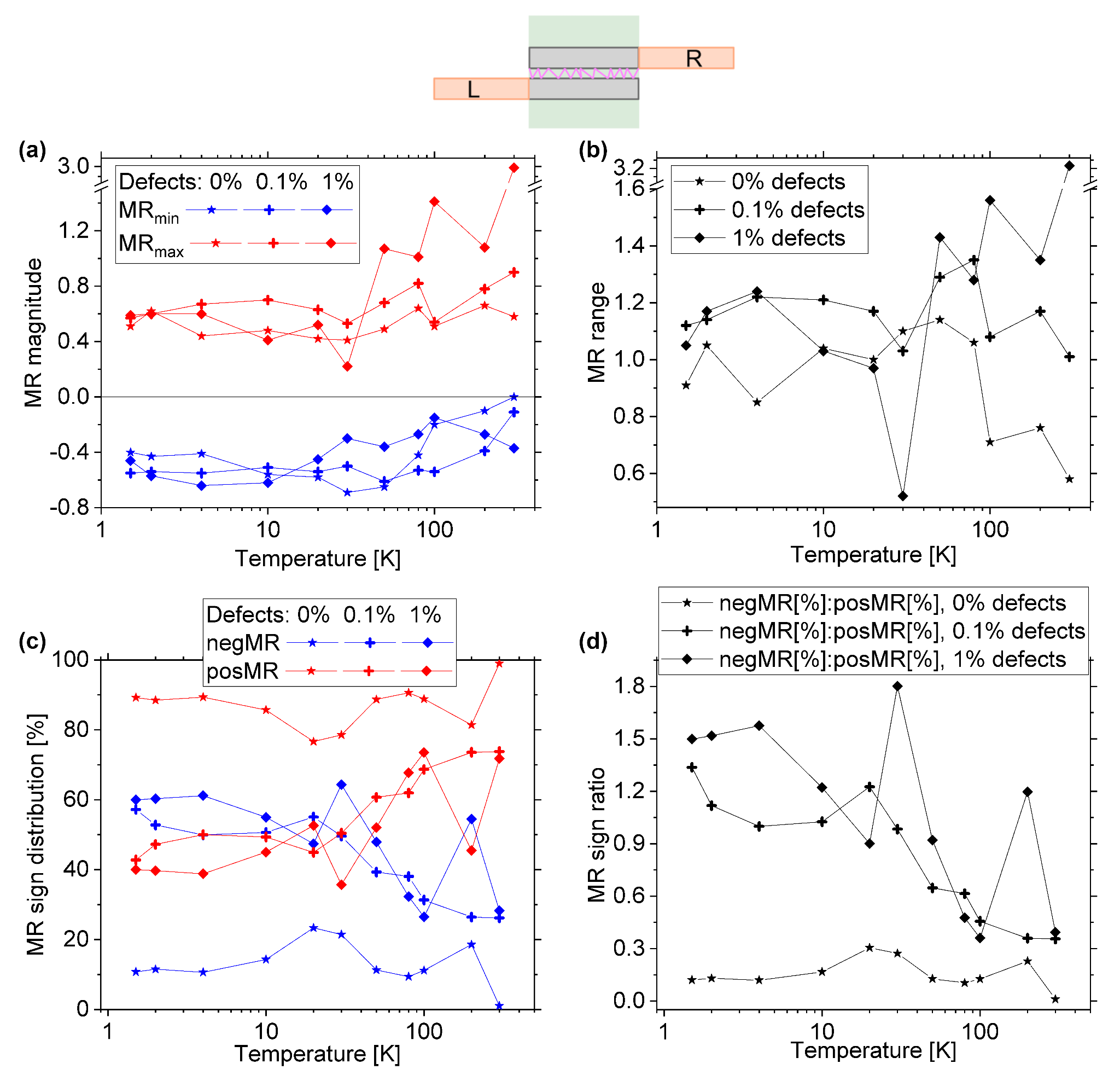}
	\caption{Magnetoresistance (MR) trends for (9,9)+(9,9) metallic-metallic CNT junctions (z$_\mathrm{initial}$= 14.3\,nm) without defects and with 0.1\,\% and 1\,\% Stone-Wales defects, at various temperatures, doping levels, and under external perpendicular magnetic fields. The MR trends were extracted from the MR maps presented in Supplementary Fig.~\ref{FigS28}, which were computed for doping levels in the range [-0.6, 0.0]\,eV, sampled at 0.001\,eV intervals. Panel (a) shows the minimum (blue) and maximum (red) MR values for each system, while panel (b) presents the MR range, defined as a difference between maximum and minimum MR value, illustrating the spread of MR values across different temperatures and defect concentrations. Panel (c) displays the percentage of negative (blue) and positive (red) MR values within the energy range [-0.6, 0.0]\,eV. Panel (d) shows the ratio of the percentage of negative MR values to positive MR values in the same energy range.}
	\label{FigS29}
\end{figure}
	
\FloatBarrier 	
	
\textbf{Comparison with experimental trends.}
	
To assess how well the model reproduces experiment, we analysed the temperature dependence of minimum MR, maximum MR, and the percentage of negative MR for (9,9)+(9,9) junctions with 0\,\%, 0.1\,\%, and 1\,\% of 5--7 defects (Supplementary Fig.~\ref{FigS29}) and compared them with the experimental data in Extended Data Fig.~\ref{FigE1}.
	
The temperature dependence of the minimum MR resembles the experimental trend (Extended Data Fig.~\ref{FigE1} c). Both perfect and defected junctions follow a parabolic-like evolution: MR$_{\mathrm{min}}$ becomes more negative at low T, reaches a minimum, and then increases again at higher T. The position of this extremum, however, depends on defect density. In the perfect case, the minimum occurs at 30\,K, whereas for 1\,\% defects it shifts to \,K, indicating that structural disorder changes the 	temperature at which negative MR is strongest.
	
The maximum MR shows partial disagreement with experiment (Extended Data Fig.~\ref{FigE1}a)). Experimentally, MR at 50\,T decreases sharply from the lowest temperatures. In contrast, in our simulations MR$_{\mathrm{max}}$ remains nearly constant at low temperatures for all three junction types, lacking this strong sensitivity. This difference can be explained by the fact that experimental fibres were not 	exclusively metallic, and other types of junctions in the sample likely enhance the low-T increase of MR at 50\,T. At higher temperatures, however, our results capture the experimental trend of increasing MR$_{\mathrm{max}}$, consistent with the behaviour of SWCNT\,1, SWCNT\,2, and SWCNT\,2 doped samples.
	
For the percentage of negative MR as a function of temperature (Supplementary Figure~\ref{FigS29}c), simulations yield non-monotonic but overall decreasing trends. This differs from the parabolic dependence seen in experiment (Extended Data Fig.~\ref{FigE1}b), where the fraction of negative MR first increases with T, reaches a maximum, and then decreases.
	
The observed suppression of negative MR with increasing temperature parallels behaviour in disordered semimetals such as V$_{1/3}$TaS$_2$\cite{wang2024}, where magnetic and structural scattering jointly weaken coherent interference.
	
\textbf{Additional numerical observation.}
	
For the perfect (9,9)+(9,9) junction, the MR range, defined as the difference between maximum and minimum MR, decreases with temperature (Supplementary Fig.~\ref{FigS29}b), indicating convergence of MR(B) curves for different doping levels at high T. For 1\,\% defects, the MR range increases with temperature, consistent with a growing spread of MR (B) responses across doping levels. At 0.1\,\% defects, the MR range oscillates with temperature, suggesting an intermediate regime where temperature does not strongly affect the diversity of MR responses.
	
\subsection{Extended statistical analysis of field and temperature scaling in MR response}\label{secStatistics2}

\begin{figure}[h!]
	\centering
	\includegraphics[width=0.95\textwidth]{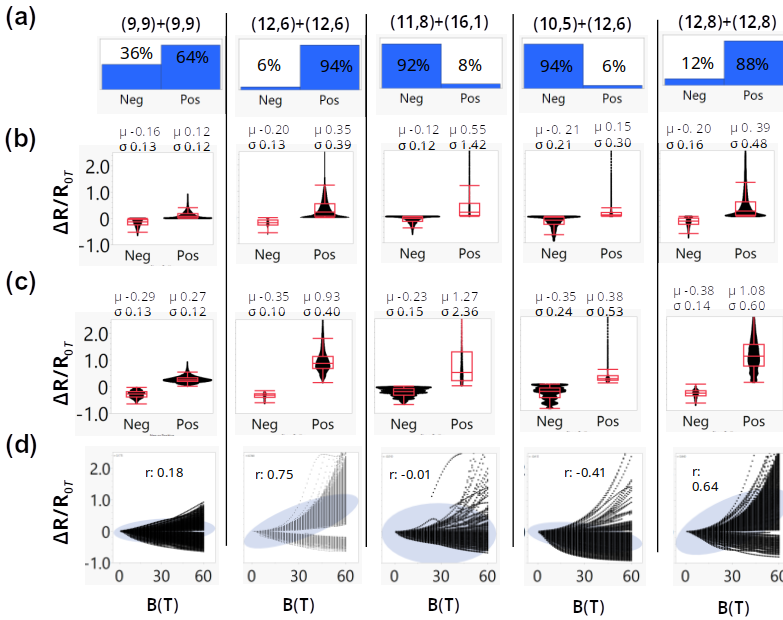}
	\caption{Statistical analysis of the MR distributions for each chirality/ junction system 
		represented in columns. MR values with relative error greater than 50\,\% are eliminated from analysis; for the  (11,8)+(16,1) system, this threshold is 100\,\%. a), the percentage of datapoints with negative or positive MR. b) 
		the distribution of MR values with average $\upmu$ and standard deviation $\upsigma$. c) the same as b except restricted between a field of 50 
		to 60 T. d) MR versus field with correlation r.   }
	\label{FigS46}
\end{figure}

In this section, we discuss the results of the statistical analysis of the numerical tightbinding magneto-resistance from various CNT junction systems. Supplementary Figure~\ref{FigS46}a-d shows the range of possible MR values that was calculated for each chirality junction system, showing that the MR can be positive or negative; Supplementary Figure~\ref{FigS46}a shows most of the MR data tends to be positive for the homojunctions ((9,9)+(9,9); (12,6)+ (12,6); (12,8)+(12,8)) and negative for the heterojunctions ((11,8)+(16,1); (10,5)+(12,6)). Supplementary Figure~\ref{FigS46}b shows that, generally, the distribution of positive MR magnitudes skew to more extreme values than negative MR. Supplementary Figure~\ref{FigS46}c shows these MR distributions while only considering high field (50 to 60\,T), showing a greater skew to more positive MR values. Supplementary Figure~\ref{FigS46}d plots MR versus field with calculated correlation. The homojunctions have positive correlation: ((9,9)+(9,9) r=0.18; (12,6)+ (12,6) r=0.75; (12,8)+(12,8) r=0.64)). The heterojunctions have zero to no negative correlation: ((11,8)+(16,1) r=-0.01; (10,5)+(12,6) r=-0.41). This analysis glosses over other input factors such as temperature, defect density, and doping level; multivariate linear regression can better account for the contributions of other input factors and is discussed below.

\subsubsection{MR for (9,9)+(9,9)}

\begin{figure}[h!]
	\centering
	\includegraphics[width=0.95\textwidth]{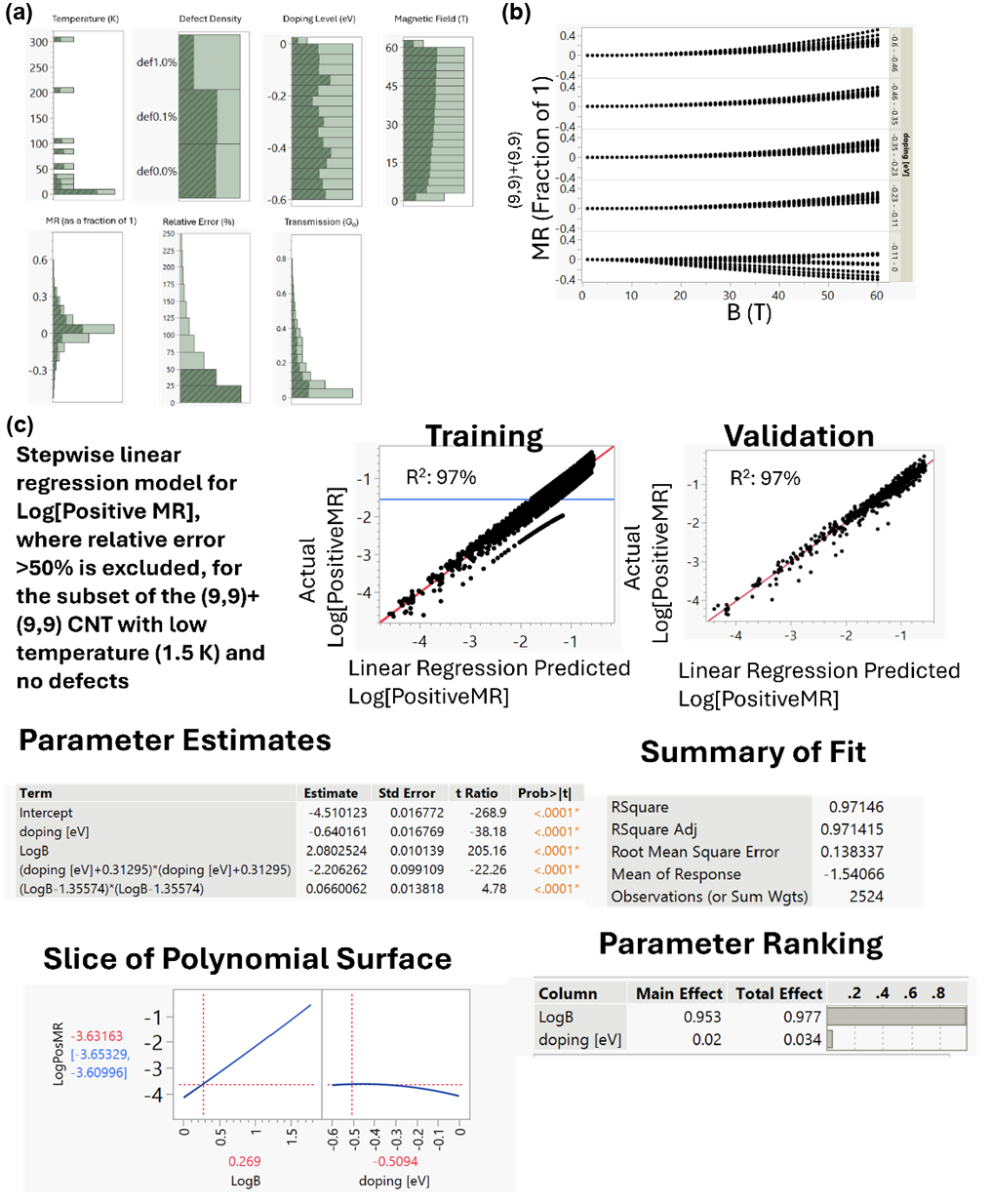}
	\caption{a) Distribution of input factors and output responses of the (9,9)+(9,9) dataset, consisting of 121\,000 different observations/rows. The shaded region represents relative error $\le$ 50\,\%, which is the subset that this analysis mostly considers. b) As an introduction, MR versus magnetic field B for various doping (rows) for the for (9,9)+(9,9) with relative error $<$ 50\,\%-- in the particular subset of coldest temperature (1.5\,K) and no defects. This subset shows the effect of doping without the complications yet from defects and temperature, showing how doping moves MR from negative to positive.  c) Linear regression response surface model of positive MR of the subset of (9,9)+(9,9), with relative error $\le$ 50\,\% and only the coldest temperature (1.5\,K) and no defects are considered. The fitted parameter for Log[B] is 2.08, which is close to a quadratic response. Positive MR is most impacted by field B with a small contribution from doping.}
	\label{FigS30}
\end{figure}
	
We now analyze the dataset that just addresses the (9,9) to (9,9) system where magnetic field, defect density, doping level, and temperature are varied. See Supplementary Figure~\ref{FigS30}a below.
	
As a starting point for simplicity, we first consider a subset of the (9,9)+(9,9) dataset that only 	includes the coldest temperature and zero defects. In Supplementary Figure~\ref{FigS30}b is shown  a simple bivariate MR vs B plot showing how doping level can affect the nature of the MR response, although for most doping levels the response is positive/ up-ward trending.
	
Now we build a linear regression response surface model for this case (the low temperature, no defect subset of (9,9)+(9,9) with relative error$<$50\,\%). Simply using MR, which can be positive or negative and varies over orders of magnitude, results in multivariate models with poor predictive capability (R$^2$$<$60\,\%). Taking the log of MR results in better models, although excludes the parameter space yielding negative MR values.

\begin{figure}[h!]
	\centering
	\includegraphics[width=0.95\textwidth]{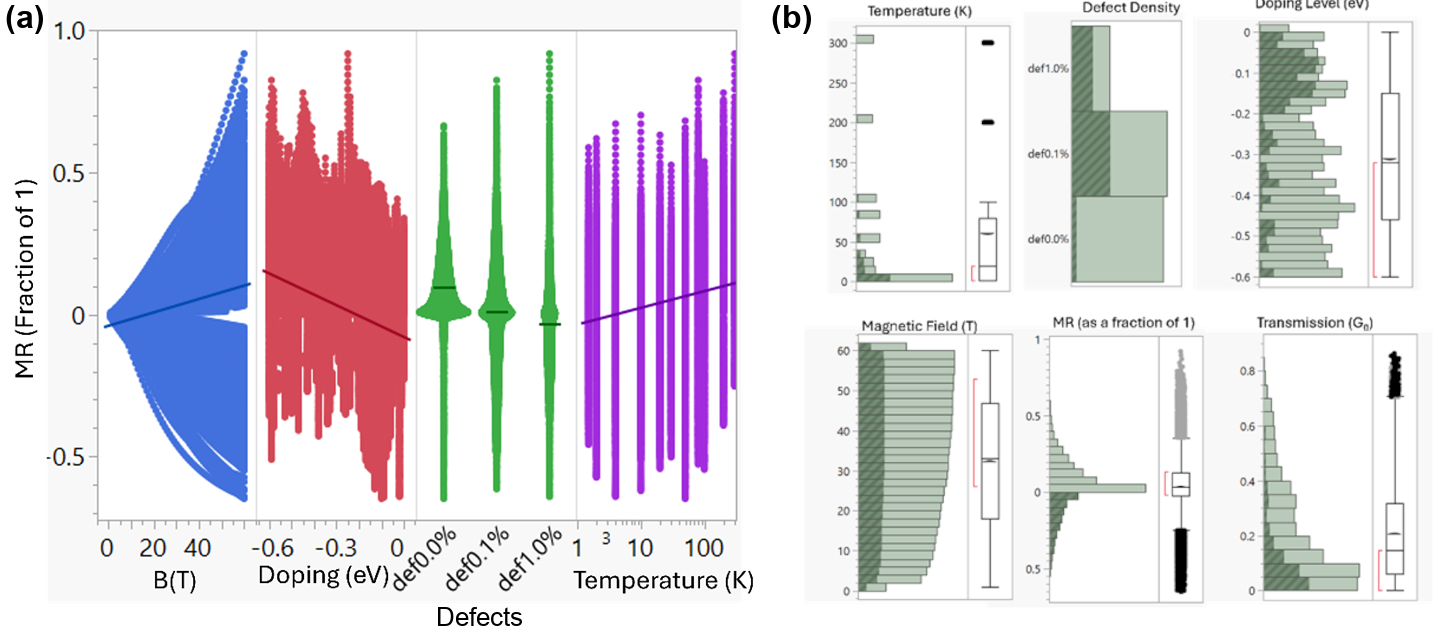}
	\caption{a) MR versus input factors, for (9,9)+(9,9) with relative error $\le$ 50\,\%, with best fit lines. Note that each datapoint has other input factors that are not held constant, which causes significant noise in these 2D plots. From these plots it is apparent all input factors have some impact on MR. As shown with the simpler subset, MR versus magnetic field B has a bifurcation between positive and negative MR. b) To understand the bifurcation in MR in the previous panel, consider the distribution of the (9,9)+(9,9) dataset with relative error $<$ 50\,\% where all negative MR values are highlighted. As shown, negative MR is associated with near zero doping levels and low transmission.}
	\label{FigS31}
\end{figure}

The multivariate model showed that only when positive MR is considered, the MR is most affected by field with a power law exponent of two, while there is only a small contribution from doping. This is interpreted as once the doping level moves beyond a threshold ($\approx$-0.11\,eV), the MR goes from negative to positive and no longer particularly affects the MR. We will now consider the full (9,9)+(9,9) system with variable defect density and temperature, although relative error is still $<$50\,\%. Supplementary Figure~\ref{FigS31}a below shows calculated MR versus these other input factors, each showing broad impact on the MR. The MR versus B still bifurcates between positive and negative values. Supplementary Figure~\ref{FigS31}b is a collection of distributions with the entire negative MR highlighted. It shows that the negative MR is associated with low transmission and, as shown before, doping levels close to zero.

\begin{figure}[h!]
	\centering
	\includegraphics[width=0.65\textwidth]{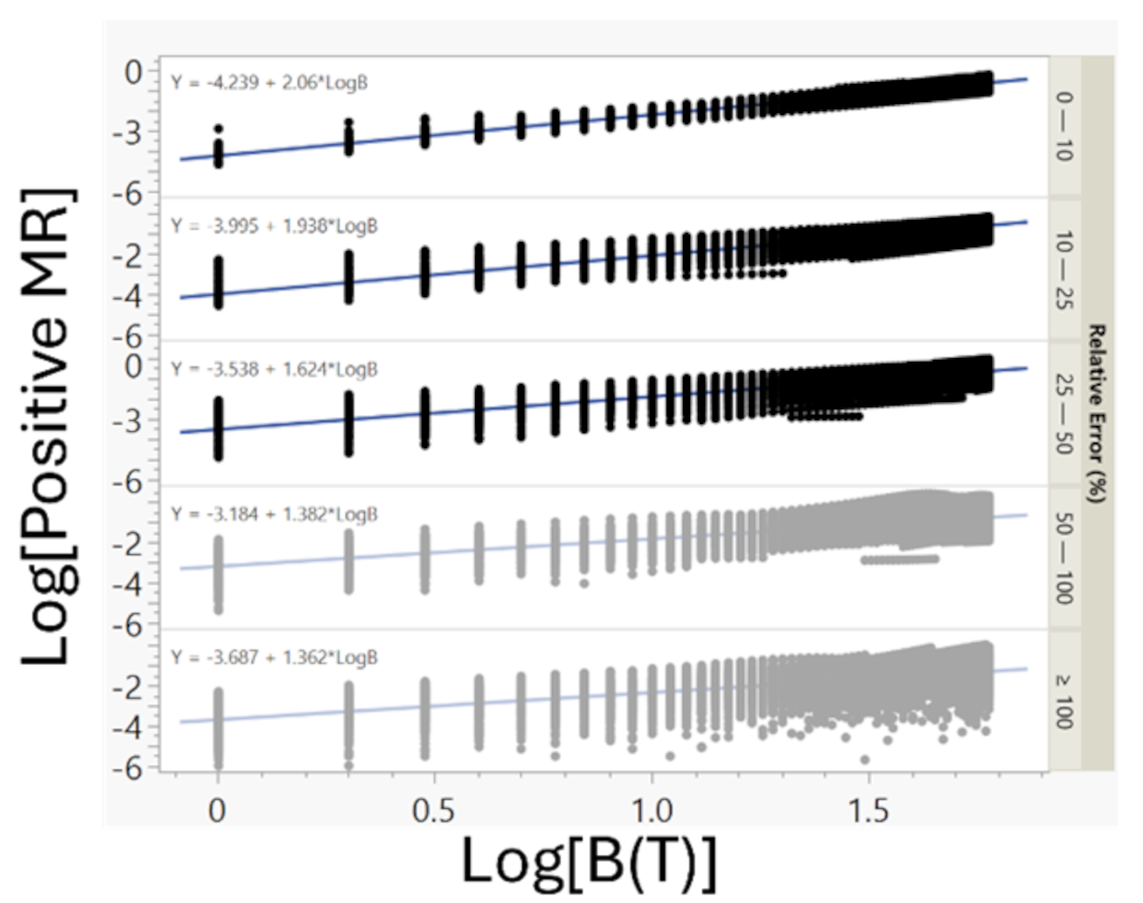}
	\caption{MR versus input factors, for (9,9)+(9,9) with relative error $\le$ 50\,\%, with best fit lines. Note that each datapoint has other input factors that are not held constant, which causes significant noise in these 2D plots. From these plots it is apparent all input factors have some impact on MR. As shown with the simpler subset, MR versus magnetic field B has a bifurcation between positive and negative MR.}
	\label{FigS32}
\end{figure}

\begin{figure}[h!]
	\centering
	\includegraphics[width=0.90\textwidth]{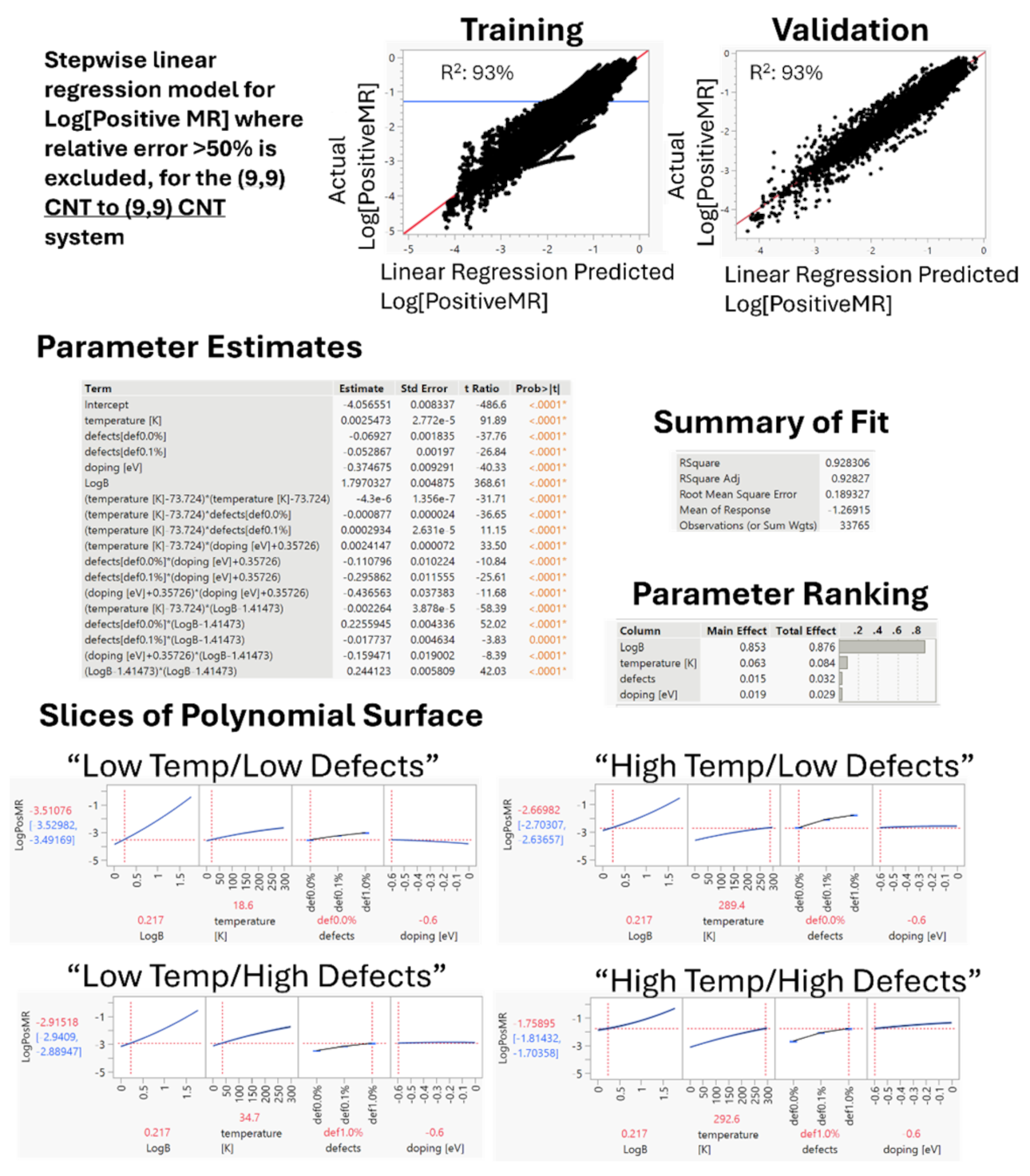}
	\caption{To understand the bifurcation in MR in the Supplementary Fig.~\ref{FigS32}, consider the distribution of the (9,9)+(9,9) dataset with relative error $<$ 50\,\% where all negative MR values are highlighted. As shown, negative MR is associated with near zero doping levels and low transmission.}
	\label{FigS33}
\end{figure}

To identify any power-laws and aid in model construction, we now take the log of MR, which 	removes negative MR from consideration. Supplementary Figure~\ref{FigS32} shows that as the relative error decreases, the best fit power law approaches an exponent of two. This trend is present despite the impact of other non-represented input factors causing spread in the response. Supplementary Figure~\ref{FigS33} addresses the multivariate impact with a linear regression response surface model, demonstrating that B is the most important parameter with exponent 1.8. Increasing temperature or defect density interacts with the MR vs B response, making it less steep.

\subsubsection{MR for other junctions}
	
\begin{figure}[h!]
	\centering
	\includegraphics[width=0.9\textwidth]{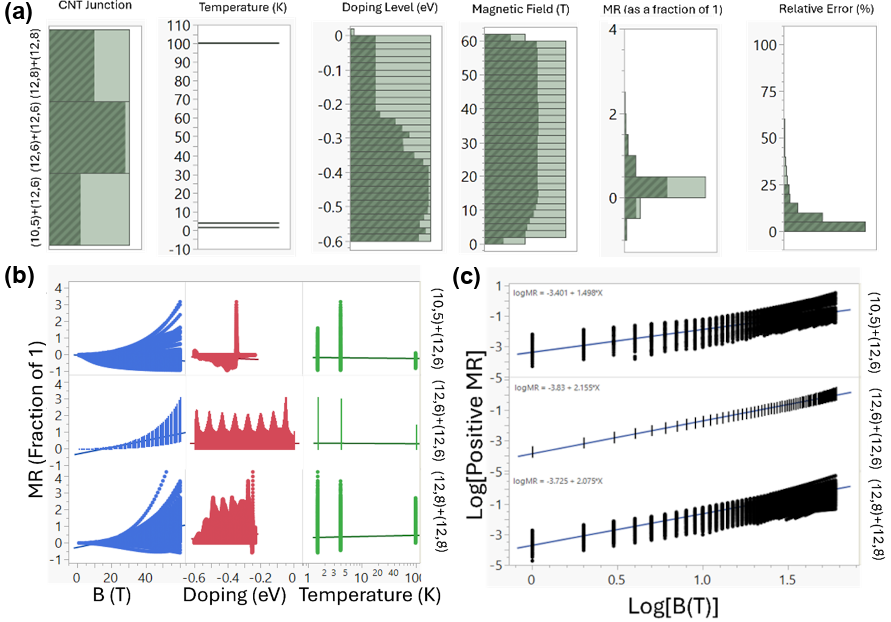}
	\caption{a) Distribution of input factors and output responses of the multiple chirality--junctions dataset, consisting of 325\,000 different observations/rows. The shaded region represents relative error $\le$ 50\,\%, which is the subset that this analysis considers. b) Bivariate plots of calculated MR versus input factors, for all three chirality junction systems (in rows), with relative error $\le$ 50\,\%, with best fit lines. Note that each datapoint has other input factors that are not held constant, which causes significant noise in these 2D plots. c) Power-law analysis for the positive MR of the chirality junction systems (in rows) with relative error $\le$ 50\,\%. The best fit slopes of these log log plots indicate the power--law exponent. The homogenous junctions here have power--law close to quadratic, while the heterogenous junction is further away from this.}
	\label{FigS34}
\end{figure}

We now analyze the dataset that considers other chirality junction systems ((12,8)+(12,8); (12,6)+(12,6); (10,5)+(12,6); (11,8)+(16,1)) where magnetic field, doping level, and temperature are varied (Supplementary Fig.~\ref{FigS34}a)).

Supplementary Figure.~\ref{FigS34}b shows the MR as a function of input factors, for three of the chirality junction systems (12,8)+(12,8); (12,6)+(12,6); (10,5)+(12,6). The MR primarily trends upward with B for all three chirality junctions, although some negative MR is present. The (12,6)+(12,6) system particularly has little negative MR. There are multiple complicated responses between MR and doping level with "rippling" structures, while temperature does not seem impactful at this level of analysis.

\begin{figure}[h!]
	\centering
	\includegraphics[width=0.80\textwidth]{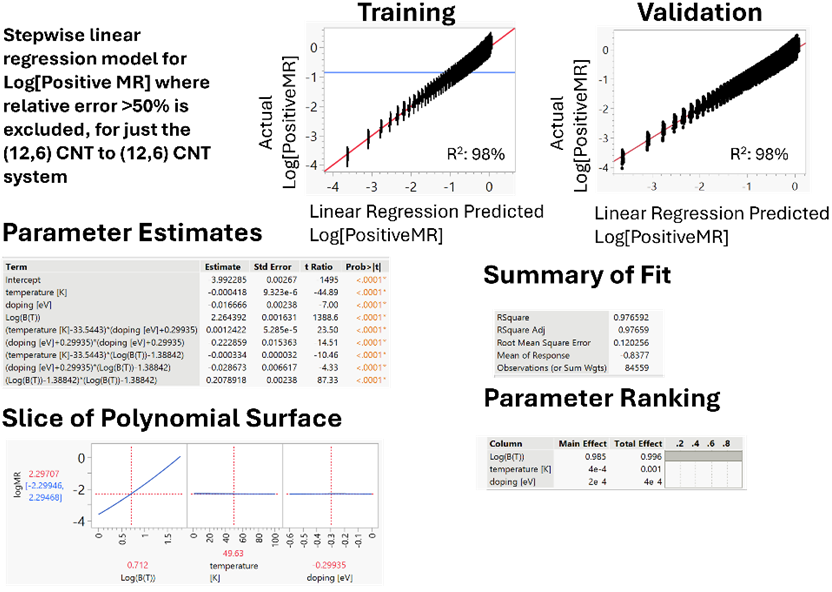}
	\caption{Linear regression response surface model of positive MR for just  (12,6)+(12,6) with relative error $\le$ 50\,\%. The fitted parameter for Log[B] is 2.26, which is close to a quadratic response. Other parameters do not impact the output meaningfully relative to magnetic field.}
	\label{FigS35}
\end{figure}

\begin{figure}[h!]
	\centering
	\includegraphics[width=0.80\textwidth]{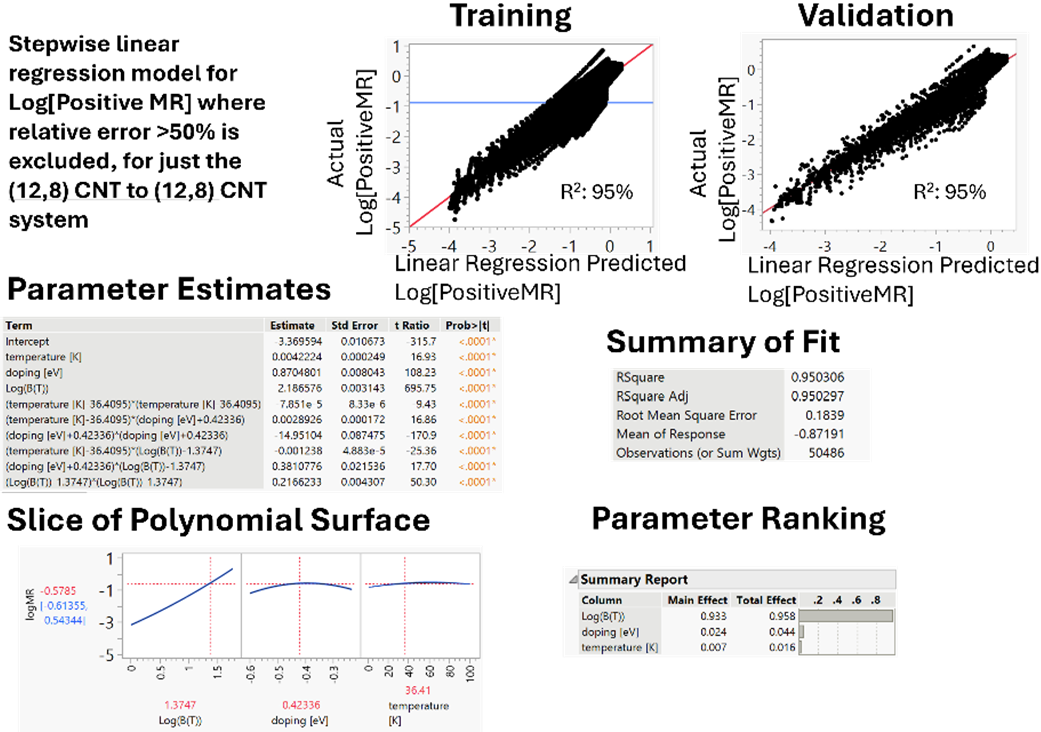}
	\caption{Linear regression response surface model of positive MR for just  (12,8)+(12,8) with relative error $\le$ 50\,\%. The fitted parameter for Log[B] is 2.19, which is close to a quadratic response. Other parameters do not impact the output meaningfully relative to magnetic field.}
	\label{FigS36}
\end{figure}

\begin{figure}[h!]
	\centering
	\includegraphics[width=0.80\textwidth]{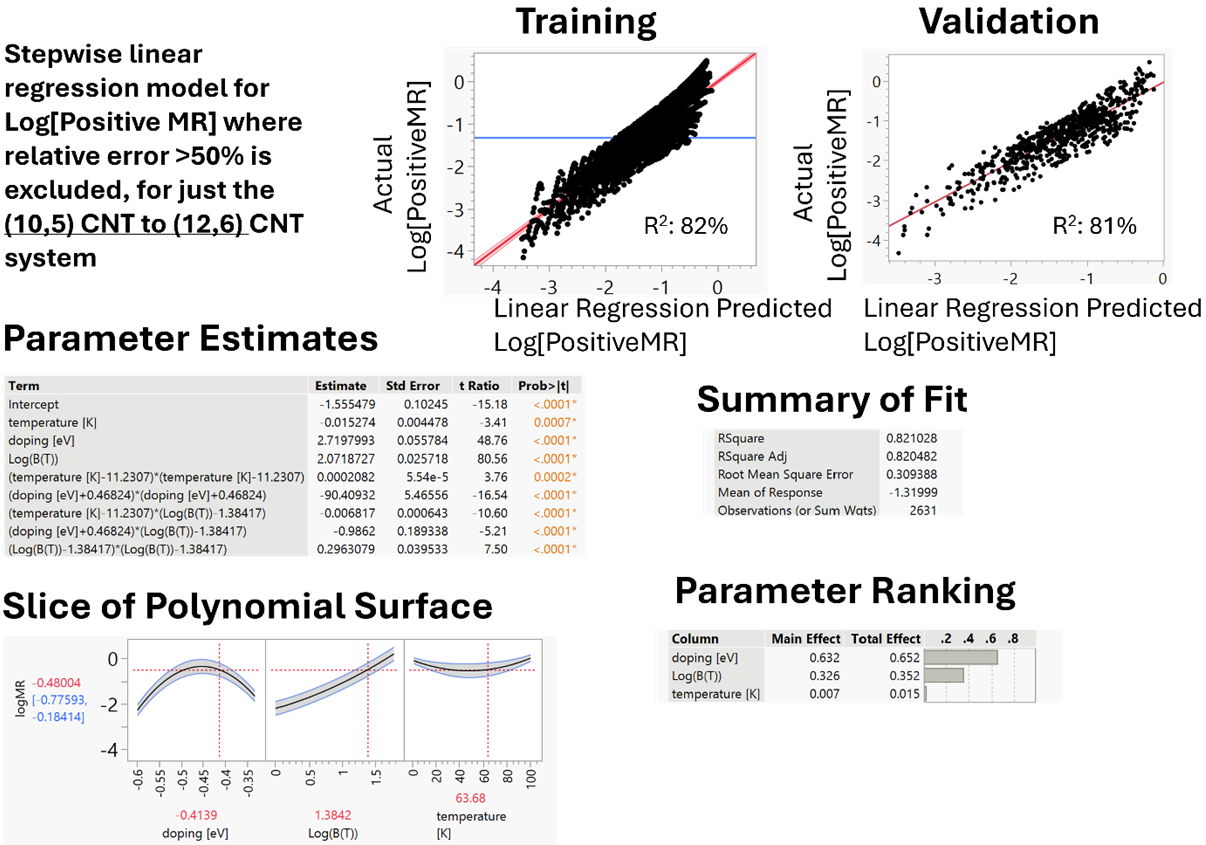}
	\caption{Linear regression response surface model of positive MR for just  (10,5)+(12,6) with relative error $\le$ 50\,\%. The fitted parameter for Log[B] is 2.07, which is close to a quadratic response. Unlike the other junctions, doping has significant influence on the MR response relative to the impact of B. Once doping is taken into account, the quadratic response is clear, which was obscured in the bivariate analysis above.}
	\label{FigS37}
\end{figure}

\begin{figure}[h!]
	\centering
	\includegraphics[width=0.95\textwidth]{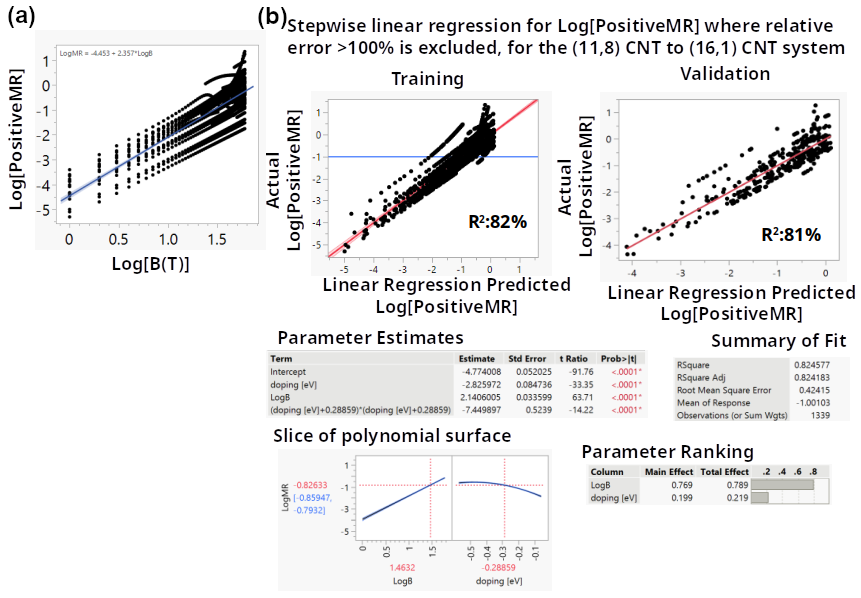}
	\caption{For just the (11,8)+(16,1) system with relative error $\le$ 10\,\%, a) log plot of positive MR verses magnetic field B, showing a fitted exponent of 2.36. b) Linear regression response surface model of positive MR. The fitted parameter for Log[B] is 2.14, which is close to a quadratic response. Unlike the homojunctions, doping has significant influence on the MR response relative to the impact of B. Once doping is taken into account, the quadratic response is clear, which was more obscured in the bivariate analysis.}
	\label{FigS45}
\end{figure}

To identify any power-laws and aid in model construction, Supplementary Fig.~\ref{FigS34}c) plots the log of MR versus the log of B, which removes negative MR from consideration. Despite the influence of non-represented input factors, the positive MR power law with B is quadratic for the homogeneous junctions, while the heterogeneous junction has a smaller exponent.

To take into account the influence of other input factors, we now build linear regression response surface models for Log[MR] for all three chirality junction systems, as shown in Supplementary Figs.~\ref{FigS35}--\ref{FigS37}. Both (12,6)+(12,6) and (12,8)+(12,8) systems have a quadratic response with B, where B is the most impactful parameter with little relative influence from doping level or temperature. We saw complex ripple structures in the bivariate data, although the ripples are likely smoothed by the log transform and are features not captured in the model. In the case of (10,5)+(12,6), doping was the most important factor; when this was taken into account a quadratic dependence between MR and B was again uncovered. Previously, in the power-law fitting this quadratic relationship was not obvious and highlights the importance of multivariate modeling.

Now we discuss the (11,8)+(16,1) junction system. Previously we had restricted analysis to relative error to less than 50\,\%. This had the effect of removing most datapoints in this particular system, so here we permitted relative error less than 100\,\% to allow some degree of analysis on its trends.  Supplementary Fig.~\ref{FigS45}a shows that the power law analysis yields a fitted exponent of 2.36. Supplementary Fig.~\ref{FigS45}b shows the multivariate linear regression model. After doping is taken into account, the new exponent is closer to quadratic at a value of 2.14.


\subsubsection{Negative MR for (9,9)+(9,9)}
	
\begin{figure}[h!]
	\centering
	\includegraphics[width=0.90\textwidth]{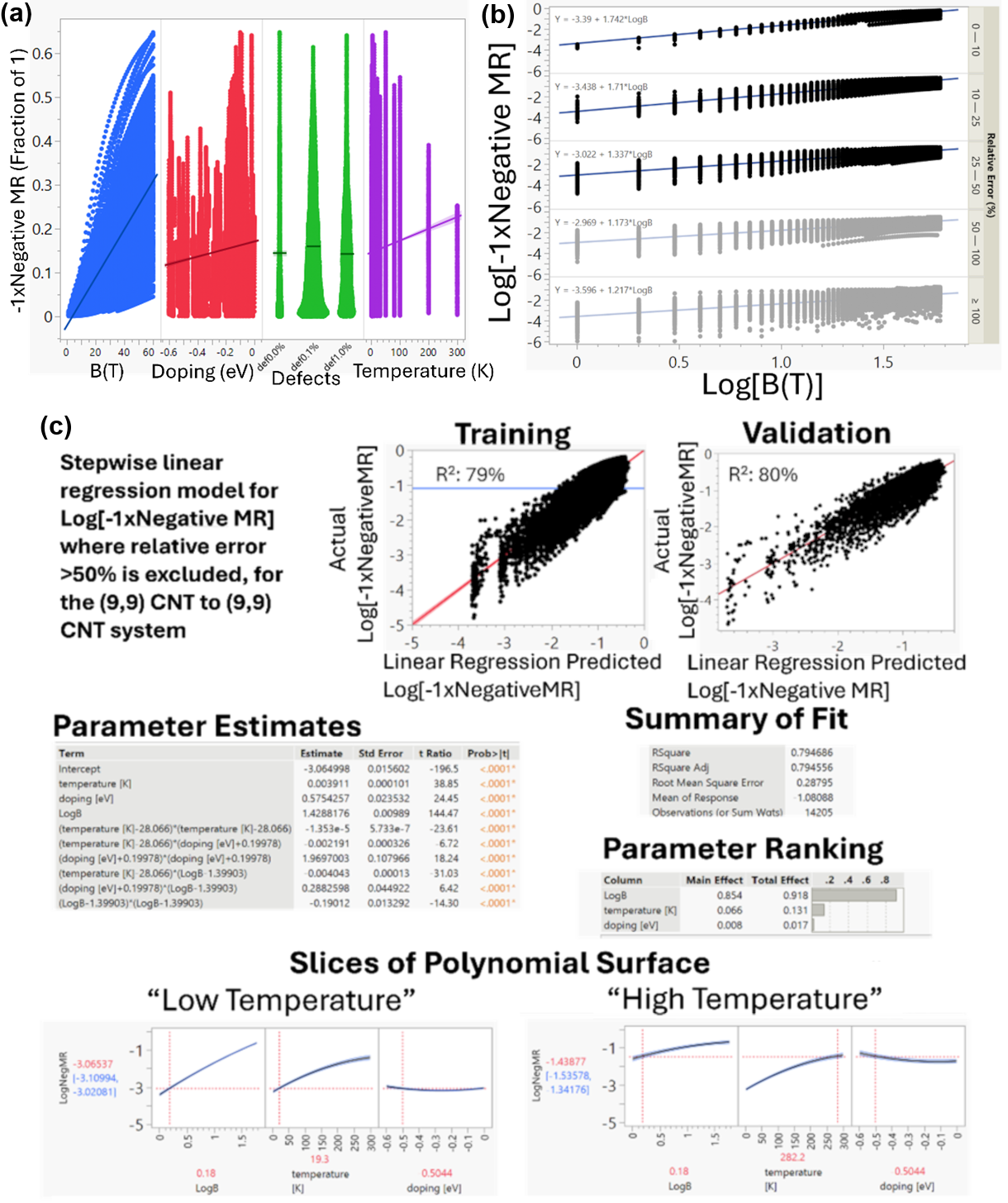}
	\caption{a) Negative MR versus input factors, for (9,9)+(9,9) with relative error $\le$ 50\,\%, with best fit lines. Note that each datapoint has other input factors that are not held constant, which causes significant noise in these 2D plots. b) Negative MR power-law analysis for (9,9)+(9,9) with unrestricted relative error, although relative error is instead binned in rows. The best fit slopes of these log log plots indicate the power-law exponent. As the relative error of each bin decreases, the exponent goes from approximately 1.2 to 1.7. c) Linear regression response surface model of negative MR for (9,9)+(9,9) with relative error$\le$ 50\,\%. The fitted parameter for Log[B] is 1.4.}
	\label{FigS38}
\end{figure}

Previously, taking the Log transform removed negative MR from consideration. Here, we discuss 	trends with strictly negative MR, where we multiply negative MR with -1 and take the log, and then conduct similar analysis. We start again with the (9,9)+(9,9) dataset. Supplementary Figure~\ref{FigS38}a below shows the calculated negative MR versus input factors, showing broad influence from all the input factors; Supplementary Figure~\ref{FigS38}b shows the power-law relationship with B, showing the exponent converging from 1.2 to 1.7 as relative error is reduced. Supplementary Figure~\ref{FigS38}c is the linear regression response surface model, showing that the best fit exponent is 1.4; there is also an interaction with temperature where higher temperature weakens the response between negative MR and B.

	
\subsubsection{Negative MR for other junctions}
	
Still considering just negative MR, we now turn back to the dataset that addresses other chirality junction systems ((12,8)+(12,8); (12,6)+(12,6); (10,5)+(12,6)) where magnetic field, doping level, and temperature are varied. See Supplementary Figure~\ref{FigS39}a for negative MR versus input factors. As shown, no data is present for (12,6)+(12,6) and little is present for (12,8)+(12,8). Supplementary Figure~\ref{FigS39}b is the best fit power-law analysis, showing negative MR vs B power-law exponents of 1.2 for (10,5)+(12,6) and 2.0 for (12,8)+(12,8). Similar exponents were uncovered for the linear regression response surface models, Supplementary Figs.~\ref{FigS40}--\ref{FigS41}.

\begin{figure}[h!]
	\centering
	\includegraphics[width=0.90\textwidth]{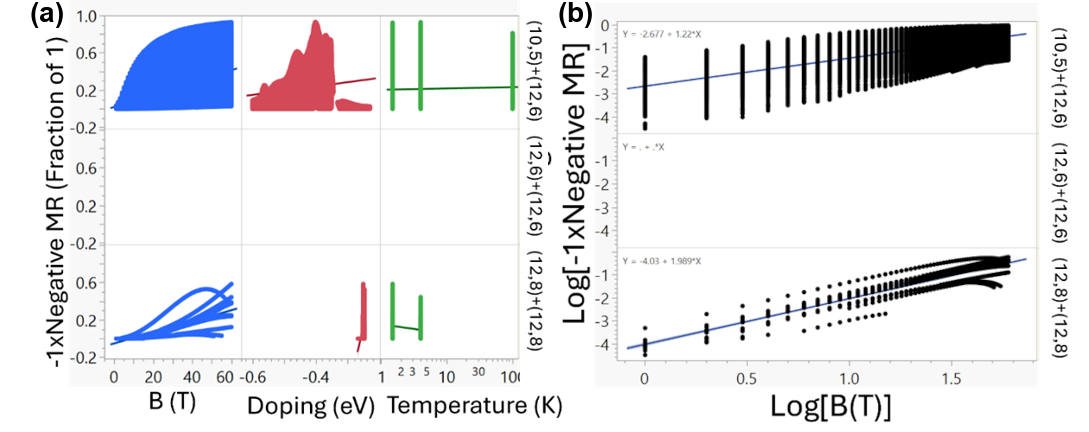}
	\caption{a) For the various chirality junction systems (in rows) with relative error $\le$ 50\,\%, bivariate negative MR versus input factors, with best fit lines are a guide for the eye. Note that each datapoint has other input factors that are not held constant, which causes significant noise in these 2D plots. As shown, along with the relative error $\le$ 50\,\% restriction, no data is available for (12,6)+(12,6) and little is available for (12,8)+(12,8). b) Negative MR power-law analysis for the chirality junction systems (in rows) with relative error $\le$ 50\,\%. The best fit slopes of these log log plots indicate the power-law exponent. (10,5)+(12,6) has exponent 1.2, while (12,8)+(12,8) has exponent 2.0. }
	\label{FigS39}
\end{figure}

\begin{figure}[h!]
	\centering
	\includegraphics[width=0.90\textwidth]{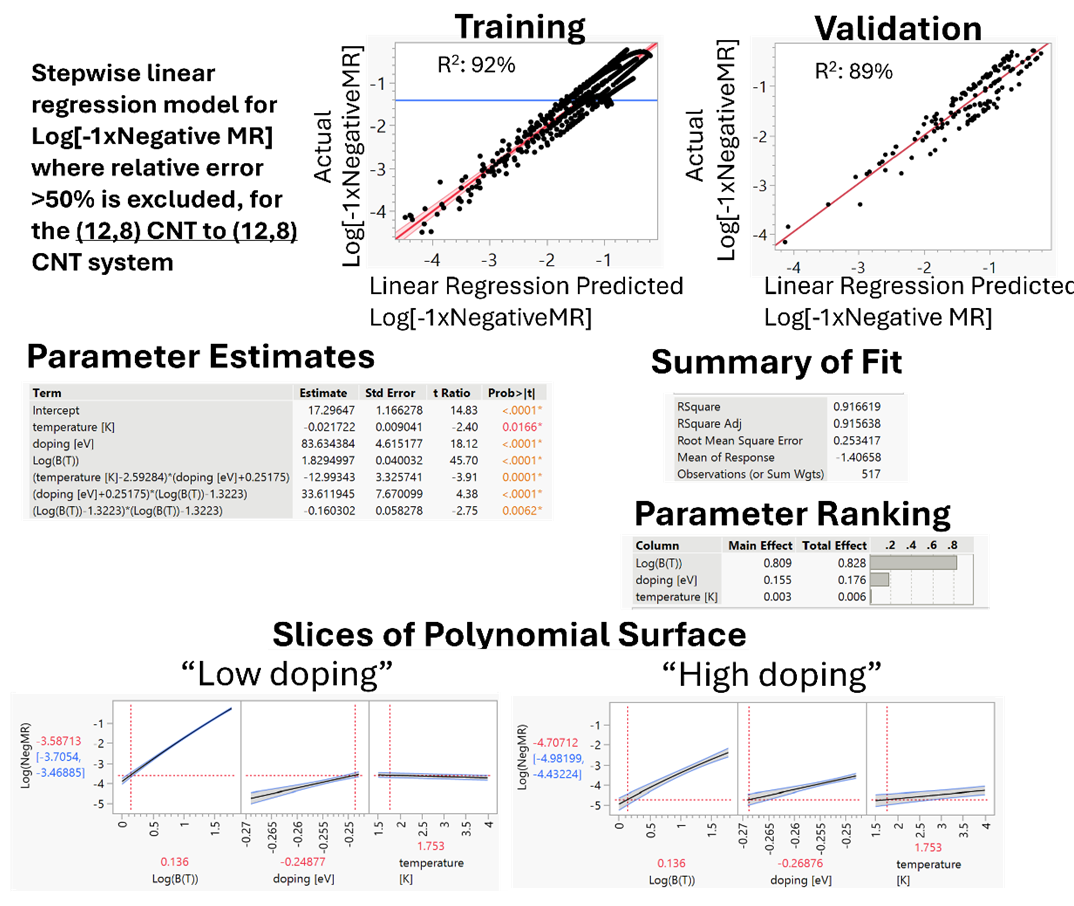}
	\caption{Linear regression response surface model of negative MR for just (12,8)+(12,8) with relative error $\le$ 50\,\%. The fitted parameter for Log[B] is 1.8.}
	\label{FigS40}
\end{figure}

\begin{figure}[h!]
	\centering
	\includegraphics[width=0.90\textwidth]{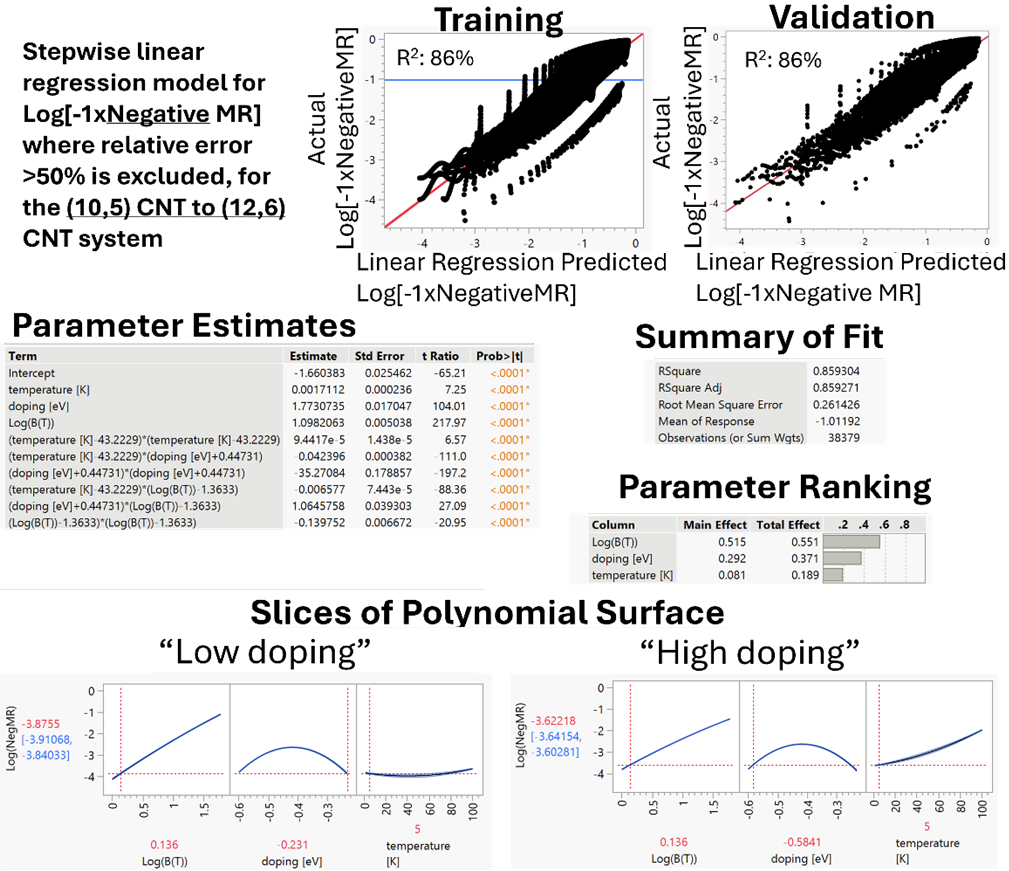}
	\caption{Linear regression response surface model of negative MR for just (10,5)+(12,6) with relative error $\le$ 50\,\%. The fitted parameter for Log[B] is 1.10. There is an interaction between temperature and doping, such that temperature becomes important at high levels of doping. }
	\label{FigS41}
\end{figure}

\FloatBarrier 
	
\subsection{Additional Modelling Results: complex connections}
	
For the loops, the data show sharp, field-selective suppressions of conductance (arrow in Supplementary Figure~\ref{FigS43}f), where G$\to$0 and MR spikes, together with multiple sign changes across doping. 	This is the hallmark of coherent multi-path interference in a cavity (Fabry--Perot-like reflections in the intermediate tubes), with the perpendicular field tuning the Peierls phases on the inter-tube hoppings; it naturally yields narrow nodes rather than a smooth cusp at B=0. In addition, loops display only a modest fraction of negative MR ($\sim$5--10\,\%), far below heterojunctions and even below simple junctions with defects, and comparable to the tiny fractions in bundles (Supplementary Table~\ref{table}). Taken together--sharp field-tuned nodes, large positive-MR excursions, and a low share of negative MR--the loop behaviour does not match the diffuse, cusp-like negative magnetoresistance expected from weak localisation in perpendicular fields. Instead, it points to phase-coherent cavity interference governed by geometry-dependent multiple scattering in the 	loop.
	
\begin{figure}[h!]
	\centering
	\includegraphics[width=0.95\textwidth]{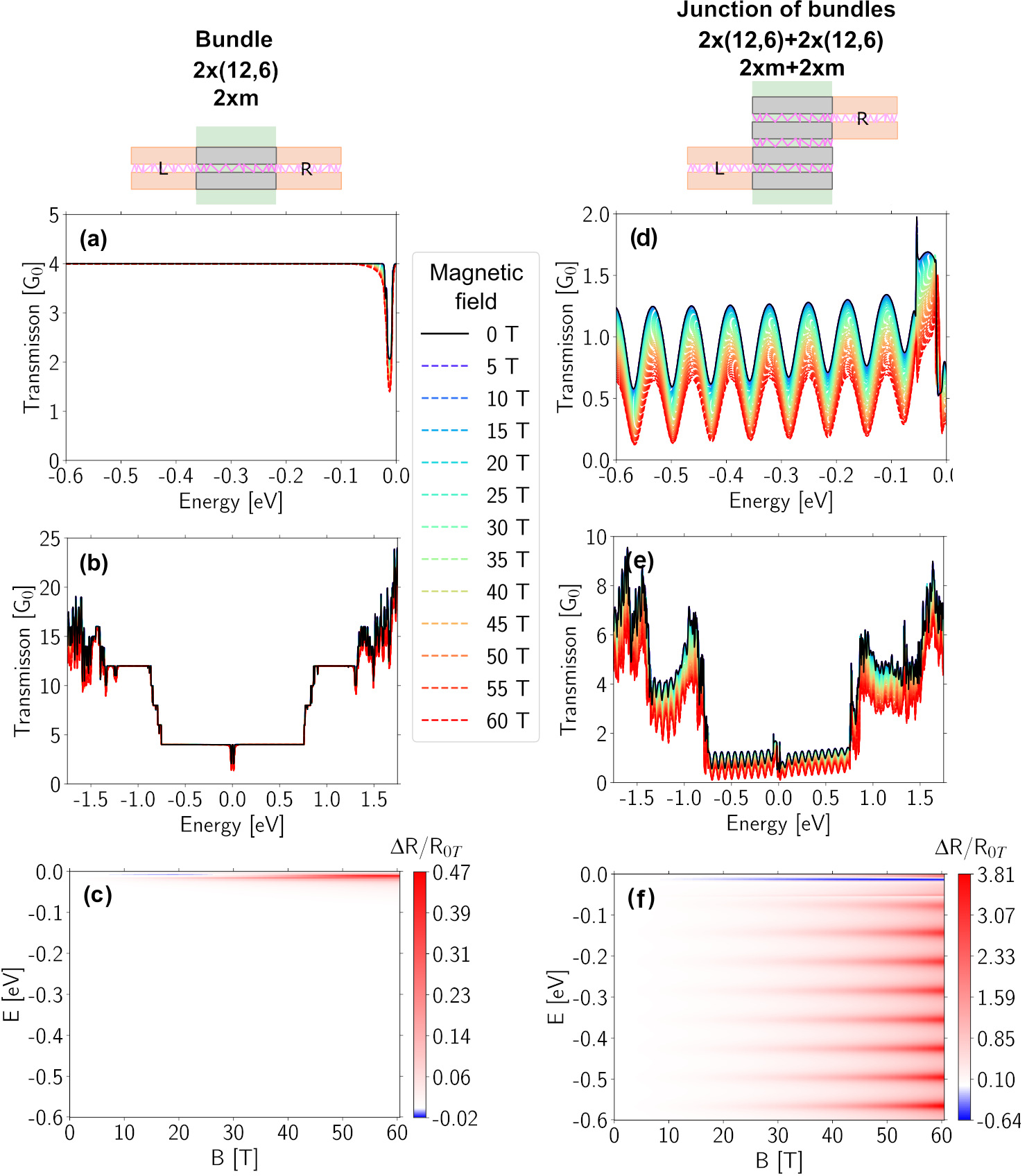}
	\caption{(a,d) The zero-bias transmission spectra, (b,e) transmission spectra in the wider energy range, and (c,f) magnetoresistance maps of (a-c) metallic CNT bundles and (d-f) metallic CNT junction of bundles computed under external perpendicular magnetic fields. The overlap regions (z) are 22.6\,nm,in both systems, corresponding to twenty unit cells of a (12,6) CNT.}
	\label{FigS42}
\end{figure}
	
\begin{figure}[h!]
	\centering
	\includegraphics[width=0.95\textwidth]{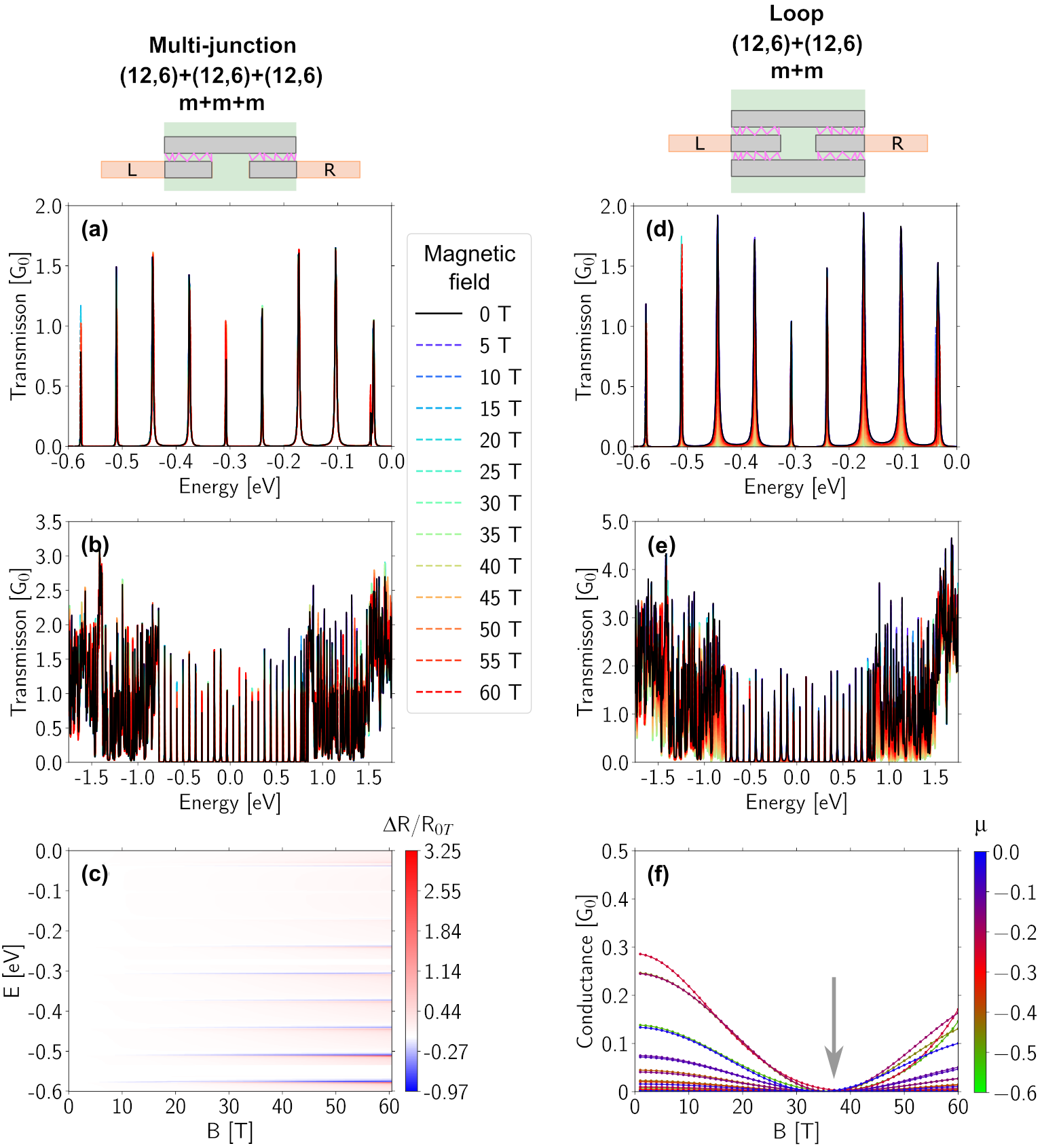}
	\caption{Zero-bias transmission (a,d), extended-range transmission spectra (b,e), and magnetoresistance maps (c) or conductance (f) of metallic homogeneous CNT multi-junctions (a--c) and metallic homogenous CNT loops (d--f) under external perpendicular magnetic fields. The overlap region (z) is 5.64\,nm in each case (five (12,6) SWCNT unit cells). The arrow in (f) marks the field at which conductance is completely suppressed.}
	\label{FigS43}
\end{figure}

Similar geometry-controlled interference signatures have also been reported in layered semimetals such as ZrSiS and MoO$_2$, where Berry-phase effects produce angular MR modulations and anisotropic suppression patterns\cite{novak2019,chen2020}. Beyond topological systems, sharp conductance features and MR sign reversals reminiscent of our loop geometries have been observed in frustrated magnetic materials such as Ce$_3$ScBi$_5$, where Kondo-lattice physics and magnetic anisotropy give rise to non-monotonic, angle-dependent MR\cite{xu2024}. Notably, similar field-sensitive MR oscillations and conductance suppression are also seen in twisted multilayer or multi-junction architectures in crystalline semimetals like ZrTe$_5$ and ReO$_3$, where multiple scattering paths and open-orbit effects modulate transport\cite{pi2024, chen2021}. While the microscopic mechanisms differ, the presence of field-tunable interference in these systems supports a broader interpretation of loops as minimal models for quantum interference in correlated networks.

\FloatBarrier 
	
\subsection{Additional Modelling Results: Key Factors Influencing Magnetotransport in the Studied Systems	}\label{secTfactors2}

\begin{figure}[h!]
	\centering
	\includegraphics[width=0.93\textwidth]{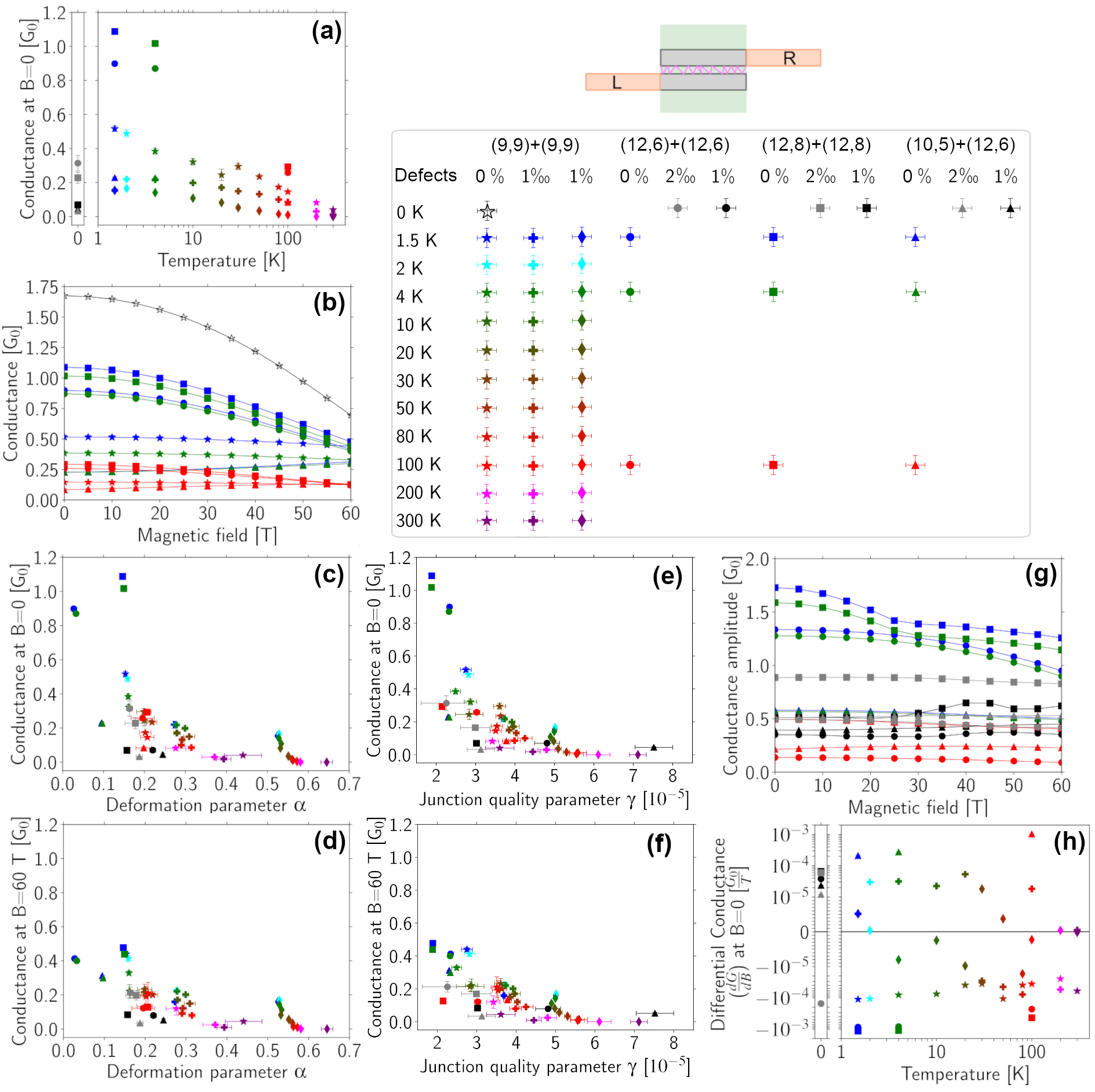}
	\caption{Key factors influencing magnetotransport in simple CNT junctions (a--h) Computed conductance and differential conductance for various CNT junctions under structural deformation, temperature variation, and magnetic field. These panels include simulations of three representative junction types with the following initial overlaps: metallic--metallic (12,6)+(12,6), z$_\mathrm{initial}$=22.6\,nm; semiconducting--semiconducting (12,8)+(12,8), z$_\mathrm{initial}$=22.3\,nm; and semiconducting--metallic (10,5)+(12,6), z$_\mathrm{initial}$=22.6\,nm. a) Temperature-dependent conductance at zero magnetic field, with temperature shown on a logarithmic scale. b) Magnetic-field-dependent conductance for pristine junctions at 1.5\,K (blue), 4\,K (green), and 100\,K (red). The empty star denotes the (9,9)+(9,9) junction at 0\,K with a 24.6\,nm overlap; filled stars indicate the same junction at finite temperatures with an initial overlap of 14.3\,nm. (c--f) Conductance as a function of the deformation parameter $\upalpha$ (c,d) and the junction quality parameter $\upgamma$ (e,f), shown without external magnetic field (c,e) and under a perpendicular field of 60,T (d,f). The junction quality parameter remains unchanged between field conditions, as structural modifications induced by the magnetic field, including possible variations in intertube separation, were not considered. The conductance shown here is averaged transmission value over the energy region extending from the nitric acid doping\cite{hayashi2020} (-0.6\,eV) to the Fermi level (0.0\,eV) for all defected CNTs, while for CNT junctions at finite temperature it is averaged over first transmission step as depicted on Supplementary Fig.~\ref{FigS5}c,d. (g) Averaged conductance amplitudes as a function of magnetic field for pristine CNT junctions at 0\,K, 1.5\,K, 4\,K, and 100\,K. h) Differential conductance dG/dB as a function of temperature at B=0\,T. For T$>$0\,K, both axes use logarithmic scaling; at T=0\,K, only the vertical axis is logarithmic, allowing comparison from absolute zero. All data in a) c), d), f) include both pristine and defected CNT junctions.}
	\label{FigS44}
\end{figure}

To enable comparison across junctions, we analysed the field dependence of the averaged conductance. It drops with B in homojunctions but rises in the heterojunction (Supplementary Fig.~\ref{FigS44}b). Because junction transmission exhibits oscillations, we also consider a conductance amplitude, defined as the amplitude of the transmission oscillations within the energy window of the first transmission step. In pristine homojunctions, it decreases monotonically with B and temperature (Supplementary Fig.~\ref{FigS44}g). With defects, the trend becomes non-monotonic: in (9,9)+(9,9), amplitude initially rises with B at low T for 0.1\% defects, then flattens at 1\% except at the lowest T (Supplementary Fig.~\ref{FigS26}g). Similar behaviour appears in other homo junctions with 1\% defects (Supplementary Fig.~\ref{FigS44}g), while at 0.2\%, non-monotonicity is limited to the metallic-metallic case. By contrast, the heterojunction shows a steady increase in amplitude with B at 100\,K. Oscillatory magnetoconductance is evident in CNT films reported by Gao et al.\cite {gao2021}, although it was not discussed; inspection of their data reveals clear B-dependent oscillations.
	
We used the differential conductance, $\frac{dG}{dB}$, to track dominant MR sign with temperature (Supplementary Fig.~\ref{FigS44}h). Positive  $\frac{dG}{dB}$ indicates negative MR, and vice versa. In perfect metallic-metallic $\frac{dG}{dB}$ is consistently positive for chiral (12,6)+(12,6) and negative for achiral (9,9)+(9,9). Introducing 0.1\% defects into the latter reverses the sign at low temperature and at 100\,K. At 1\% defects, $\frac{dG}{dB}$ oscillates weakly around zero. In semiconducting-semiconducting (12,8)+(12,8) homojunction, thermal disorder yields a negative $\frac{dG}{dB}$, but defects reverse it. The heterojunction shows a positive $\frac{dG}{dB}$ under both temperature and defect effects, with its magnitude increasing with temperature.

\FloatBarrier

\begin{sidewaystable}[p]
	\caption{The impact of various factors on magnetoresistance (MR) in selected systems: nanotube
		chirality, metallicity, overlap length, temperature, and defects. The metallic character of nanotubes is denoted as m (metallic) and s (semiconducting).  For each system, the percentage of negative MR is reported, while the minimal and maximal MR values are given in parentheses (\textcolor{blue}{minimal}; \textcolor{red}{maximal}).}
	\label{table}
	
	\footnotesize
	\setlength{\tabcolsep}{3pt}
	\renewcommand{\arraystretch}{1.35}
	
	\begin{tabular*}{0.93\textheight}{@{\extracolsep\fill} l c c c c c c c c}
		\toprule
		system & type & $z$ [nm] &
		\multicolumn{4}{c}{temperature [K]} &
		\multicolumn{2}{c}{concentration of 5--7 defects [\%]} \\
		\cmidrule(lr){4-7}\cmidrule(lr){8-9}
		&  &  & 0 & 1.5 & 4 & 100 & 0.2 & 1 \\
		\midrule
		
		\begin{tabular}[c]{@{}l@{}}Junction\\[-1pt]$(9,9){+}(9,9)$\end{tabular} & $m{+}m$ & 24.6 &
		\begin{tabular}[c]{@{}c@{}}6.23\%\\[-1pt](\textcolor{blue}{-0.29};\,\textcolor{red}{13.04})\end{tabular} &
		& & & &
		\begin{tabular}[c]{@{}c@{}}68.99\%\\[-1pt](\textcolor{blue}{-0.78};\,\textcolor{red}{1})\end{tabular} \\
		
		\begin{tabular}[c]{@{}l@{}}Junction\\[-1pt]$(9,9){+}(9,9)$\end{tabular} & $m{+}m$ & 73.8 &
		\begin{tabular}[c]{@{}c@{}}45.11\%\\[-1pt](\textcolor{blue}{-0.99};\,\textcolor{red}{83.59})\end{tabular} &
		& & & & \\
		
		\begin{tabular}[c]{@{}l@{}}Junction\\[-1pt]$(12,6){+}(12,6)$\end{tabular} & $m{+}m$ & 22.6 &
		\begin{tabular}[c]{@{}c@{}}0.0\%\\[-1pt](\textcolor{blue}{0};\,\textcolor{red}{3.15})\end{tabular} &
		\begin{tabular}[c]{@{}c@{}}0.0\%\\[-1pt](\textcolor{blue}{0};\,\textcolor{red}{3.10})\end{tabular} &
		\begin{tabular}[c]{@{}c@{}}0.0\%\\[-1pt](\textcolor{blue}{0};\,\textcolor{red}{3.02})\end{tabular} &
		\begin{tabular}[c]{@{}c@{}}0.62\%\\[-1pt](\textcolor{blue}{$-3.6\times10^{-4}$};\,\textcolor{red}{1.4})\end{tabular} &
		\begin{tabular}[c]{@{}c@{}}3.16\%\\[-1pt](\textcolor{blue}{-0.21};\,\textcolor{red}{2.04})\end{tabular} &
		\begin{tabular}[c]{@{}c@{}}64.08\%\\[-1pt](\textcolor{blue}{-0.94};\,\textcolor{red}{38.48})\end{tabular} \\
		
		\begin{tabular}[c]{@{}l@{}}Junction\\[-1pt]$(12,8){+}(12,8)$\end{tabular} & $s{+}s$ & 22.3 &
		\begin{tabular}[c]{@{}c@{}}5.24\%\\[-1pt](\textcolor{blue}{-1.12};\,\textcolor{red}{125.01})\end{tabular} &
		\begin{tabular}[c]{@{}c@{}}2.05\%\\[-1pt](\textcolor{blue}{-0.69};\,\textcolor{red}{3.55})\end{tabular} &
		\begin{tabular}[c]{@{}c@{}}2.30\%\\[-1pt](\textcolor{blue}{-0.45};\,\textcolor{red}{3.59})\end{tabular} &
		\begin{tabular}[c]{@{}c@{}}0.0\%\\[-1pt](\textcolor{blue}{0};\,\textcolor{red}{2.63})\end{tabular} &
		\begin{tabular}[c]{@{}c@{}}56.92\%\\[-1pt](\textcolor{blue}{-1};\,\textcolor{red}{8.7})\end{tabular} &
		\begin{tabular}[c]{@{}c@{}}71.87\%\\[-1pt](\textcolor{blue}{-1};\,\textcolor{red}{13.51})\end{tabular} \\
		
		\begin{tabular}[c]{@{}l@{}}Junction\\[-1pt]$(10,5){+}(12,6)$\end{tabular} & $s{+}m$ & 22.6 &
		\begin{tabular}[c]{@{}c@{}}75.37\%\\[-1pt](\textcolor{blue}{-0.94};\,\textcolor{red}{13.06})\end{tabular} &
		\begin{tabular}[c]{@{}c@{}}87.79\%\\[-1pt](\textcolor{blue}{-0.92};\,\textcolor{red}{0.78})\end{tabular} &
		\begin{tabular}[c]{@{}c@{}}87.73\%\\[-1pt](\textcolor{blue}{-0.93};\,\textcolor{red}{2.53})\end{tabular} &
		\begin{tabular}[c]{@{}c@{}}94.42\%\\[-1pt](\textcolor{blue}{-0.81};\,\textcolor{red}{0.1})\end{tabular} &
		\begin{tabular}[c]{@{}c@{}}64.52\%\\[-1pt](\textcolor{blue}{-0.86};\,\textcolor{red}{3.72})\end{tabular} &
		\begin{tabular}[c]{@{}c@{}}66.04\%\\[-1pt](\textcolor{blue}{-1};\,\textcolor{red}{28.80})\end{tabular} \\
		
		\begin{tabular}[c]{@{}l@{}}Junction\\[-1pt]$(11,8){+}(16,1)$\end{tabular} & $m{+}m$ & 23.5 &
		\begin{tabular}[c]{@{}c@{}}51.34\%\\[-1pt](\textcolor{blue}{-0.84};\,\textcolor{red}{3.36})\end{tabular} &
		& & & &
		\begin{tabular}[c]{@{}c@{}}65.25\%\\[-1pt](\textcolor{blue}{-1};\,\textcolor{red}{2.96})\end{tabular} \\
		
		\begin{tabular}[c]{@{}l@{}}multi-junction\\[-1pt]$(12,6){+}(12,6){+}(12,6)$\end{tabular} & $m{+}m{+}m$ & 5.64 &
		\begin{tabular}[c]{@{}c@{}}10.39\%\\[-1pt](\textcolor{blue}{-0.97};\,\textcolor{red}{3.25})\end{tabular} &
		& & & & \\
		
		\begin{tabular}[c]{@{}l@{}}bundle\\[-1pt]$(12,6){+}(12,6)$\end{tabular} & $m{+}m$ & 22.6 &
		\begin{tabular}[c]{@{}c@{}}0.38\%\\[-1pt](\textcolor{blue}{-0.02};\,\textcolor{red}{0.47})\end{tabular} &
		& & & & \\
		
		\begin{tabular}[c]{@{}l@{}}junction of bundles\\[-1pt]$2{\times}(12,6){+}2{\times}(12,6)$\end{tabular} & $2{\times}m{+}2{\times}m$ & 22.6 &
		\begin{tabular}[c]{@{}c@{}}1.76\%\\[-1pt](\textcolor{blue}{-0.64};\,\textcolor{red}{3.81})\end{tabular} &
		& & & & \\
		
		\begin{tabular}[c]{@{}l@{}}loop\\[-1pt]$(12,6)$\end{tabular} & $m{+}m{+}m{+}m$ & 1.13 &
		\begin{tabular}[c]{@{}c@{}}10.48\%\\[-1pt](\textcolor{blue}{-0.99};\,\textcolor{red}{1368.78})\end{tabular} &
		& & & & \\
		
		\begin{tabular}[c]{@{}l@{}}loop\\[-1pt]$(12,6)$\end{tabular} & $m{+}m{+}m{+}m$ & 3.38 &
		\begin{tabular}[c]{@{}c@{}}9.19\%\\[-1pt](\textcolor{blue}{-0.89};\,\textcolor{red}{6936.00})\end{tabular} &
		& & & & \\
		
		\begin{tabular}[c]{@{}l@{}}loop\\[-1pt]$(12,6)$\end{tabular} & $m{+}m{+}m{+}m$ & 5.64 &
		\begin{tabular}[c]{@{}c@{}}4.98\%\\[-1pt](\textcolor{blue}{-0.91};\,\textcolor{red}{5990.86})\end{tabular} &
		& & & & \\
		
		\botrule
	\end{tabular*}
\end{sidewaystable}

\FloatBarrier

\end{appendices}


\bibliography{sn-bibliography}

\end{document}